\documentclass[twoside,12pt,a4paper]{article}
\usepackage[a4paper]{geometry}
\usepackage{epsfig}

\def\Journal#1#2#3#4{{#1} {#2} (#4) #3 }

\def\NPHYS{\em Nucl. Phys.}
\def\NPA{{\em Nucl. Phys.} A}
\def\NPB{{\em Nucl. Phys.} B}

\def\PLB{{\em Phys. Lett.} B}

\def\PREV{\em Phys. Rev.}

\def\PRD{{\em Phys. Rev.} D}
\def\PRC{{\em Phys. Rev.} C}

\def\ZPHY{\em Z. Phys.}
\def\ZPC{{\em Z. Phys.} C}

\def\ANNP{\em Ann. Phys. (N.Y.)}

\def\EURA{{\em Eur. Phys. J} A}

\def\ss{\mbox{\boldmath $\sigma$}}
\def\se{\mbox{\boldmath $\epsilon$}}
\def\sg{\mbox{\boldmath $\gamma$}}
\newcommand{\be}{\begin{equation}}
\newcommand{\ee}{\end{equation}}
\newcommand{\bea}{\begin{eqnarray}}
\newcommand{\eea}{\end{eqnarray}}
\newcommand{\nn}{\nonumber}
\begin{document}
\title{ \vspace{1cm} Electroexcitation of 
nucleon resonances}
\author{I.G.\ Aznauryan$^{1,2}$ and V.D.\ Burkert$^1$ \\
\\
$^1$Thomas Jefferson National
Accelerator Facility, Newport News, Virginia 23606, USA\\
$^2$Yerevan Physics Institute, Yerevan 0036, Armenia}
\maketitle
\begin{abstract} 
We review recent progress in the investigation of the
electroexcitation of nucleon resonances, both in experiment
and in theory. We describe current experimental facilities, the experiments 
performed on $\pi$ and $\eta$ electroproduction off
protons, and theoretical approaches used for the extraction
of resonance contributions from the experimental data. 
The status of $2\pi$, $K\Lambda$, and $K\Sigma$ electroproduction  
is also presented. The most accurate results have been obtained
for  the electroexcitation amplitudes of
the four lowest excited states, 
which have been measured in a range of $Q^2$
up to $8$ and $4.5~$GeV$^2$ 
for the $\Delta(1232)P_{33}$, $N(1535)S_{11}$ and
$N(1440)P_{11}$, $N(1520)D_{13}$, respectively.
These results have been confronted with calculations based
on lattice QCD, large-$N_c$ relations,
perturbative QCD (pQCD), and QCD-inspired models. 
The amplitudes for the $\Delta(1232)$ indicate large pion-cloud contributions 
at low $Q^2$ and don't show any sign of approaching the pQCD regime for
$Q^2<7~$GeV$^2$.
Measured for the first time, the electroexcitation 
amplitudes of the Roper resonance, $N(1440)P_{11}$,
provide strong evidence for this 
state as a predominantly radial excitation  
of a three-quark (3$q$) 
ground state, with additional non-3-quark
contributions needed to describe the low 
$Q^2$ behavior of the amplitudes. 
The longitudinal transition amplitude
for the $N(1535)S_{11}$ was determined and
has become a challenge for quark models. Explanations may require 
large meson-cloud contributions or
alternative representations of this state.
The $N(1520)D_{13}$ clearly shows the rapid changeover from
helicity-3/2 dominance at the real photon point to helicity-1/2
dominance at $Q^2 > 0.5~$GeV$^2$, confirming a long-standing prediction
of the constituent quark model. 
The interpretation of the
moments of resonance transition form factors
in terms of transition transverse charge distributions in 
infinite momentum frame
is presented. 
\end{abstract}

\section{Introduction}
The excitation of nucleon resonances in electromagnetic interactions
has long been recognized as an important source 
of information for understanding 
strong interactions
in the domain of quark confinement. 
Theoretical and experimental investigations of the 
electroexcitation of nucleon resonances
have a long history. Early investigations in the 1960's to 1980's were based on
experiments at the electron beam accelerators  
 DESY at Hamburg in Germany, NINA at Daresbury in the UK, 
and at the University of Bonn in Germany.  At the real photon point, 
systematic measurements
were made at these facilities (including also the electron 
accelerators at Yerevan and Char'kov), 
which included a variety of polarization experiments along with measurements 
of differential cross sections.  
Phenomenological analyses of the data were able to extract information 
on the $\gamma N\rightarrow N^*$ transition amplitudes for the well-established 
resonances with masses below $2~$GeV. The situation for virtual photons was different;
only sparse data on differential cross sections for
the reactions $\gamma^* N\rightarrow \pi N$ and
$\gamma^* p\rightarrow \eta p$ were obtained for
photon virtualities up to $Q^2=3~$GeV$^2$. 
The data provided limited information,
with large systematic differences among the various data sets,
on the magnetic-dipole
$\gamma^* N\rightarrow \Delta(1232)P_{33}$ amplitude and
on the transverse $\gamma^* N\rightarrow N(1520)D_{13}$,
$N(1535)S_{11}$, and $N(1680)F_{15}$ amplitudes.
One of the most interesting questions for the $\Delta(1232)P_{33}$
is its possible 
quadrupole deformation, which can be revealed 
through measurements of the 
electric-quadrupole and
scalar $\gamma^* N\rightarrow \Delta(1232)P_{33}$ amplitudes.
The sensitivity of the data to these amplitudes was 
very limited, as was their sensitivity to the 
$\gamma^* N\rightarrow N(1520)D_{13}$,
$N(1535)S_{11}$, and $N(1680)F_{15}$
longitudinal amplitudes; in fact, these quantities have 
not been determined. The theoretical scheme for the interpretation of the 
$\gamma(\gamma^*) N\rightarrow N^*$ amplitudes extracted
from experimental data in the 1960's to the 1980's was based on the
constituent quark model (CQM) and the single quark transition model (SQTM).
A review of these early data, the extracted amplitudes, and their theoretical 
interpretation at this stage of electroproduction
experiments can be found in Refs. \cite{PDG82,Foster}.

The experimental situation changed dramatically with the advent of the
new generation of electron beam facilities - the Continuous
Electron Beam Accelerator Facility (CEBAF) at the 
Thomas Jefferson National Accelerator Facility
(JLab),  Mainz Microtron (MAMI) at Mainz,
and the MIT/Bates out-of-plane
scattering (OOPS) facility.
Large amounts of significantly more precise and complete data 
were collected, in both pion and eta electroproduction off protons
in the first, second, and third resonance regions 
in the range of $Q^2 < 8$~GeV$^2$.
For pion electroproduction, measurements 
of differential cross sections along with a variety of polarization experiments
were performed. 
The list of new measurements of pion and eta electroproduction 
is given in Table \ref{tab:data}.
The majority of new data was obtained at JLab,
in particular with the CEBAF Large Acceptance Spectrometer (CLAS)
in Hall B. 
The MAMI and MIT/Bates experiments consist of measurements of $ep\rightarrow 
ep\pi^0$ and 
$\vec{e}p\rightarrow e\vec{p}\pi^0$  
in the vicinity of the
$\Delta(1232)P_{33}$ resonance at small 
$Q^2 < 0.2$~GeV$^2$. 
Due to the new measurements, for the first time, electrocoupling amplitudes
of the Roper resonance $N(1440)P_{11}$
have been extracted from experimental data, as well
the electric-quadrupole and
scalar $\gamma^* N\rightarrow \Delta(1232)P_{33}$ amplitudes
and the 
$\gamma^* p\rightarrow N(1520)D_{13}$ and
$N(1535)S_{11}$ 
longitudinal amplitudes. Overall, accurate results
have been obtained for the  amplitudes of the
$\gamma^* p\rightarrow \Delta(1232)P_{33}$ and 
$N(1535)S_{11}$ transitions up to $Q^2=8~$GeV$^2$,
and of
the $\gamma^* p\rightarrow N(1440)P_{11}$ and $N(1520)D_{13}$ 
transitions up to $Q^2=4.5~$GeV$^2$.
Experimental and theoretical advances on early stages
of these investigations 
are reviewed in Refs. \cite{BurkertLee,SmithLee}.

Progress in the experimental investigation of the
electroexcitation of nucleon resonances 
was accompanied by significant developments in
understanding of QCD, including the domain of quark confinement.
This made it possible in some cases
to set relations between 
the properties of QCD found from first principles and 
the amplitudes extracted from experimental data.
Below we list those relations that are directly 
connected to the results on the $\gamma^* N\rightarrow N^*$
amplitudes discussed in this review.

Spontaneous chiral
symmetry breaking in QCD leads to the existence
of nearly massless Goldstone bosons (pions). As a consequence, there can be 
significant pion-loop contributions to the electromagnetic form factors 
at relatively small
momentun transfer. These contributions are crucial for the description of
the neutron electric form factor
in CQM and bag models
\cite{Lu,Miller,Agbakpe,Faessler,Cloet} and are essential
for the $\gamma^*N\rightarrow \Delta(1232)P_{33}$ transition
amplitudes \cite{Bermuth,Fiolhais,Thomas,Faessler1}.
The importance of the pion-cloud contributions
to the transition form factors
has been confirmed by lattice QCD calculations \cite{Alexandrou},
where at small $Q^2$ they modify the quenched results
in agreement with expectations from chiral perturbation
theory \cite{Pascalutsa}.
The meson-cloud
contribution is also identified as a
source of the long-standing discrepancy
between the data and CQM
predictions for the $\gamma^*N\rightarrow \Delta(1232)P_{33}$
magnetic-dipole amplitude 
within dynamical reaction models \cite{Kamalov1999,Kamalov2001,Sato2001,Sato2007}. 
From the results presented in this review it will be seen that
complementing of the quark core contribution
by that of the pion cloud can be necessary also for the correct
description of  the 
$\gamma^* p\rightarrow N(1440)P_{11}$, 
$N(1520)D_{13}$, and
$N(1535)S_{11}$ 
amplitudes extracted
from experimental data.

\begin{table}
\begin{center}
\begin{minipage}[t]{16.5 cm}
\caption{List of $ep\rightarrow eN\pi,eN\eta$ measurements
at JLab, MAMI, and MIT/Bates.}
\label{tab:data}
\end{minipage}
\begin{tabular}{llllll}
\hline
Facility&Observable &$Q^2$ (GeV$^2$)&$W$ (GeV)&Ref.\\
\hline
JLab/Hall A
&$\frac{d\sigma}{d\Omega}~(\pi^0 p)~$&$1.0$
&1.1 - 1.95&\cite{Laveis}\\
&Response functions&&&\\
&for $\vec{e}p\rightarrow e\vec{p}\pi^0$&$1.0$
&1.17 - 1.35&\cite{KELLY1,KELLY2}\\
\hline
JLab/Hall B&
$\frac{d\sigma}{d\Omega}~(\pi^0 p,\pi^+ n)$&0.16 - 0.36 
&1.1 - 1.38&\cite{Cole}\\
&$\frac{d\sigma}{d\Omega}~(\pi^0 p)$&0.4 - 1.8 
&1.1 - 1.68&\cite{Joo1}\\
&$\frac{d\sigma}{d\Omega}~(\pi^0 p)$&3.0 - 6.0
&1.1 - 1.39&\cite{Ungaro}\\
&$A_{LT'}~(\pi^0 p)$& 0.4, 0.65 
&1.1 - 1.66&\cite{Joo2}\\
&$A_t,~A_{et}~(\pi^0 p)$&0.252, 0.385, 0.611 
&1.12 - 1.55&\cite{Biselli}\\
&$\frac{d\sigma}{d\Omega}~(\pi^+ n)$&0.3 - 0.6 
&1.1 - 1.55&\cite{Egiyan}\\
&$\frac{d\sigma}{d\Omega},A_{LT'}~(\pi^+ n)$&1.7 - 4.5&1.11 - 1.69&\cite{Park}\\
&$A_{LT'}~(\pi^+ n)$& 0.4, 0.65 
&1.1 - 1.66&\cite{Joo3}\\
&$\frac{d\sigma}{d\Omega}~(\eta p)$&0.375 - 1.385 
&1.5 - 1.86&\cite{Thompson}\\
&$\frac{d\sigma}{d\Omega}~(\eta p)$&0.17 - 3.1
&1.5 - 2.3&\cite{Denizli}\\
\hline
JLab/Hall C&$\frac{d\sigma}{d\Omega}~(\pi^0 p)~$&2.8, 4.2 
&1.115 - 1.385&\cite{Frolov}\\
&$\frac{d\sigma}{d\Omega}~(\pi^0 p)~$&6.4,  7.7 
&1.11 - 1.39&\cite{Vilano}\\
&$\frac{d\sigma}{d\Omega}~(\eta p)$&2.4 3.6
&1.49 - 1.62&\cite{Armstrong}\\
&$\frac{d\sigma}{d\Omega}~(\eta p)$&5.7 7.0&1.5 - 1.8&\cite{Dalton}\\
\hline
MAMI
&$\frac{d\sigma}{d\Omega},A_{LT'}~(\pi^0 p)~$&$0.06 - 0.2$
&1.22 - 1.3&\cite{Stave2006,Sparveris2007,Stave2008}\\
&$P~(\vec{e}p\rightarrow ep\pi^0)$&$0.121$
&1.23&\cite{Pospishil}\\
&$A_{LT'}~(\pi^0 p)$&$0.2$
&1.232&\cite{Bartsch}\\
&$\sigma_{LT}~(\pi^0 p)$&$0.2$
&1.232&\cite{Elsner}\\
\hline
MIT/Bates
&$\frac{d\sigma}{d\Omega}~(\pi^0 p)~$&$0.127$
&1.23&\cite{Mertz,Kunz,Sparveris2005}\\
&$P~(ep\rightarrow ep\pi^0)$&$0.126$
&1.232&\cite{Warren}\\
\hline
\end{tabular}
\begin{minipage}[t]{16.5 cm}
\vskip 0.5cm
\noindent
$A_{LT'}$ is a longitudinally polarized beam
asymmetry for $\vec{e}p\rightarrow eN\pi$,
$A_t$ and $A_{et}$ are longitudinal-target and
beam-target asymmetries for $\vec{e}\vec{p}\rightarrow ep\pi^0$,
$P$ is a polarization of the
final proton in the corresponding reactions,
and $\sigma_{LT}$ is a longitudinal-transverse structure function.
\end{minipage}
\end{center}
\end{table}

The $1/N_c$ expansion introduced by 't Hooft \cite{Hooft}
and Witten \cite{Witten} has been shown to be a powerful tool
for exposing properties of QCD in the non-perturbative domain.
It led to the understanding of
baryon properties, such as ground-state and excited
baryon masses, as well as their magnetic moments and electromagnetic
transitions (see Refs. \cite{Stancu,Goity1,Goity2,Jenkins}
 and references therein). In this 
review we will demonstrate good agreement
between the $\gamma^*N\rightarrow \Delta(1232)P_{33}$
amplitudes extracted from experimental data and recent predictions
obtained in the large $N_c$ limit
\cite{Pascalutsa1,Pascalutsa2,Grigoryan}. 
The predictions are made for a wide range
of $Q^2$. In particular, for the magnetic-dipole
$\gamma^* N\rightarrow \Delta(1232)P_{33}$ amplitude,  
they extend up to $Q^2=6-8~$GeV$^2$. 

In recent years there has been significant progress 
in lattice QCD calculations by using a number of different 
fermion discretization schemes and pion masses 
reaching closer to the physical pion mass (the review can be found
in Ref. \cite{AlexBeij}). Significant effort has been made to
get consistent results for the benchmark $\gamma^*N\rightarrow 
\Delta(1232)P_{33}$ transition. Recent predictions
have been shown to be quite definite and in qualitative agreement
with experimental data \cite{Alexandrou}.
There are also first exploratory calculations of the 
$\gamma^* p\rightarrow N(1440)P_{11}$ amplitudes \cite{Lin},
which need improvement using smaller pion mass values 
and employ an unquenched approximation.

Another approach, which can be considered as a tool
that relates the first-principles properties of QCD
to the $\gamma^* N\rightarrow 
N^*$ amplitudes, is presented in Ref. \cite{Braun}.
In this approach, the 
$N(1535)S_{11}$ light-cone distribution amplitudes
found through lattice calculations have been used to calculate
the $\gamma^* p\rightarrow N(1535)S_{11}$ transition amplitudes
by utilization of light-cone sum rules. At $Q^2>2~$GeV$^2$,
the predictions are in quite good agreement with the amplitudes 
extracted from experimental data.

The CQM remains a useful tool
for understanding of the internal structure 
of hadrons and of their interactions.
The majority of 
experimentally observed hadrons
can be classified according to the group $SU(6)\otimes O(3)$.
The string model for confinement forces plus the associated spin-orbit
interactions, as well as the interactions expected
from the one-gluon exchange between quarks, approximately describe
the mass spectrum of hadrons \cite{Godfrey_Isgur,Capstick_Isgur} 
and 
their widths \cite{Godfrey_Isgur,Capstick_Roberts}.
However, there are well known shortcomings of this picture.
These include the wrong mass ordering
between the $N(1440)P_{11}$ and $N(1535)S_{11}$, 
difficulties in the description of
large width of the $N(1440)P_{11}$, and the large 
branching ratio of the $N(1535)S_{11}$ to the $\eta N$ channel.
It was demonstrated in Refs. \cite{Riska,An} that extension of the quark model
by inclusion of the lowest lying $qqqq\bar{q}$ components 
can in principle overcome these problems. For example, agreement
with the empirical value of the $\pi N$ decay width for the $N(1440)P_{11}$ 
can be reached with an $\sim 30\%$ $qqqq\bar{q}$ component in this
state \cite{Riska}. For the $N(1535)S_{11}$, it was found that the 
most likely lowest energy configuration is given by
the $qqqs\bar{s}$ component \cite{An}. This could solve the problem
of mass ordering between the $N(1440)P_{11}$ and $N(1535)S_{11}$, 
and explain the large couplings of the $N(1535)S_{11}$
to $\eta N$, as well as the
recently observed large couplings of this state  
to the $\phi N$ and $K\Lambda$ channels \cite{Xie,Liu}.

To deal with the shortcomings of the CQM in the case of 
the $N(1440)P_{11}$, an alternative description of this resonance
was proposed by treating it as a
hybrid $q^3G$ state \cite{Li1,Li2}. 
This possibility  was motivated by the fact that in the bag model
the lightest hybrid state has quantum numbers of
the Roper resonance, and its mass can be $< 1.5~$GeV \cite{Close}.
Another alternative representation of the nucleon
resonances, including the $N(1440)P_{11}$ and $N(1535)S_{11}$,
is the possibility that they are meson-baryon molecules
generated in chiral coupled-channel
dynamics \cite{Weise,Krehl,Nieves,Oset1,Lutz}.

In this review we present and discuss the predictions from alternative 
approaches 
for the $\gamma^* p\rightarrow N(1440)P_{11}$ and $N(1535)S_{11}$ 
transitions,  as well as the results of
extended versions of  the CQM. This will allow us to draw 
some conclusions as to the internal structure  
and nature of these resonances.

The information on the $\gamma^* p\rightarrow 
\Delta(1232)P_{33}$, $N(1440)P_{11}$, $N(1520)D_{13}$, and 
$N(1535)S_{11}$ 
transition amplitudes, extracted from experimental data
in a wide range of $Q^2$, is of great interest for
understanding of the $Q^2$ scale where the asymptotic
domain of QCD may set in for these transitions.
QCD in the asymptotic limit puts clear
restrictions on the $Q^2$ behavior of the transition
amplitudes. They follow from hadron helicity
conservation \cite{Brodsky1} and dimensional
counting rules \cite{Matveev,Brodsky2,Brodsky3,Brodsky4,Efremov}.
We compare the $Q^2$ dependence of the amplitudes extracted
from experimental data with the predictions of 
pQCD. 

Empirical knowledge of the transition amplitudes 
in a wide range of $Q^2$ also allows 
mapping out of the quark transverse charge distributions
that induce these transitions \cite{Carlson_Van,Tiator_Van,Tiator_Van1}.
These distributions will be presented and discussed
in the review.

The results presented in this review are related
mostly to the $\gamma^* p\rightarrow
\Delta(1232)P_{33}$, $N(1440)P_{11}$, $N(1520)D_{13}$, and
$N(1535)S_{11}$
transition amplitudes extracted in $\pi$ and $\eta$ electroproduction.
Recently published CLAS measurements 
\cite{Ripani,Fedotov09} present significant
progress in the investigation of two-pion electroproduction, which  
is one of the biggest contributors to the process 
of electroproduction in the resonance energy region.
This channel becomes increasingly important 
for high-lying resonances with masses above $1.6~$GeV.
Evaluation of the $\gamma^*NN^*$ electrocouplings from the CLAS
two-pion electroproduction data is now in progress.
There are already preliminary results that
may be found in Refs. \cite{Mo091,Mo10,Mokeev_2010} and will be shown
when presenting the results 
extracted from $\pi$ and $\eta$ electroproduction.
Two-pion electroproduction as well electroproduction of
$K\Lambda$ and $K\Sigma$ are intensively investigated
with CLAS at JLab \cite{Carman2003,Carman2008,Carman2009,Ambrozewicz}. 
These are 
channels with potential for the
discovery of some of the so-called ``missing" resonances,
the states that are predicted by the CQM, however, are weakly
coupled to $\pi N$ and $\eta N$ \cite{Koniuk_Isgur}, 
and by this reason are not observed in
$\pi N$ and $\eta N$ production. According to the
quark model predictions \cite{Capstick_Roberts,Capstick_Roberts1},
some of these resonances may be more efficiently studied  
in the photo- and electroproduction of $\pi\pi N$,
$K\Lambda$ and $K\Sigma$ systems. 

The paper is organized as follows. In section \ref{sec:facilities}, we 
present the facilities and setups
where the electroexcitation of nucleon resonances reported
in this review have been investigated. In section \ref{sec:definitions}, we present 
the definitions related
to the kinematics and formalism of the reaction $eN\rightarrow e N\pi$.
Special attention is paid to the relations between different definitions
of the $\gamma^* N\rightarrow N^*$ helicity amplitudes:
through the $\gamma^*N\rightarrow N\pi$
multipole amplitudes, through the matrix elements of the electromagnetic
current, and through the $\gamma^* N\rightarrow N^*$ form factors.
It is known that the $\gamma^* N\rightarrow N^*$ helicity amplitudes
extracted from experimental data include the
sign of the $N^*\rightarrow N\pi$ vertex.
We present explicit relations that account for this sign. 
In section \ref{sec:theory}, we give
a brief review of theoretical approaches that are employed in the analyses
of photo- and electroproduction reactions in the resonance energy region.  
Approaches that have been used in the extraction of the electroexcitation
amplitudes reported in this review are presented in more detail.
In section \ref{sec:data}, we describe the experiments
performed on the new generation of electron accelerators,  
list the approaches used in the analyses of the experimental data, and
present examples of the theoretical description of the data. 
The main results are discussed in section \ref{sec:results}. Here we present 
the $\gamma^* p\rightarrow
\Delta(1232)P_{33}$, $N(1440)P_{11}$, $N(1520)D_{13}$, and
$N(1535)S_{11}$ transition amplitudes as determined in the most recent analyses 
of the new data, and discuss the progress
achieved due to the new experiments. We also perform some detailed comparison with 
theoretical models, including  developments 
in understanding of QCD in the domain
of quark confinement. We also discuss results related to the 
quark transverse charge distributions in the transitions
and conclusions on the approach to the pQCD asymptotic regime. 
Finally, in section \ref{sec:third}, we present and discuss results
related to the third resonance region, before we conclude with some future prospects 
in section \ref{sec:outlook}. 

\section{Experimental Facilities
\label{sec:facilities}}

\subsection{\it Thomas Jefferson National Accelerator Facility 
\label{sec:jlab}}
\noindent
The Thomas Jefferson National Accelerator Facility 
in Newport News, Virginia, 
operates a CW electron accelerator with energies in the range up to 6 
GeV~\cite{grunder}. 
Three experimental Halls receive highly polarized electron beams with 
the same energies or with 
different but correlated energies. Beam currents in the range from 
0.1 nA to 
150 $\mu$A can be delivered to the experiments simultaneously. 
In addition, the development of polarized nucleon targets that
can be used in fairly intense electron beams, as well as use of
recoil polarimeters in magnetic spectrometers, has provided access
to a previously unavailable set of observables that are sensitive to
the interference of resonant and non-resonant processes.

\subsubsection{\it Experimental Hall A - HRS$^2$
\label{sec:halla}}
Hall A houses a pair of identical focusing 
high resolution magnetic spectrometers (HRS$^2$) \cite{halla}, each 
with a momentum resolution of $\Delta p /p \sim 2\times 10^{-4}$; one of them 
is instrumented with a gas \^Cerenkov counter and a shower counter for 
the identification of electrons. The hadron arm is instrumented with 
a proton recoil polarimeter. The detector package allows 
identification of charged pions, kaons, and protons.  
The pair of spectrometers can be operated at very high beam currents of up to 
100 $\mu$A.
The HRS$^2$ spectrometers have been used to measure the 
reaction $\vec{e} p \rightarrow e\vec{p}\pi^0$ 
in the $\Delta(1232)P_{33}$ resonance region \cite{KELLY1,KELLY2}.
The excellent momentum resolution allows efficient
use of the ``missing mass" technique, where the undetected $\pi^0$ is inferred 
from 
the overdetermined kinematics. 
Due to the small angle and momentum acceptance, the angle and momentum settings 
have to be changed many times to cover the full kinematical range of interest. 
These data have been used to extract a large number of single and double polarization 
response functions for specific kinematics. 
    
\subsubsection{\it Experimental Hall B - CLAS
\label{sec:clas}}

\begin{figure}[tb]
\begin{center}
\begin{minipage}[l]{120pt}
\epsfig{file=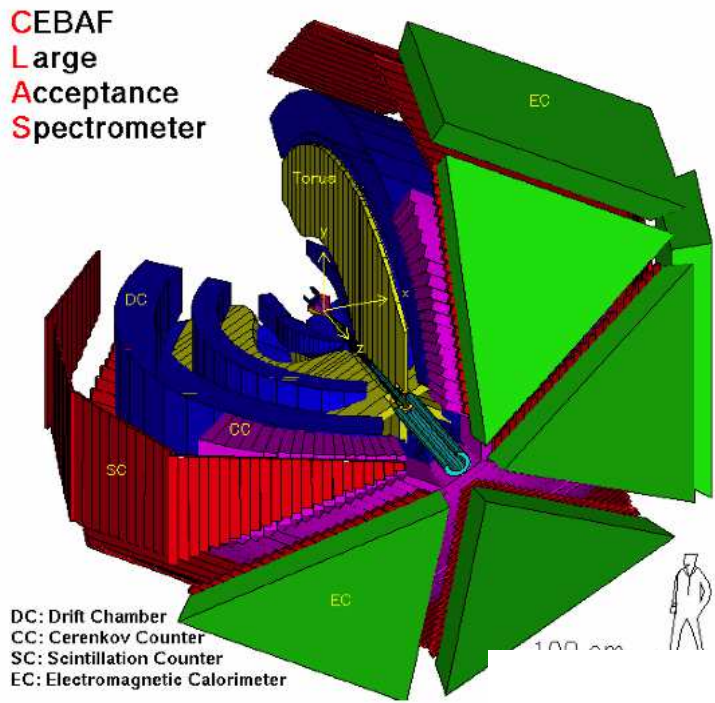,scale=0.5}
\end{minipage}
\hskip 220pt
\begin{minipage}[r]{150pt}
\caption{View of the CLAS detector at JLab. Several detector elements 
have been omitted for clarity.  
\label{clas1}}
\end{minipage}
\end{center}
\end{figure}

Hall B houses the CEBAF Large Acceptance Spectrometer (CLAS)
detector and a photon energy tagging facility~\cite{clas}.
CLAS can be operated with electron beams and with energy tagged photon beams.
The detector system was designed with the detection of 
multiple particle final states in mind. The driving motivation for the construction 
of CLAS 
was the nucleon resonance ($N^*$) program, with emphasis on the
study of the $\gamma^* N\rightarrow N^*$ 
transition form factors and the 
search for {\it missing resonances}.
Figure \ref{clas1} shows 
the CLAS detector. 
At the core of the detector is a toroidal magnet consisting of six superconducting 
coils 
symmetrically arranged around the beam line. Each of the six sectors is instrumented 
as an independent 
spectrometer with 34 layers of tracking chambers allowing
for the full reconstruction of the charged particle 3-momentum vectors. 
Charged hadron identification is accomplished by combining momentum and 
time-of-flight with the measured path length from the target to the 
plastic scintillation counters that surround the entire tracking region.
The wide range of particle identification allows for study of the complete 
range of reactions relevant to the $N^*$ program. In the polar angle range 
of up to 70$^{\circ}$, photons and neutrons can be 
detected using the electromagnetic calorimeters.
The forward angular range from about 10$^{\circ}$ to 50$^{\circ}$ is 
instrumented with gas \^Cerenkov counters for the identification of electrons.  

\begin{figure}[ht!]
\begin{center}
\begin{minipage}[t]{16 cm}
\epsfig{file=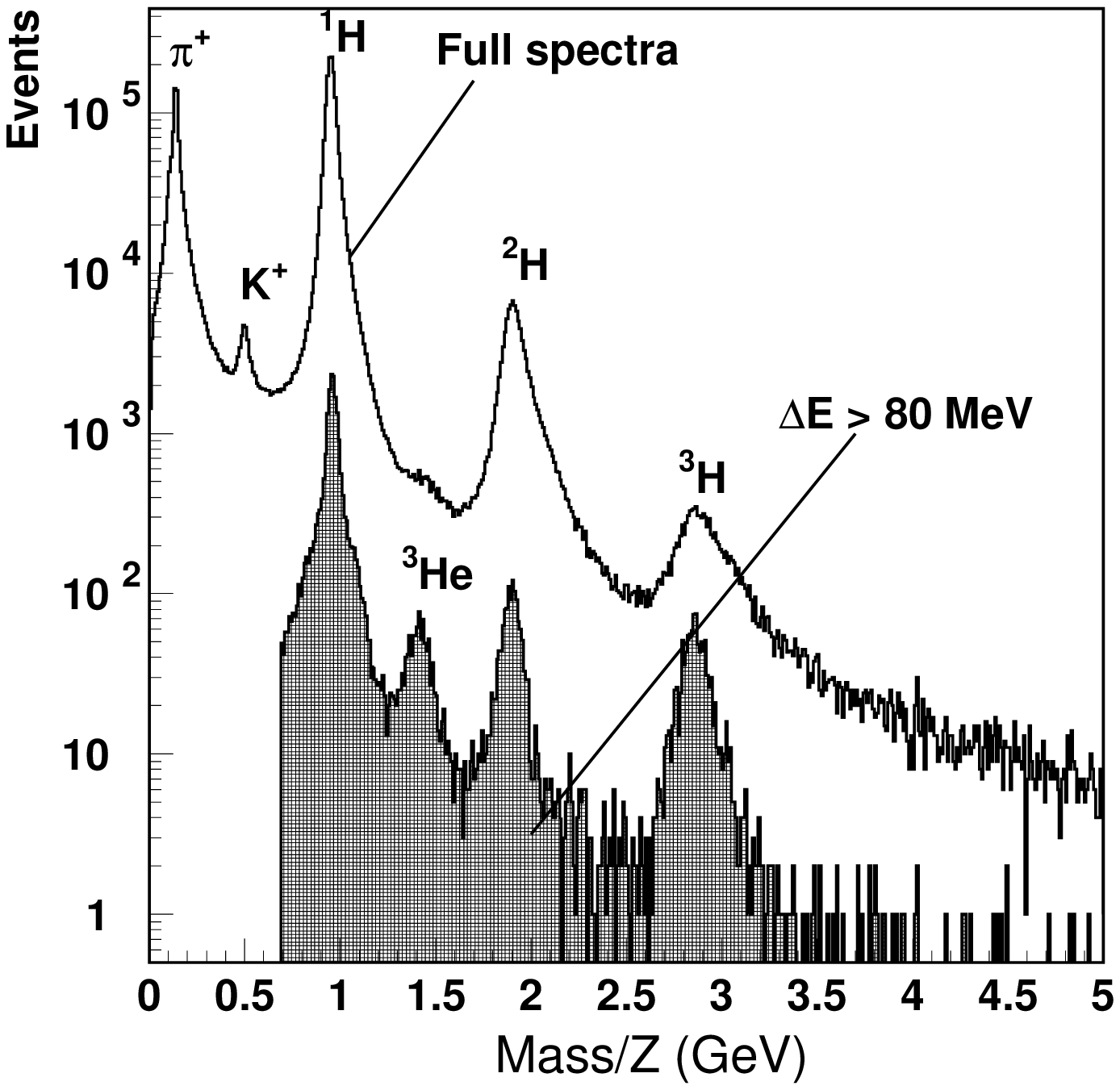,scale=0.5}
\epsfig{file=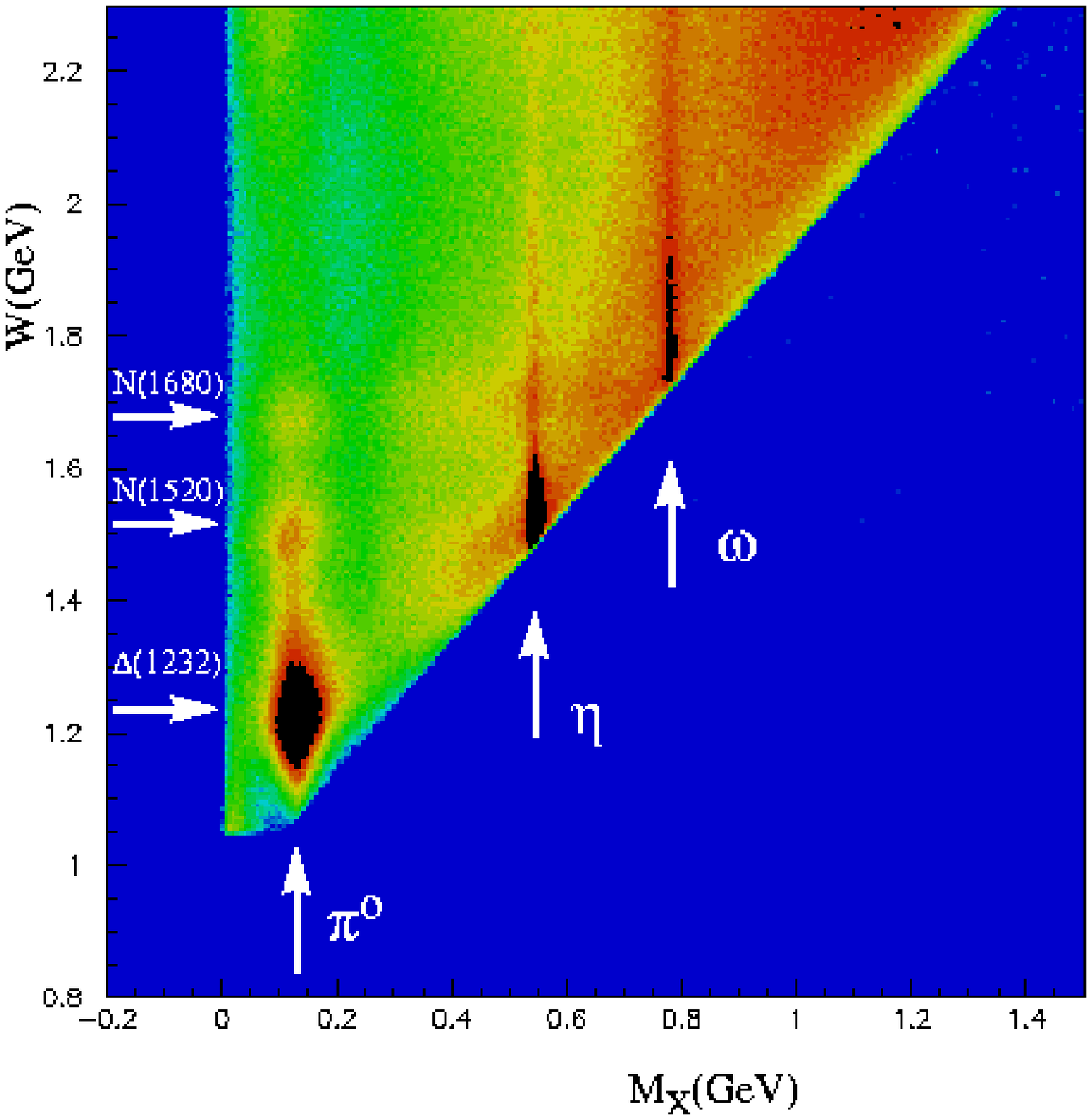,scale=0.43}
\end{minipage}
\begin{minipage}[t]{16.5 cm}
\caption{Left panel: 
Charged particle identification in CLAS. 
The reconstructed mass/Z (charge number) 
for positive tracks from a carbon target is shown. Additional sensitivity to 
high-mass particles is obtained by requiring large energy loss in the 
scintillators (shaded histogram). Right panel: Invariant mass versus missing mass for 
$ep\rightarrow epX$ at an electron beam energy of 4 GeV. 
\label{clas2}}
\end{minipage}
\end{center}
\end{figure}

In the $N^*$ program, CLAS is often used as a ``missing mass'' spectrometer, 
where all final state particles except one particle are detected. The undetected 
particle 
is inferred through the overdetermined kinematics, making use of the good
momentum ($\Delta p/p \approx 1\%$) and angle ($\Delta \Theta \approx 1-2^\circ$) 
resolution. Figure \ref{clas2} shows an 
example of the kinematics covered in the reaction $ep \rightarrow epX$.
It shows the invariant hadronic mass $W$  versus the missing mass $M_X$. 
The undetected
particles $\pi^0$, $\eta$, and $~\omega$ are clearly visible as 
bands of constant $M_X$. The
correlation of certain final states with specific resonance excitations
is also seen.   

\subsubsection{\it Experimental Hall C - HMS and SOS
\label{section:hallc}}

Hall C houses the high momentum spectrometer (HMS) and the short orbit 
spectrometer (SOS). The HMS reaches a maximum momentum of 7 GeV, while 
the SOS is limited to about 1.8 GeV. 
The spectrometer pair has been used to measure the $\gamma^* N\rightarrow
\Delta(1232)P_{33}$ \cite{Frolov,Vilano} and 
$\gamma^*p\rightarrow N(1535)S_{11}$ transitions
at high $Q^2$ \cite{Armstrong,Dalton}. For these kinematics the SOS was 
used as an electron spectrometer and the HMS to detect the proton. To achieve 
a large kinematic coverage, the spectrometers have to be moved in angle, and 
the spectrometer optics have to be adjusted to accommodate different particle momenta. 
This makes such a two-spectrometer setup most useful for studying meson production 
at high momentum transfer, or close to threshold. 
In either case, the Lorentz boost guarantees that particles are produced in a 
relatively 
narrow cone around the virtual photon, and can be detected in magnetic spectrometers 
with relatively small solid angles.

\subsection{\it MAMI
\label{sec:mami}}

The MAMI-B microtron electron accelerator~\cite{mami} at Mainz in Germany reaches a 
maximum beam 
energy of 850 MeV, and produces a highly polarized and stable electron beam with 
excellent beam properties. The recently upgraded MAMI-C machine reaches a 
maximum electron energy of 1.55~GeV. 
There are experimental areas for electron scattering experiments  with
three focusing magnetic spectrometers with high resolution~\cite{mami_a1,mami_a2}. 
A two-spectrometer configuration has been 
used in cross section and polarization asymmetry measurements of $\pi^0$ 
electroproduction from protons in the $\Delta(1232)P_{33}$ 
region  \cite{Stave2006,
Sparveris2007,Stave2008,Pospishil,Bartsch,Elsner}.

\subsection{\it MIT-Bates
\label{sec:mit-bates}}

The Bates 850 MeV linear electron accelerator has been used to study 
$\pi^0$ production in the 
$\Delta(1232)P_{33}$ resonance region using an out-of-plane spectrometer 
setup~\cite{bates_oops}. A 
set of 
four independent focusing spectrometers was used to measure various response 
functions, including the beam helicity-dependent out-of-plane response 
function. 
Because of the small solid angles covered by this setup, a limited 
range of the polar angles in the center-of-mass frame of the
$p\pi^0$ subsystem  could be covered. These spectrometers are no longer in use, 
but results recently published from earlier data taking 
are included in this review \cite{Mertz,Kunz,Sparveris2005,Warren}.

\section{Definitions and Conventions
\label{sec:definitions}}

The results on the electroexcitation of nucleon
resonances reported in this review are based mostly on 
the experiments
on pion and eta electroproduction
off nucleons.
We therefore only present the definitions that are important
for extraction and presentation of the results
for these reactions.
Throughout we use natural units, $h=c=1$, so that momenta
and masses are expressed in units of GeV
(rather than GeV$/c$ or GeV$/c^2$). We also use the
following conventions for the metric and $\gamma$-matrices:
$g^{\mu\nu}=$diag(1,-1,-1,-1), $\epsilon_{0123}=1$,
$a^{\mu}=(a_0,\bf{a})$,
$\{\gamma^{\mu},\gamma^{\nu}\}$=2$g^{\mu\nu}$,
$\gamma_5=i\gamma^0\gamma^1\gamma^2\gamma^3$. More explicitly,
$\gamma$-matrices have the following form:
\be
\sg =
\left(\begin{array}{cc}0&\ss\\-\ss&0\end{array}\right),
~~\gamma_0 =
\left(\begin{array}{cc}1&0\\0&-1\end{array}\right),~~
\gamma_5 =
-\left(\begin{array}{cc}0&1\\1&0\end{array}\right).
\label{eq:def1}
\ee
\subsection{\it  Kinematics
\label{sec:kinematics}}

\begin{figure}[ht!]
\begin{center}
\begin{minipage}[l]{200pt}
\epsfig{file=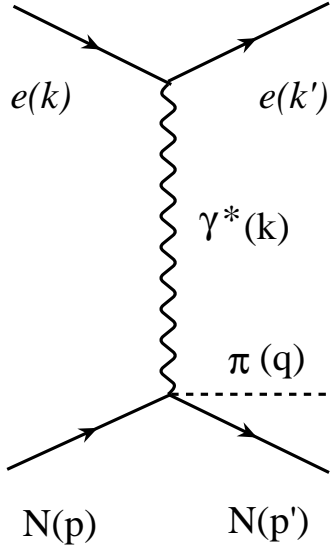,scale=0.8}
\end{minipage}
\hskip 20pt
\begin{minipage}[r]{200pt}
\caption{Electroproduction of pions off
nucleons in the one-photon approximation.
The four-momenta of the particles are given in parentheses. 
\label{fig1}}
\end{minipage}
\end{center}
\end{figure}

The differential cross section of the electroproduction of pions off 
nucleons in the one-photon approximation (Fig.~\ref{fig1}) is related to 
the differential
cross section of the production of pions by virtual photons
in the standard way (see e.g. Refs. \cite{tiator_def,tiator_def1}) 
through the virtual 
photon flux $\Gamma$ as:
\be
\frac{d\sigma}{dE_f d\Omega_e 
d\Omega}=\Gamma\frac{d\sigma}{d\Omega},
\label{eq:def2}
\ee
where
\be
\Gamma=\frac{\alpha}{2\pi^2 Q^2}\frac{(W^2-m^2)E_f}{2mE_i}
\frac{1}{1-\epsilon}, ~~~
\epsilon=\left[1+2\left(1+\frac{\nu^2}{Q^2}\right)\tan^2
\frac{\theta_e}{2}\right]^{-1}, 
\label{eq:def4}
\ee
$\alpha$ is the fine structure constant, $E_i$ and $E_f$ are the initial 
and 
final electron energies
in the laboratory frame, 
$\nu=E_i-E_f$,
$\epsilon$ is the polarization
factor of the virtual photon,
$\Omega_e=\Omega_e(\theta_e,\phi_e)$ is the laboratory
solid angle of the scattered electron, and
$\Omega=\Omega(\theta,\phi)$ is the pion solid angle 
in the c.m. system of the reaction $\gamma^* N\rightarrow N\pi$, where
$\theta$ is the angle between the pion and virtual photon in
this system, and $\phi$ is the angle between the electron
scattering and hadron production planes. 
The virtuality of the photon is given by $k^2=\nu^2-\bf{k}^2$.
Since the photon is spacelike, i.e. $k^2 < 0$, it is convenient to work
with the positive quantity $Q^2\equiv -k^2$. The invariant
mass squared of the final hadronic state (here, $\pi$ and $N$)
is 
$W^2=(p+k)^2=m^2+2m\nu-Q^2$,
where $p$ and $k$ are the target nucleon and virtual
photon four momenta, and $m$ is the nucleon mass.

For unpolarized particles
and for longitudinally polarized electron beam, 
the $\phi$-dependence of the 
$\gamma^* N\rightarrow N\pi$ cross section can be specified
in the following way:
\be
\frac{d\sigma}{d\Omega}=\sigma_T+\epsilon\sigma_L
+\epsilon\sigma_{TT}\cos{2\phi}
+\sqrt{2\epsilon(1+\epsilon)}\sigma_{LT}\cos{\phi}
+h\sqrt{2\epsilon(1-\epsilon)}\sigma_{LT'}\sin{\phi}.
\label{eq:def6}
\ee
Here we use notations of Ref. \cite{Park},
$\sigma_T$, $\sigma_L$, $\sigma_{TT}$,
and $\sigma_{LT}$ are the so-called structure functions
of the reaction $\gamma^* N\rightarrow N\pi$ that
depend on $W,Q^2,\cos \theta$, and
$h$ describes the longitudinal polarization of
the incident electron: $h=+1(-1)$ if electrons are polarized
parallel (anti-parallel) to the beam 
direction.
For longitudinally polarized electron beam and
polarized target and recoil nucleons, the relevant formulas can be found
in Refs. \cite{tiator_def,tiator_def1}.

\subsection{\it Expansion over multipole amplitudes
\label{sec:multipoles}}

In order to extract resonance contributions from the data on the reaction
$\gamma^* N\rightarrow N\pi$, the observables should
be defined through the multipole amplitudes.
These are transverse amplitudes
$M_{l\pm}(W,Q^2)$ and $E_{l\pm}(W,Q^2)$
and scalar(longitudinal) amplitudes $S_{l\pm}(W,Q^2)$ 
($L_{l\pm}=k_0S_{l\pm}/|\mathbf{k}|$);
they are related, respectively, to the photons
of the magnetic, electric, and Coulombic type; 
$l$ is the angular monentum of pion in the 
c.m. system of the reaction $\gamma^* N\rightarrow N\pi$.
For this purpose, 
it is convenient to introduce 
transverse
partial wave helicity amplitudes: 
\bea
&&A_{l+}=\frac{1}{2}\left[(l+2){E}_{l+}
+l{M}_{l+}\right], 
~~~~B_{l+}={E}_{l+}-{M}_{l+},
\label{eq:def25}\\
&&A_{(l+1)-}=\frac{1}{2}\left[
(l+2){M}_{(l+1)-}-l{E}_{(l+1)-}\right],
~~~~B_{(l+1)-}={E}_{(l+1)-}+{M}_{(l+1)-}.
\label{eq:def27}
\eea
The amplitudes 
$A_{l\pm},B_{l\pm},S_{l\pm}$
are related to the
$\gamma^*N\rightarrow \pi N$ helicity amplitudes
in the center-of-mass system (c.m.s.) of the reaction
in the following way:
\bea
H_1=&&\frac{1}{\sqrt{2}}
\sin\theta
\cos\frac{\theta}{2}
\sum(B_{l+}-B_{(l+1)-})[P''_{l}(\cos 
\theta)-P''_{l+1}(\cos
\theta)],
\label{eq:def28}\\
H_2=&&\sqrt{2}
\cos\frac{\theta}{2}
\sum(A_{l+}-A_{(l+1)-})[P'_{l}(\cos 
\theta)-P'_{l+1}(\cos
\theta)],
\label{eq:def29}\\
H_3=&&\frac{1}{\sqrt{2}}
\sin\theta
\sin\frac{\theta}{2}
\sum(B_{l+}+B_{(l+1)-})[P''_{l}(\cos 
\theta)+P''_{l+1}(\cos
\theta)],
\label{eq:def30}\\
H_4=&&\sqrt{2}
\sin\frac{\theta}{2}
\sum(A_{l+}+A_{(l+1)-})[P'_{l}(\cos 
\theta)+P'_{l+1}(\cos
\theta)],
\label{eq:def31}\\
H_5=&&
\frac{Q}{|\bf{k}|}
\cos\frac{\theta}{2}
\sum(l+1)(S_{l+}+S_{(l+1)-})[P'_{l}(\cos 
\theta)-P'_{l+1}(\cos
\theta)],
\label{eq:def32}\\
H_6=&&
\frac{Q}{|\bf{k}|}
\sin\frac{\theta}{2}
\sum(l+1)(S_{l+}-S_{(l+1)-})[P'_{l}(\cos 
\theta)+P'_{l+1}(\cos
\theta)],
\label{eq:def33}
\eea
where $H_1,H_2,...H_6$ are the elements of the matrices
$H_{\mu_2\mu_1}$,  
\be
\lambda_{\gamma}=1:
\left(\begin{array}{cc}H_4&H_3\\H_2&H_1\end{array}\right);
~~~\lambda_{\gamma}=-1:
\left(\begin{array}{cc}H_1&-H_2\\-H_3&H_4\end{array}\right);
~~~\lambda_{\gamma}=0:
\left(\begin{array}{cc}-H_5&H_6\\H_6&H_5\end{array}\right). 
\label{eq:def19}
\ee
Here $\mu_1$ and $\mu_2$ are
the initial and final nucleon
helicities, and
$\lambda_{\gamma}$ is the photon helicity.

The structure functions given in Eq. (\ref{eq:def6})
are related to the helicity amplitudes 
$H_{1,2,...6}(W,\cos \theta,Q^2)$ by:
\bea
\sigma_T&&=~\frac{|\bf{q}|}{2K}(|H_1|^2+|H_2|^2+|H_3|^2+|H_4|^2),
\label{eq:def20}\\
\sigma_L&&=~\frac{|\bf{q}|}{K}(|H_5|^2+|H_6|^2),
\label{eq:def21}\\
\sigma_{TT}&&=~\frac{|\bf{q}|}{K}Re(H_3H_2^*-H_4H_1^*),
\label{eq:def22}\\
\sigma_{LT}&&=~-\frac{|\bf{q}|}{\sqrt{2}K}
Re[(H_1-H_4)H_5^*+(H_2+H_3)H_6^*],
\label{eq:def23}\\
\sigma_{LT'}&&=~-\frac{|\bf{q}|}{\sqrt{2}K}
Im[(H_1-H_4)H_5^*+(H_2+H_3)H_6^*],
\label{eq:deflt}
\eea
where $K=\frac{W^2-m^2}{2W}$ and 
$\bf{k}$ and $\bf{q}$
are, respectively, the photon 
equivalent energy and the virtual photon and pion 3-momenta in the
$\gamma^*N\rightarrow \pi N$ c.m.s.

The 
$\gamma^*N\rightarrow \pi N$ total cross section can be written
through partial wave helicity amplitudes in the compact way:
\be
\sigma^{tot}= \sigma_{1/2}+\sigma_{3/2}+\epsilon\sigma_L^{tot},
\label{eq:def34}
\ee
\bea
\sigma_{1/2}=&&2\pi\frac{|\bf{q}|}{K}
\sum 2(l+1)[|A_{l+}|^2+|A_{(l+1)-}|^2],
\label{eq:def35}\\
\sigma_{3/2}=&&2\pi\frac{|\bf{q}|}{K}
\sum \frac{l}{2}(l+1)(l+2)[|B_{l+}|^2+|B_{(l+1)-}|^2],
\label{eq:def36}\\
\sigma_{L}^{tot}=&&4\pi\frac{|\bf{q}|}{K}
\sum \frac{Q^2}{\bf{k}^2}(l+1)^3[|S_{l+}|^2+|S_{(l+1)-}|^2].
\label{eq:def37}
\eea

\subsection{\it Definition of the $\gamma^*N\rightarrow N^*$
helicity amplitudes 
\label{sec:amplitudes}}

\begin{figure}[ht!]
\begin{center}
\begin{minipage}[t]{16. cm}
\epsfig{file=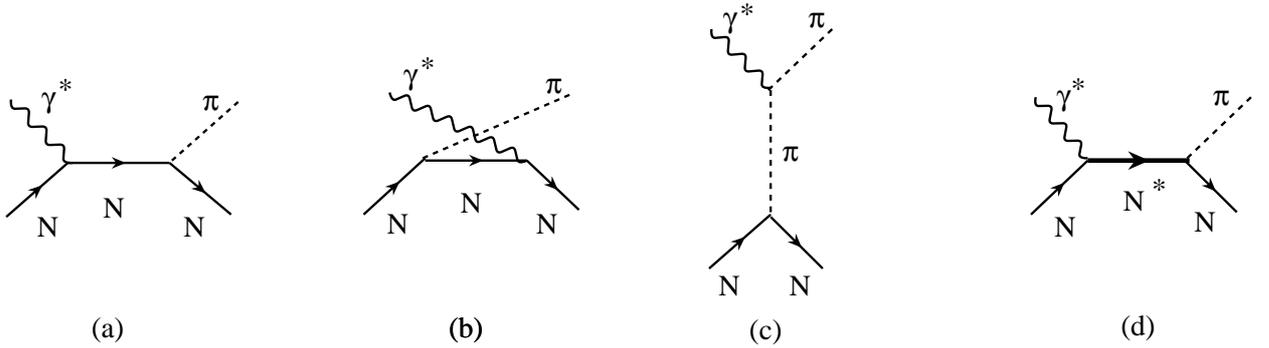,scale=0.6}
\end{minipage}
\begin{minipage}[t]{16.5 cm}
\caption{
The diagrams corresponding to the Born terms (a,b,c)
and resonance (d) contributions
to $\gamma^*N\rightarrow N\pi$.
\label{fig2}}
\end{minipage}
\end{center}
\end{figure}

Experimental results on the $\gamma^*N\rightarrow N^*$
helicity amplitudes  (transverse amplitudes $A^N_{1/2}$ and $A^N_{3/2}$ 
and scalar (or longitudinal) amplitude $S^N_{1/2}$), 
extracted from the data on
$\gamma^*N\rightarrow N\pi$,  correspond to
the contribution of diagram (d) in Fig.~\ref{fig2}
to this reaction. 
They are related to
the resonant portions of the 
corresponding multipole
amplitudes at the resonance positions
in the following way:
\bea
&&A^N_{1/2}=\mp \hat {A}_{l\pm}, 
\label{eq:def38}
\\
&&A^N_{3/2}=\pm \sqrt{\frac{(2J-1)(2J+3)}{16}}\hat {B}_{l\pm}, 
\label{eq:def39}
\\
&&S^N_{1/2}=-\frac{2J+1}{2\sqrt{2}} \hat {S}_{l\pm}, 
\label{eq:def40}
\eea
where 
\bea
{\hat {A}}_{l\pm}({\hat {B}}_{l\pm},{\hat {S}}_{l\pm})&&\equiv~
aImA^R_{l\pm}(B^R_{l\pm},S^R_{l\pm})(W=M),
\label{eq:def41}
\\
a&&\equiv~\frac{1}{C_I}
\left[(2J+1)\pi\frac{|{\bf q}|_{r}}{K_r}\frac{M}{m}
\frac{\Gamma}{\beta_{\pi N}}\right]^{1/2},
\label{eq:def42}
\eea
$\Gamma$, $M$, $J$ and $I$ are, respectively, the total width, mass,
spin and isospin of the
resonance,  $J=l\pm\frac{1}{2}$
for $l\pm$ amplitudes, $\beta_{\pi N}$ is  
the branching ratio of the resonance
to the $\pi N$ channel,
$K_r$ and $|{\bf{q}}_r|$  are the photon equivalent
energy and the pion 3-momentum
at the resonance position in the c.m.s. 
of $\gamma^*N\rightarrow N\pi$, and
$C_I$ are the isospin
Clebsch-Gordon coefficients in the decay $N^*\rightarrow \pi N$:
\bea
&C_{1/2}=\mp\sqrt{\frac{1}{3}},~C_{3/2}=\sqrt{\frac{2}{3}}
~~{\rm for}~~\gamma^* p\rightarrow \pi^0 p~~(\gamma^* n\rightarrow \pi^0 n),
\label{eq:def24}\\
&C_{1/2}=-\sqrt{\frac{2}{3}},~C_{3/2}=\mp\sqrt{\frac{1}{3}}
~~{\rm for} ~~\gamma^* p\rightarrow \pi^+ n~~(\gamma^* n\rightarrow \pi^- p),
\label{eq:def0}
\eea
where we have taken into account that the pion isomultiplet
is $\pi=(\pi^-,\pi^0,-\pi^+)$.

At the photon point, the helicity amplitudes (\ref{eq:def38},\ref{eq:def39})
are related to the $N^*\rightarrow N\gamma$ decay width by:
\be
\Gamma(N^*\rightarrow N\gamma)=
\frac{2K_r^2}{\pi(2J+1)}\frac{m}{M}\left(|A^N_{1/2}|^2
+|A^N_{3/2}|^2\right).
\label{eq:def43}
\ee

\noindent For the transverse amplitudes $A^N_{1/2}$ and $A^N_{3/2}$,
the relations (\ref{eq:def38},\ref{eq:def39}) were introduced by Walker 
\cite{Walker};
for the longitudinal amplitude, the relation (\ref{eq:def40})
coincides with that
from Refs. \cite{Arndt,Kamalov_def}.

According to the definitions (\ref{eq:def38}-\ref{eq:def40}),
the $\gamma^*N\rightarrow N^*$
helicity amplitudes extracted from the data on the $\gamma^*N\rightarrow 
N\pi$ reaction contain the sign of the $\pi N N^*$ vertex;
it defines the relative sign of the diagrams
that correspond to the resonance (Fig.~\ref{fig2}d)
and Born terms (Figs.~\ref{fig2}a,b,c) contributions
to $\gamma^*N\rightarrow N\pi$. The situation is analogous
in other reactions. For example, 
the $\gamma^*N\rightarrow N^*$
helicity amplitudes extracted from the data on the $\gamma^*N\rightarrow 
N\eta$ reaction contain the sign of the $\eta N N^*$ vertex.

In the calculations of the  $\gamma^*N\rightarrow N^*$
helicity amplitudes in theoretical approaches,
the commonly used definition relates these amplitudes
to the matrix elements of the electromagnetic current:
\bea
&&{\tilde A}^N_{\frac{1}{2}}=\sqrt{\frac{2\pi \alpha}{K_r}}\frac{1}{e}
<N^*,S_z^*=\frac{1}{2}|\epsilon^{(+)}_{\mu}J_{em}^{\mu}|N,S_z=-\frac{1}{2}>,
\label{eq:def44}\\
&&{\tilde A}^N_{\frac{3}{2}}=\sqrt{\frac{2\pi \alpha}{K_r}}\frac{1}{e}
<N^*,S_z^*=\frac{3}{2}|\epsilon^{(+)}_{\mu}J_{em}^{\mu}|N,S_z=\frac{1}{2}>,
\label{eq:def45}\\
&&{\tilde S}^N_{\frac{1}{2}}=\sqrt{\frac{2\pi \alpha}{K_r}}
\frac{1}{e}
<N^*,S_z^*=\frac{1}{2}|\frac{|{\bf{k}}|}{Q}
\epsilon^{(0)}_{\mu}J_{em}^{\mu}|N,S_z=\frac{1}{2}>,
\label{eq:def46}
\eea
where $\frac{e^2}{4\pi}=\alpha$, and it is assumed 
that the $z$-axis is directed along the photon 3-momentum
(${\bf{k}}$) in the $N^*$ rest frame,  
$S_z$ and $S^*_z$ are the projections of the nucleon
and resonance spins on this axis, and
\be
\epsilon^{(0)}_{\mu}=\frac{1}{Q}(|{\bf{k}}|,0,0,-k_0),
~~\epsilon^{(+)}_{\mu}=(0,~-~{\se}^{(+)}),
~~{\se}^{(+)}=-\frac{1}{\sqrt{2}}(1,i,0).
\label{eq:def47}
\ee
To distinguish the amplitudes (\ref{eq:def44}-\ref{eq:def46})
from those extracted from experiment (\ref{eq:def38}-\ref{eq:def40}),
they are labeled by tildes. The
amplitudes (\ref{eq:def38}-\ref{eq:def40}) and (\ref{eq:def44}-\ref{eq:def46}) 
are related by
\be
A^N_{\frac{1}{2},\frac{3}{2}}
= \zeta {\tilde A}^N_{\frac{1}{2},\frac{3}{2}},
~~~S^N_{\frac{1}{2}}=\zeta {\tilde S}^N_{\frac{1}{2}},
\label{eq:def48}
\ee
where $\zeta$ is the sign that reflects the presence
of the $\pi NN^*$ vertex in Fig.~\ref{fig2}d.
The relation between $\zeta$
and the sign of the ratio of the $\pi NN$ and $\pi NN^*$
coupling constants 
was found in Ref. \cite{DEFINITIONS}
using the results of covariant calculations
of the Fig.~\ref{fig2} diagrams in Ref.  \cite{Devenish}.
With the definitions
\be
<p|J_{\pi}(0)|p>=
g\bar{u}(p')\gamma_5 u(f),
\label{eq:def49}
\ee
\be
<p|J_{\pi}(0)|N^{*+}>=\pm C_I
g^*\bar{u}(p')
\left(\begin{array}{c}1\\ \gamma_5\end{array}\right)
u(p^*),~~~~J^P={\frac{1}{2}}^{\mp},
\label{eq:def50}
\ee
and
\be
<p|J_{\pi}(0)|N^{*+}>=\pm C_I
g^*\bar{u}(p')p'_{\nu_1}
...p'_{\nu_l}
\left(\begin{array}{c}1\\ \gamma_5\end{array}\right)
u^{\nu_1...\nu_l}(p^*),~~~~J^P=\frac{3}{2}^{\pm},\frac{5}{2}^{\mp}...,
\label{eq:def51}
\ee
we have
\be
\zeta=-{\rm sign}(g^*/g).
\label{eq:def52}
\ee

In Eqs. (\ref{eq:def49}-\ref{eq:def51}), 
$f$ and $p^*$ are, respectively, the 4-momenta of the 
 intermediate proton and  resonance
in the diagrams of Figs. \ref{fig2}a and \ref{fig2}d,
$u(p)$ is the Dirac spinor,
and $u^{\nu_1...\nu_l}(p^*)$ with $l=J-\frac{1}{2}$
is the generalized Rarita-Schwinger spinor.

\subsection{{\bf \it The $\gamma^* N\rightarrow N^{*}$
helicity amplitudes in terms of
the $\gamma^* N\rightarrow N^{*}$ form factors}
\label{sec:formfactors}}

Many theoretical approaches, e.g. light-front 
relativistic quark models, QCD sum rules, and lattice QCD, use
the definition of the $\gamma^* N\rightarrow N^{*}$
helicity amplitudes in terms of
the $\gamma^* N\rightarrow N^{*}$ form factors.
In this section, we present the definition
introduced in Ref. \cite{Devenish}, where
the $\gamma^* N\rightarrow N^{*}$ 
form factors were defined in a unified
way for all resonances with $J^P=\frac{1}{2}^{\pm},\frac{3}{2}^{\pm}...$
The definition of Ref. \cite{Devenish} for the $J^P=\frac{3}{2}^{+}$ resonances
coincides with the widely used definition 
by Jones and Scadron \cite{Scadron}. 

For the $J^P={\frac{1}{2}}^{\pm}$
resonances, the definition of Ref. \cite{Devenish} 
for the $<N^{*}|J_{em}^{\mu}|N>$ matrix element
is:
\bea
&&<N^{*}|J^{em}_{\mu}|N>\equiv e\bar{u}(p^*)
\left(\begin{array}{c}1\\\gamma_5\end{array}\right)
\tilde{J}_{\mu}u(p),
\label{eq:def53}
\\
&&{\tilde{J}}_{\mu} =
\left[k\hspace{-1.8mm}\slash k_{\mu}-k^2\gamma_{\mu}\right]G_1(Q^2)
+\left[k\hspace{-1.8mm}\slash P_{\mu}-(Pk)\gamma_{\mu}\right]G_2(Q^2),
\label{eq:def54}
\eea
where $P\equiv \frac{1}{2}(p^*+p)$.
Using the definitions (\ref{eq:def44}-\ref{eq:def46}), we 
find the following relations between   
the $\gamma^* N\rightarrow N^{*}$
helicity amplitudes and the form factors $G_1(Q^2),G_2(Q^2)$:
\bea
&&{\tilde A}^N_{\frac{1}{2}}=
b\left[2Q^2G_1(Q^2)-(M^2-m^2)G_2(Q^2)\right],
\label{eq:def55}
\\
&&{\tilde S}^N_{\frac{1}{2}}=\pm  
b\frac{|{\bf{k}}|}{\sqrt{2}}\left[2(M\pm m)G_1(Q^2)+(M\mp m)G_2(Q^2)\right],
\label{eq:def56}
\\
&&b\equiv e\sqrt{\frac{(M\mp m)^2+Q^2}{16mMK_r}}.\nonumber
\eea

For the $J^P=\frac{3}{2}^{\pm},\frac{5}{2}^{\pm}$,... resonances
the definitions are more combersome:
\bea
&&<N^{*}|J^{em}_{\mu}|N>\equiv e\bar{u}
^{\nu_1...\nu_{l-1}\nu}(p^*)
k_{\nu_1}...k_{\nu_{l-1}}
\left(\begin{array}{c}\gamma_5\\1\end{array}\right)
\Gamma_{\nu\mu}(Q^2)
u(p),
\label{eq:def57}\\
&&\Gamma_{\nu\mu}(Q^2)= G_1(Q^2){\cal H}^1_{\nu\mu}+
G_2(Q^2){\cal H}^2_{\nu\mu}+
G_3(Q^2){\cal H}^3_{\nu\mu},
\label{eq:def58}\\
&&{\cal H}^1_{\nu\mu}=k\hspace{-1.8mm}\slash g_{\nu\mu}-k_{\nu}\gamma_{\mu},~~
{\cal H}^2_{\nu\mu}=k_{\nu}p^*_{\mu}-(kp^*)g_{\nu\mu},
~~{\cal H}^3_{\nu\mu}=k_{\nu}k_{\mu}-k^2g_{\nu\mu},
\label{eq:def61}
\eea

\bea
&&{\tilde A}^N_{1/2}=h_3X,
~~~{\tilde A}^N_{3/2}=\pm \sqrt{3}\frac{h_2}{l}X,
~~~{\tilde S}^N_{1/2}=h_1\frac{|\bf{k}|}{\sqrt{2}M}X,
\label{eq:def64}
\\
&&X\equiv |{\bf k}|^{l-1}\sqrt{\pi\alpha\frac{(M\mp m)^2+Q^2}
{16MmK_r\tau_{l+1}}},
~~~\tau_{l}\equiv\frac{(2l)!}{2^l(l!)^2},~~l=J-\frac{1}{2},
\label{eq:def66}
\eea
and
\bea
&&h_1(Q^2)=\pm 4MG_1(Q^2)+4M^2G_2(Q^2)+2(M^2-m^2-Q^2)G_3(Q^2),
\label{eq:def67}\\
&&h_2(Q^2)=-2(\pm M+ m)G_1(Q^2)-(M^2-m^2-Q^2)G_2(Q^2)+2Q^2G_3(Q^2),
\label{eq:def68}\\
&&h_3(Q^2)=\mp\frac{2}{M}[Q^2+m(\pm 
M+m)]G_1(Q^2)+(M^2-m^2-Q^2)G_2(Q^2)-2Q^2G_3(Q^2).
\label{eq:def69}
\eea
In Refs. 
\cite{Devenish}
and 
\cite{Scadron}, the
$\gamma^* N\rightarrow N^{*}$
matrix elements are also defined through the form factors
$G_E(Q^2)$, $G_M(Q^2)$, and $G_C(Q^2)$; in Ref. \cite{Scadron} for 
the resonances
with $J^P=\frac{3}{2}^{+}$ and in Ref. \cite{Devenish}
for the $J^P=\frac{3}{2}^{\pm},\frac{5}{2}^{\pm}$,... resonances.
These form factors are related to 
the $\gamma^* N\rightarrow N^{*}$
helicity amplitudes by:
\bea
\left(\begin{array}{c}{G_M}\\ G_E\end{array}\right)
&&=~-F_l\left(\frac{l+2}{\sqrt{3}}{\tilde 
A}^N_{3/2}+{\tilde A}^N_{1/2}\right)
\frac{2l}{l+1},
\label{eq:def70}
\\
\left(\begin{array}{c}{G_E}\\ G_M\end{array}\right)
&&=~-F_l\left(\frac{l}{\sqrt{3}}{\tilde A}^N_{3/2}-
{\tilde A}^N_{1/2}\right)
\frac{2}{l+1},
\label{eq:def71}
\\
G_C&&=~2\sqrt{2}\frac{M}{|{\bf{k}}|}F_l{\tilde S}^N_{1/2},
\label{eq:def72}\\
F_l&&\equiv~\frac{m}{|{\bf{k}}|^l}
\sqrt{\tau_{l+1}\frac{mK_r}{6\pi\alpha M}
\left[1+\frac{Q^2}{(M\pm m)^2}\right]},
\label{eq:def73}
\eea
where the upper symbols correspond to 
the $J^P=\frac{3}{2}^{+},\frac{5}{2}^{-}$,... resonances 
and the lower ones to
the $J^P=\frac{3}{2}^{-},\frac{5}{2}^{+}$,... resonances.
The relations 
(\ref{eq:def70}-\ref{eq:def73}) coincide with similar relations
found in Ref. \cite{Capstick} for the  
$\frac{3}{2}^{+}$ resonances.

\section{Theoretical Approaches for the Analysis of Pion and Eta Electroproduction
off Nucleons
\label{sec:theory}}
The investigation of pion photo- and electroproduction started in the
1950's with the seminal work of Chew, Goldberger, Low, and Nambu (CGLN)
\cite{Chew}, where the common formalism for pion
photoproduction on nucleons was developed, and fixed-$t$ dispersion
relations (DR) were used as a tool for the analysis
of the reaction. Postulates underlying 
the dispersion relations approach
are the analyticity, unitarity, and crossing symmetry
of the $S$-matrix. 
The CGLN formalism and DR were extended to 
pion electroproduction \cite{Soloviev1,Fubini,Soloviev2},
and later DR were used in the analyses
of experimental data \cite{Ball,Adler,Schwela1,Berends1,Schwela2,
Devenish1,Devenish2,Devenish3,Crawford1,Crawford2,Berends2,
Crawford3,Gerhardt,Arai,Crawford4}.
Starting in the late 1990's,
the approach was applied 
\cite{Aznauryan1998,Aznauryan1999,Drechsel_DR,
Drechsel2002,Aznauryan2003,Aznauryan2005,
Pasquini2005,Pasquini2006,Aznauryan2008,Aznauryan2009}
to analyze
the $\pi$ production data from the new
generation of electron accelerators.
It was recently extended to $\eta$ production and
used in the analysis of new data on $\eta$ 
photo- and electroproduction \cite{Aznauryan2005,Aznauryaneta}.

In the late 1960's,
the basis of other widely used approaches  
was given by the isobar model 
and the effective Lagrangian description
introduced, respectively, in Refs. \cite{Walker} and  
 \cite{Peccei}. In the isobar model \cite{Walker},
 a Breit-Wigner form with 
energy-dependent partial widths $\Gamma_{\gamma}$
and $\Gamma_{\pi}$, was used to parameterize the resonance contributions
to the partial waves.  The effective Langragian approach 
for pion photoproduction at the threshold was derived
in Ref. \cite{Peccei} from Weinberg's low-energy $\pi N$ scattering Lagrangian 
\cite{Weinberg1,Weinberg2}.
These approaches were later combined and extended, and gave 
rise to the  effective Langragian
description in the $\Delta(1232)P_{33}$ resonance region
\cite{Olsson,David1,David2,David3}
and the Unitary Isobar Model (UIM) \cite{Drechsel_UIM}.
In the approach of Refs. \cite{Arndt,GWU0,GWU1}, based on
parameterization inspired
by a unitary $K$-matrix ansatz (see below Sec. \ref{sec:said}),
the Breit-Wigner formula is used 
on the stage of extraction of resonance
contributions. 
The latter two approaches have been widely used in the 
analyses of pion 
production data with invariant masses up to $W\approx 2~$GeV. 
They are available through the dial-in codes 
MAID \cite{MAID_code}
and SAID \cite{SAID}.

Isobar models, which include the effective Lagrangian description, 
have also been developed for $\eta$ photoproduction
in Refs. \cite{Feuster,Saghai2001,Saghai2008}
and for $\eta$ photo- and electroproduction 
in Ref. \cite{etaMAID}. 
The model in \cite{etaMAID} was used in the analysis
of new $\eta$ data in Refs. \cite{Aznauryan2005,Aznauryaneta,etaMAID}; 
it is also employed through the dial-in code $\eta$-MAID 
\cite{MAID_code}.

The approaches \cite{Olsson}-\cite{Drechsel_UIM},
\cite{Feuster}-\cite{etaMAID}  based on the effective 
Langragian
description use essentially the $K$-matrix approximation. 
A rather different theoretical point of view has 
been taken in the development of dynamical models.
Early dynamical models 
\cite{Kamalov1999,Kamalov2001,Sato2001,Sato1996,
Tanabe,Yang1985,Nozawa} were
limited to the $\Delta(1232)P_{33}$ resonance region. They
used the Bethe-Salpeter formulation
and were applied to account for the off-shell rescattering effects.
The Dubna-Mainz-Taipei (DMT) 
\cite{Kamalov1999,Kamalov2001} and Sato, Lee (SL) 
\cite{Sato2001,Sato1996} dynamical models were widely used
for the analyses of experimental data on pion electroproduction 
on protons with the goal  to
separate the contribution created
by off-shell effects and associated with the pion-cloud contribution
to $\gamma^* N\rightarrow \Delta(1232)P_{33}$ from the `bare' contribution.
Recently developed dynamical models \cite{Sato2007,
Sato2008,Kamano2009,Sato2009,Kamano2010,Sirca2009,Doring2010,Sirca2011}
are based on the Hamiltonian formulation of the multi-channel and
multi-resonance reactions. These are essentially dynamical
coupled-channel approaches resulting from the unitarity
condition. They also account for the off-shell rescattering effects.
The approaches \cite{Sato2007,
Sato2008,Kamano2009,Sato2009,Kamano2010} were developed
at the Excited Baryon Analysis Center (EBAC) established
at JLab in 2006 to provide theoretical support to the excited
baryon program.

The overwhelming majority of new data 
on the electroexcitation of nucleon resonances
in single pion electroproduction
($ep\rightarrow eN\pi$) 
was obtained at JLab
with the CLAS spectrometer.
Independent analyses of these data 
were performed by the JLab group \cite{
Aznauryan2008,Aznauryan2009} associated with CLAS, and at 
Mainz \cite{MAID2007,MAID_China}.
The JLab group employed two substantially different approaches, DR and the UIM 
of Refs. \cite{Aznauryan2003,Aznauryan2009}. This enabled study of the 
model sensitivity of the results.  
The Mainz analysis is based on the UIM of Ref. \cite{Drechsel_UIM}.
As mentioned earlier, the DMT and SL models have also been used
in the analyses of pion 
electroproduction data in the $\Delta(1232)P_{33}$ resonance region.
The SAID code was applied to analyze low $Q^2$ pion electroproduction data
in the $\Delta(1232)P_{33}$ resonance region
\cite{Stave2008,Sparveris2005}.
In this section we present the approaches used in the analyses
of the new pion electroproduction data for the extraction of   
electroexcitation amplitudes of nucleon resonances that
are reported in this review. The corresponding approaches
for the analyses of the new $\eta$ data will be also discussed.

\subsection{\it Dispersion relations \label{sec:dr}}

There are two ways of utilization of dispersion relations. 
One is based on fixed-$t$ dispersion relations for 
invariant amplitudes and was successfully used for the analysis of data
throughout the nucleon resonance region
\cite{Devenish1,Devenish2,Devenish3,Crawford1,Crawford2,Berends2,
Crawford3,Gerhardt,Arai,Crawford4,
Aznauryan2003,Aznauryan2005,Aznauryan2008,Aznauryan2009}.
Another way developed in Refs. \cite{Schwela1,Schwela2} is based on DR 
for the multipole amplitudes of the $\Delta(1232)P_{33}$ resonance and
allows to get functional form of these amplitudes with one free
parameter for each of them.
It was employed for the analyses of new data in Refs. 
\cite{Aznauryan1998,Aznauryan1999,Drechsel_DR,Aznauryan2003,Aznauryan2005,
Aznauryan2008,Aznauryan2009}.  
Below we 
discuss both ways of utilizing the dispersion relations.
  
\subsubsection{\it Fixed-$t$ dispersion relations for invariant
amplitudes \label{sec:dr1}}
We define invariant 
amplitudes for the reaction 
$\gamma^* N\rightarrow \pi N$  
according to the definition of the hadron
electromagnetic current
$I^{\mu}$ for this process in Refs. \cite{Ball,Devenish1}:
\bea
&&I^{\mu}\equiv\bar{u}(p')\gamma _5 
{\cal{I}}^{\mu}u(p)\phi_{\pi},
\label{eq:dr1}\\
&&{\cal{I}}^{\mu} =\frac{B_1}{2}\left[ \gamma
^\mu k\hspace{-1.8mm}\slash-
k\hspace{-1.8mm}\slash\gamma ^\mu \right]+2{\cal P} ^\mu
B_2+2q^\mu B_3 
+2k^\mu B_4-\gamma ^\mu B_5+k\hspace{-1.8mm}\slash {\cal P}^\mu B_6+
k\hspace{-1.8mm}\slash k^\mu B_7+
k\hspace{-1.8mm}\slash q^\mu B_8,
\label{eq:dr2}
\eea
where  $B_i(s,t,Q^2),~i=1,2,...8$,
are the invariant amplitudes that are functions
of the invariant variables 
$s=(k+p)^2$, $t=(k-q)^2$, $Q^2$;
${\cal P}\equiv \frac{1}{2}(p+p')$, 
$u(p)$, $u(p')$ are the Dirac spinors 
of the initial and final nucleons, 
and $\phi_{\pi}$ is the pion field.
            
The conservation of  $I^{\mu}$
leads to the relations
\begin{eqnarray}
&&4Q^2B_4=(s-u)B_2 -2(t+Q^2-m_{\pi} ^2)B_3, 
\label{eq:dr3}
\\
&&2Q^2B_7=-2B'_5 -(t+Q^2-m_{\pi} ^2)B_8,
\label{eq:dr4}
\end{eqnarray}
where
$B'_5\equiv B_5-\frac{1}{4}(s-u)B_6$. Therefore,
only six of the eight invariant amplitudes
are independent. 
These amplitudes are usually chosen as follows:
$B_1$, $B_2$, $B_3$, $B'_5$, $B_6$, $B_8$. The relations between 
these amplitudes and the $\gamma^*N\rightarrow N\pi$ helicity  amplitudes 
are rather lengthy and can be found in Ref. \cite{Aznauryan2003}.

The amplitudes have
the following isotopic structure:
\begin{eqnarray}
&&B_i(\gamma^*~+~p\rightarrow p~+~\pi^0)=
B_i^{(+)}~+~B_i^{(0)},
\label{eq:dr5}
\\
&&B_i(\gamma^*~+~n\rightarrow n~+~\pi^0)=B_i^{(+)}~-~B_i^{(0)},
\label{eq:dr6}
\\
&&B_i(\gamma^*~+~p\rightarrow n~+~\pi^+)=
2^{1/2}(B_i^{(0)}~+~B_i^{(-)}),
\label{eq:dr7}
\\
&&B_i(\gamma^*~+~n\rightarrow p~+~\pi^-)=
2^{1/2}(B_i^{(0)}~-~B_i^{(-)}),
\label{eq:dr8}
\end{eqnarray}
where
\begin{equation}
B_i^{(+)}=\frac{1}{3}(B_i^{1/2}+2B_i^{3/2}),
~~B_i^{(-)}=\frac{1}{3}(B_i^{1/2}-B_i^{3/2}),
\label{eq:dr9}
\end{equation}
$B_i^{(0)}$ correspond to an isoscalar photon, and
$B_i^{1/2}$ and $B_i^{3/2}$ correspond to an isovector photon
with total isospin in the $s$-channel of $\frac{1}{2}$ and $\frac{3}{2}$,
respectively. 

From the high-energy Regge-pole behavior 
of the amplitudes (see, for example, Ref. \cite{Devenish1}), it follows that 
at $s\rightarrow \infty$, $B_i \sim s^{\alpha(t)-1}~
(i=1,2,5,6,8)$
and $B_3 \sim s^{\alpha(t)}$, where $\alpha$ corresponds
to trajectories $\rho$, $\omega$... with low interceptions: 
$\alpha(0)<1$.
The form of
dispersion relations depends 
also on the crossing symmetry of the amplitudes.
It is determined by the crossing symmetry in the ordinary
space, 
$\eta_1=\eta_2=\eta_6=1,~\eta_3=\eta'_5=\eta_8=-1$,
and the crossing symmetry in the isotopic space, 
$\eta^{(+)}=\eta^{(0)}=1$, $\eta^{(-)}=-1$.
As a result, for
all amplitudes, except $B_3^{(-)}$,
the unsubtracted dispersion relations at fixed $t$ can be written as:
\begin{eqnarray}
Re~ B_i^{(\pm,0)}(s,t,Q^2)=&&R_i^{(v,s)}(Q^2)
\left(\frac{1}{s-m^2}+
\frac{\eta_i \eta^{(+,-,0)}}{u-m^2}\right)\nonumber\\
&&+\frac{P}{\pi }\int \limits_{s_{thr}}^{\infty}
Im~ B_i^{(\pm,0)}(s',t,Q^2)
\left(\frac{1}{s'-s}+ \frac{\eta_i\eta^{(+,-,0)} }{s'-u}\right) ds',
\label{eq:dr10}
\end{eqnarray}
where
$s_{thr}=(m+m_{\pi})^2$ and
$R_i^{(v,s)}(Q^2)$ are residues in
the nucleon poles,
corresponding to diagrams (a) and (b) of Fig. \ref{fig2}.

The amplitude $B_3^{(-)}$ requires a subtraction:
\begin{eqnarray}
Re~ 
B_3^{(-)}(s,t,Q^2)=&&f_{sub}(t,Q^2)
-ge\frac{F_{\pi}(Q^2)}{t-m_{\pi}^2}
-\frac{ge}{4}\left(F_{1p}(Q^2)- F_{1n}(Q^2)\right)
\left(\frac{1}{s-m^2}+\frac{1}{u-m^2}\right)\nn \\
&&+\frac{P}{\pi } \int \limits_{s_{thr}}^{\infty}
Im~ B_3^{(-)}(s',t,Q^2)
\left(\frac{1}{s'-s}+\frac{1 }{s'-u}\right)ds',
\label{eq:dr12}
\end{eqnarray}
where $g^2/4\pi=14.2$, $F_{1N}(Q^2)$ is the nucleon
Pauli form factor, and
$F_{\pi}(Q^2)$ is the pion form factor.
At $Q^2=0$, using the relation $B_3=B_2\frac{s-u}{2(t-m_{\pi}^2)}$,
which follows from Eq. (\ref{eq:dr3}), and DR
for the amplitude $B_2(s,t,Q^2)$, 
one obtains
\begin{equation}
f_{sub}(t,Q^2)=
4\frac{P}{\pi } \int \limits_{s_{thr}}^{\infty}
\frac{Im~ B_3^{(-)}(s',t,Q^2)}{u'-s'}ds',
\label{eq:dr13}
\end{equation}
where $u'=2m^2+m_{\pi}^2-Q^2-s'-t$.
This expression for $f_{sub}(t,Q^2)$ was successfully used
in the analyses of pion photoproduction 
\cite{Aznauryan2003}
and 
low $Q^2=$0.4, 0.65 GeV$^2$ 
pion electroproduction  
on protons
\cite{Aznauryan2005,Aznauryan2009}.  
The functional form of the subtraction  
(Eq. \ref{eq:dr13}) does not allow, however, to describe 
the $\pi^+$ electroproduction data at
higher $Q^2=1.7-4.5~$GeV$^2$ \cite{Park}.
Instead, as is shown in
Ref. \cite{Aznauryan2009},
a simple parameterization,
$f_{sub}(t,Q^2)=f_1(Q^2)+f_2(Q^2)t$,
provides 
a  suitable subtraction function
at these $Q^2$.
The coefficients $f_1(Q^2),f_2(Q^2)$ that were
found at low 
$Q^{2}< 0.7~$GeV$^{2}$  using Eq. (\ref{eq:dr13})
are related smoothly to the coefficients found
at large $Q^{2}=1.7-4.5~$GeV$^{2}$ 
from the description of experimental data \cite{Aznauryan2009}.

The dispersion relations (\ref{eq:dr10},\ref{eq:dr12})
define the real parts of the amplitudes through integrals
over their imaginary parts, thus reducing 
the construction of the $\gamma^* N\rightarrow \pi N$
amplitudes to the construction of their imaginary parts.
According to the SAID analysis of the world data on pion
photoproduction on nucleons, the imaginary parts
of the multipole amplitudes in the energy region below
$W=2~$GeV are determined dominantly by resonance contributions.
In Ref. \cite{Aznauryan2003} it was shown that, with the exception of the 
mass region $W<1.3~$GeV, a good description of
the imaginary parts of the amplitudes can be obtained using 
resonance parameterizations in the Breit-Wigner form.
In Sec. \ref{sec:dr2} we discuss the parameterization of the 
multipole amplitudes for the $\Delta(1232)P_{33}$ resonance.

At $W<1.3~$GeV, the imaginary parts of the amplitudes
$E_{0+}^{0,1/2,3/2}$, $S_{0+}^{0,1/2,3/2}$, $M_{1-}^{3/2}$,
$S_{1-}^{3/2}$, $M_{1+}^{0,1/2}$, $E_{1+}^{0,1/2}$, and $S_{1+}^{0,1/2}$ can
contain significant non-resonant contributions. This is due
to the large $\pi N$ phases  $\delta_{0+}^{1/2,3/2}$,
$\delta_{1-}^{3/2}$, and 
$\delta_{1+}^{1/2}$.
In Ref. \cite{Aznauryan2003} it was shown 
that a good description of these contributions
is achieved when calculating their
real parts via DR and using the Watson theorem \cite{Watson}
for the subsequent construction of the imaginary parts:
\be
Im {\cal M}(W,Q^2)=\frac{\sin\delta}{\cos\delta}Re{\cal M}(W,Q^2),
\label{eq:dr11}
\ee
where ${\cal M}$ denotes any of the above-listed amplitudes and
$\delta$ is the corresponding $\pi N$ phase.

The dispersion integrals over the high energy region $W>2~$GeV
were estimated \cite{Aznauryan2003} 
using a gauge invariant Regge-trajectory-exchange
model developed in  Refs. \cite{Laget1,Laget2}. This model
gives a good description of the pion photoproduction data above
the resonance region and can be extended to
finite $Q^2$  \cite{Laget3}. The 
contribution of these integrals is negligible in the first and 
second resonance regions, and is small in the 
third resonance region.

\subsubsection{\it The $\Delta(1232)P_{33}$ resonance \label{sec:dr2}}

According to the $\pi N$ partial-wave analyses 
(see, for example, the results of the SAID analysis \cite{SAID}),
the amplitude corresponding to
the $\Delta(1232)P_{33}$ resonance is elastic
up to $W\approx 1.5~$GeV and at these energies
can be written
in the form
\begin{equation}
h_{1+}^{3/2}(W)=\delta_{1+}^{3/2}(W) {\rm exp}[i \delta_{1+}^{3/2}(W)].
\label{eq:dr151}
\end{equation}
\noindent 
Here the $\pi N$ partial-wave amplitudes are defined as 
$h_{l\pm}^{I}(W)\equiv (\eta^I_{l\pm}e^{2i\delta^I_{l\pm}}-1)/2i$
and $\eta^I_{l\pm}$ and $I$ are the elasticity
and isospin of the amplitude.
Therefore, for the multipole amplitudes
$M_{1+}^{3/2}$,
$E_{1+}^{3/2}$, and   
$S_{1+}^{3/2}$, corresponding to the
$\Delta(1232)P_{33}$ resonance,
the Watson theorem can be used up to energies that are much
higher than the energies in the $\Delta(1232)P_{33}$ resonance
region. In this case, the dispersion relations for 
$M_{1+}^{3/2}$,
$E_{1+}^{3/2}$,  
$S_{1+}^{3/2}$
turn into linear integral equations \cite{Schwela1,Schwela2} as:
\begin{equation}
{\cal M}(W,Q^2)={\cal M}^B(W,Q^2)+
\frac{1}{\pi}\int\limits_{W_{thr}}^{\infty}
\frac{h^*(W'){\cal M}(W',Q^2)}{W'-W-i\varepsilon}dW'
+\frac{1}{\pi}\int\limits_{W_{thr}}^{\infty} K(W,W',Q^2)h^*(W')
{\cal M}(W',Q^2)dW'.
\label{eq:dr15}
\end{equation}
Here ${\cal M}(W,Q^2)$
denotes any of the amplitudes 
$M_{1+}^{3/2}/|{\bf k}| |{\bf q}|$,
$E_{1+}^{3/2}/|{\bf k}| |{\bf q}|$,  and 
$S_{1+}^{3/2}/|{\bf k}| |{\bf q}|$;
$h\equiv h_{1+}^{3/2}$. In the integrands
the Watson theorem is used:
$Im {\cal M}(W',Q^2)=h^*(W'){\cal M}(W',Q^2)$.
${\cal M}^B(W,Q^2)$ is the contribution of the Born terms to ${\cal 
M}(W,Q^2)$, and 
$K(W,W',Q^2)$ is a non-singular kernel arising from the
$u$-channel contribution into the dispersion integral and the
non-singular part of the $s$-channel contribution.  
Here we have neglected the contributions of other multipole
amplitudes in the dispersion integrals that were estimated
\cite{Aznauryan2003} to be small.

At $K(W,W',Q^2)=0$, the integral equation (\ref{eq:dr15})
has a solution in the analytical form:
\begin{equation}
{\cal M}_{K=0}(W,Q^2)={\cal M}_{part,K=0}^{B}(W,Q^2)+
c_{{\cal M}}{\cal M}_{K=0}^{hom}(W,Q^2),
\label{eq:dr16}
\end{equation}
where
${\cal M}_{part,K=0}^{B}(W,Q^2)$ is the
particular solution of Eq. (\ref{eq:dr15}) generated by $M^B$:
\begin{equation}
{\cal M}_{part,K=0}^{B}(W,Q^2)={\cal M}^{B}(W,Q^2)+
\frac{1}{\pi}\frac{1}{D(W)}
\int\limits_{W_{thr}}^{\infty}
\frac{D(W')h(W'){\cal M}^{B}(W',Q^2)}{W'-W-i\varepsilon}dW',
\label{eq:dr17}
\end{equation}
and
\begin{equation}
{\cal M}_{K=0}^{hom}(W,Q^2)=\frac{1}{D(W)}=
\exp\left[\frac{W}{\pi}
\int\limits_{W_{thr}}^{\infty}\frac{\delta
(W')}{W'(W'-W-i\varepsilon)}dW'\right]
\label{eq:dr18}
\end{equation}
is
the solution of the homogeneous equation (\ref{eq:dr15}) with 
${\cal M}^B=0$. It
enters the solution (\ref{eq:dr16}) with an arbitrary weight factor $c_{\cal M}$.

At $K(W,W',Q^2)\neq~0$, one can transform the singular integral
equation (\ref{eq:dr15}) into the non-singular integral equation 
\cite{Aznauryan1998}.
The solution of this equation also has the form (\ref{eq:dr16}),
where both parts ${\cal M}_{part}^{B}(W,Q^2)$ and 
${\cal M}^{hom}(W,Q^2)$ are
very close to those at $K(W,W',Q^2)=0$
\cite{Aznauryan1998,Aznauryan2003}.
In the DR analyses
of $\gamma^* N\rightarrow \pi N$, the factors $c_M$, $c_E$, and
$c_S$ are fitting parameters that correspond to the
contribution of the $\Delta(1232)P_{33}$ resonance.
For other resonances, the fitting parameters are
the $\gamma^* N\rightarrow N^*$ helicity amplitudes.

\subsection{\it Unitary Isobar Model 
\label{sec:uim}}

The Unitary Isobar Model
was developed in Ref. \cite{Drechsel_UIM}.
As in the original 
effective Lagrangian approach
for pion photoproduction  \cite{Peccei}, the background of
the UIM is constructed
from the contributions of nucleon exchanges in the $s$- and
$u$-channels (Figs. \ref{fig2}a,b) and $t$-channel $\pi$ exchange (Fig. \ref{fig2}c).
The $\pi NN$ coupling is pure pseudovector at the threshold.
Such coupling follows 
from the derivation of the effective Lagrangian
in Ref. \cite{Peccei} and gives a good
description of the $E_{0+}$ amplitude at the threshold.
However, in the framework of the UIM \cite{Drechsel_UIM},
the pseudovector $\pi NN$ coupling does not provide a good
description at higher energies, and a mixed type of the
$\pi NN$ coupling is utilized, where being pure
pseudovector at the threshold, it transforms into a pseudoscalar 
coupling with increasing energy:
\be
L_{\pi NN}=\frac{\Lambda^2}{\Lambda^2+|\bf{q}|^2}L^{PV}_{\pi NN}    
+\frac{|\bf{q}|^2}{\Lambda^2+|\bf{q}|^2}L^{PS}_{\pi NN}.
\label{eq:uim1}
\ee    
In addition to these contributions,
the $t$ channel $\rho$
and $\omega$ exchanges are introduced. The background, constructed in this
way, is unitarized for each multipole amplitude in the
$K$-matrix approximation:
\be
Unitarized(M_{l\pm},E_{l\pm},S_{l\pm})_{background}=
(M_{l\pm},E_{l\pm},S_{l\pm})_{background}(1+ih_{l\pm}).
\label{eq:uim2}
\ee

The resonance contributions to multipole
amplitudes are written assuming a Breit-Wigner energy dependence of the form
\be
aA^R_{l\pm}(B^R_{l\pm},S^R_{l\pm})=
\hat{A}_{l\pm}(\hat{B}_{l\pm},\hat{S}_{l\pm})
\frac{M\Gamma_{tot} e^{i\phi}}{M^2-W^2-iM\Gamma_{tot}}
f_{\gamma N}(W),
\label{eq:uim3}
\ee
where $a$ and the
$\gamma^* N\rightarrow N^*$ helicity
amplitudes
$\hat{A}_{l\pm},\hat{B}_{l\pm},\hat{S}_{l\pm}$ 
are defined by Eqs. (\ref{eq:def41},\ref{eq:def42}),
$\phi\equiv\phi(W,Q^2)$ are the 
phases, which are found empirically for each resonance, and
$f_{\gamma N}(W)$ defines the $W$ dependence of
the $\gamma^* NN^*$ vertex beyond the resonance peak:
\be
f_{\gamma N}(W)=\left(\frac{|\bf{k}|}{|{\bf{k}}_r|}\right)^n
\left(\frac{X^2+|{\bf{k}}_r|^2}{X^2+|\bf{k}|^2}\right),
\label{eq:uim4}
\ee
$X$ is a damping parameter, assumed to be $X=0.5~$GeV for
all resonances, and $n\geq l_{\gamma}$, with
$l_{\gamma}$ the orbital angular momentum of the photon
in the $N^*$ rest frame.

The total width $\Gamma_{tot}$ is taken as the sum of $\Gamma_{\pi N}$
and the ``inelastic" width $\Gamma_{inel}$:
\bea
&&\Gamma_{\pi N}=\beta_{\pi N}\Gamma
\left(\frac{|\bf{q}|}{|{\bf{q}}_r|}\right)^{2l+1}
\left(\frac{X^2+|{\bf{q}}_r|^2}{X^2+|\bf{q}|^2}\right)^l\frac{M}{W},
\label{eq:uim5}\\
&&\Gamma_{inel}=(1-\beta_{\pi N})\Gamma
\left(\frac{|{\bf{q}}_{2\pi}|}{|{\bf{q}}_{2\pi,r}|}\right)^{2l+4}
\left(\frac{X^2+|{\bf{q}}_{2\pi,r}|^2}{X^2+|{\bf{q}}_{2\pi}|^2}\right)^{l+2},
\label{eq:uim6}
\eea
where ${\bf{q}}_{2\pi}$ is the momentum of the compound $2\pi$ system
with the mass $2m_{\pi}$. An exception is made for the 
$N(1535)S_{11}$ resonance, where the $\eta N$ channel is also taken
into account with a width similar to 
$\Gamma_{\pi N}$ (Eq. \ref{eq:uim5}).

The fitting parameters of the model are
the $\gamma N\rightarrow N^*$ helicity
amplitudes 
$\hat{A}_{l\pm},\hat{B}_{l\pm},\hat{S}_{l\pm}$ 
in Eq. (\ref{eq:uim3}), and the parameters that
define the phases $\phi(W,Q^2)$.
The $\rho NN$ and $\omega NN$ 
coupling constants, as well the parameter $\Lambda$
in Eq. (\ref{eq:uim1}),
are adjustable parameters.
At $Q^2=0$ these parameters  
were found from the description
of the SAID multipole amplitudes for $l\leq 3$ up to $W\simeq 1.7~$GeV.
The $Q^2$ dependence of the parameters is presented
in MAID2007 \cite{MAID2007}.

The UIM of Ref. \cite{Aznauryan2003} was developed on the basis of MAID
\cite{Drechsel_UIM}. One of the modifications is the incorporation of 
Regge poles with increasing energies.
This allowed the description of the SAID
pion photoproduction multipole amplitudes
up to $W=2$~GeV 
with a unified Breit-Wigner parameterization of the
resonance contributions without the energy-dependent phases $\phi(W,Q^2)$.
The phases were also assumed to be zero ($\phi(W,Q^2)=0$) 
for the electroproduction data.
The incorporation of Regge poles into the background amplitudes of UIM
was made in the following way:
\begin{eqnarray}
&&Background
=[N+\pi+\rho+\omega]_{UIM}~at~s<s_0,\nonumber\\
&&=[N+\pi+\rho+\omega]_{UIM}\frac{1}{1+(s-s_0)^2}+
Re[\pi+\rho+\omega+b_1+a_2]_{Regge}
\frac{(s-s_0)^2}{1+(s-s_0)^2}~at~s>s_0.
\label{eq:uim7}
\end{eqnarray}
Here the background of UIM is built as in Ref. \cite{Drechsel_UIM} from
the nucleon exchanges in the $s$-
and $u$-channels with a mixed type of 
$\pi NN$ coupling (Eq. \ref{eq:uim1}) and the $t$-channel
$\pi$, $\rho$, and $\omega$ exchanges.
The Regge-pole amplitudes are constructed
using the Regge-trajectory-exchange
model \cite{Laget1,Laget2,Laget3} 
and consist of
reggeized $\pi$, $\rho$, $\omega$,  $b_1$,
and $a_2$ $t$-channel 
contributions.
The background (\ref{eq:uim7})  is unitarized in the $K$-matrix 
approximation.
The value of $s_0\simeq 1.2~$GeV$^2$
was found \cite{Aznauryan2003}
from the description of the SAID multipole amplitudes.
With $s_0=1.2~$GeV$^2$, a good description of
$\pi$ electroproduction data
was obtained
at $Q^2=0.4$ and $0.65~$GeV$^2$
in the first, second and
third resonance regions \cite{Aznauryan2005,Azn065}.
When the relation (\ref{eq:uim7}) was applied 
to the analysis of data
at $Q^2 \geq 0.9~$GeV$^2$ and $W<1.8~$GeV,
the best data description was obtained with
$\sqrt s_0 > 1.8~$GeV \cite{Aznauryan2008,Aznauryan2009}, i.e.
the background of UIM is built at these $Q^2$ just from
the nucleon exchanges in the $s$-
and $u$-channels and $t$-channel
$\pi$, $\rho$, and $\omega$ exchanges.

\subsection{\it Dispersion relations and isobar model for $\eta$ photo-
and electroproduction  
\label{sec:eta_dr}}

The DR approach for $\eta$ production
was developed and used in Refs. \cite{Aznauryan2005,Aznauryaneta}.
In $\gamma^*~+~N\rightarrow N~+~\eta$, 
there are only two amplitudes in
the isotopic space, and
both have positive crossing symmetry
similar to $B_i^{(+,0)}(\gamma^*~+~N\rightarrow N~+~\pi)$.
Therefore, unlike pion production, in $\eta$ production none of the
invariant amplitudes
$B_i(\gamma^*~+~N\rightarrow N~+~\eta)$
needs a subtraction.
Another distinctive feature in $\eta$ production
is the presence of the
unphysical region from $s=s_{cut}=(m+m_{\pi})^2$ to
$s=s_{thr}=(m+m_{\eta})^2$  in the dispersion integrals. 
This region is approximated by the contribution of the Roper
resonance $N(1440)P_{11}$.

The isobar model for $\eta$ photo- and electroproduction 
was developed in Ref. \cite{etaMAID}. The amplitudes 
include non-resonant background built from
the nucleon exchanges in the $s$-
and $u$-channels and $t$-channel
$\rho$ and $\omega$ contributions.
The resonance contributions are parameterized
in the Breit-Wigner
form similar to that in Eq. (\ref{eq:uim3}).
The model was applied to the analysis of data in Refs.
\cite{Aznauryan2005,Aznauryaneta,etaMAID}.

\subsection{\it SAID  
\label{sec:said}}
The model employed by SAID for the analysis of pion photoproduction
data is presented in Refs. \cite{GWU0,GWU1}.
The $T$ matrix for pion photoproduction is parameterized 
in the form:
\be
T_{\gamma N,\pi N}=A_I(1+iT_{\pi N,\pi N}) +A_R T_{\pi N,\pi N},
\label{eq:said}
\ee 
where $T_{\pi N,\pi N}$ is the empirical
$\pi N$ amplitude available in SAID, 
and $A_I$ and $A_R$ are polynomial functions of 
the pion and photon momenta with the coefficients,
which are fitting parameters in the analyses of experimental data.
$A_I$ also includes a part that corresponds to the Born term
contribution with pseudoscalar $\pi NN$ coupling and to
the $t$-channel $\rho$ and $\omega$ exchanges.
The $N^*$ parameters are extracted by fitting the 
resulting amplitude $T_{\gamma N,\pi N}$ 
near the resonance positions using the Breit-Wigner parameterization
for the resonance contributions
similar to that in Eq. (\ref{eq:uim3}).

\subsection{\it Sato-Lee dynamical model 
\label{sec:sato-lee}}

The SL model was developed in Ref. \cite{Sato1996}, 
and later applied to investigate new data on
$ep\rightarrow e p\pi^0$ in the $\Delta(1232)P_{33}$ resonance region
(see, for example, Ref. \cite{Sato2001}).
The essential feature of the model is the consistent simultaneous description
of the $\pi N$ scattering and the pion electroproduction
on nucleons. 
The starting Hamiltonian is $H=H_0+H_I$, where $H_0$ is the free
Hamiltonian and $H_I$ is built from the $\gamma BB'$,
$\gamma MM'$, and $MBB'$ vertices with 
$B,B'=N,\Delta$ and $M,M'=\pi,\rho,\omega$. 
Using the unitary transformation method
an effective Hamiltonian is derived, where
the unphysical vertex interactions $MB\rightarrow B'$
with $m_B+m_M<m_{B'}$ are eliminated by absorbing
their effects into $MB\rightarrow M'B'$ two-body interactions.
The resulting  effective
Hamiltonian has the following form:
\be 
H_{eff}=H_0+v_{\pi N} +v_{\gamma N}
+\Gamma_{\pi N\rightarrow \Delta}
+\Gamma_{\gamma N\rightarrow \Delta}.
\label{eq:sl1}
\ee 
Here $v_{\pi N}$ is a non-resonant $\pi N$ potential,
and $v_{\gamma N}$ describes the non-resonant
$\gamma N\leftrightarrow \pi N$ transition that
consists of the contributions of 
the Born terms (Figs. \ref{fig2}a,b,c) with 
pseudovector $\pi NN$ coupling, the $t$-channel $\rho$
and $\omega$  exchanges, and the crossed $\Delta$ term.
The $\Delta$ excitation is described by the bare vertices
$\Gamma_{\gamma N\rightarrow \Delta}$ 
and $\Gamma_{\pi N\rightarrow \Delta}$.

The pion electroproduction amplitude derived from
the effective Hamiltonian (\ref{eq:sl1}) can be decomposed
into two parts:    
\be 
t_{\gamma \pi} (E)=t^b_{\gamma \pi} (E)+
\frac{\bar{\Gamma}_{\Delta\rightarrow \pi N}
\bar{\Gamma}_{\gamma N\rightarrow \Delta}}{E-m_{\Delta}-\Sigma_{\Delta}(E)},
\label{eq:sl2}
\ee 
where $t^b_{\gamma \pi} (E)$ is the non-resonant amplitude
calculated from $v_{\gamma\pi}$ by
\be 
t^b_{\gamma \pi} (E)=v_{\gamma\pi} 
+t^b_{\pi N}(E)G_{\pi N}(E)v_{\gamma\pi},
\label{eq:sl3}
\ee 
$G_{\pi N}(E)$ is the $\pi N$ free propogator,
and $t^b_{\pi N}(E)$ is calculated from the non-resonant 
$\pi N$ interaction $v_{\pi N}$ by solving an 
equation similar to Eq. (\ref{eq:sl3}) for 
the $\pi N$ scattering.
The dressed vertices $\bar{\Gamma}_{\gamma N\rightarrow \Delta}$, 
and $\bar{\Gamma}_{\Delta\rightarrow \pi N}$,  
and the $\Delta$ self-energy $\Sigma_{\Delta}(E)$ in 
Eq. (\ref{eq:sl2}), have the following form:
\bea 
&&\bar{\Gamma}_{\gamma N\rightarrow \Delta}=
\Gamma_{\gamma N\rightarrow \Delta}+
\bar{\Gamma}_{\Delta\rightarrow \pi N}G_{\pi N}(E)v_{\gamma\pi},
\label{eq:sl4}\\
&&\bar{\Gamma}_{\Delta\rightarrow \pi N}=
\left[1+t^b_{\pi N}(E)G_{\pi N}(E)\right]\Gamma_{\Delta\rightarrow \pi N},
\label{eq:sl5}\\
&&\Sigma_{\Delta}(E)=\Gamma_{\pi N \rightarrow \Delta}
G_{\pi N}(E)\bar{\Gamma}_{\Delta\rightarrow \pi N}.
\label{eq:sl6}
\eea 
In Eqs. (\ref{eq:sl3}-\ref{eq:sl6}), the integrals over the momenta
of the intermediate particles are written in the terms that
contain $G_{\pi N}$. These terms reflect the
effects arising from the $\pi N$ final state interaction.  
As is seen, 
the model explicitly identifies the influence of the
final state interaction
on the resonance properties; in particular,
the bare vertex $\Gamma_{\gamma N\rightarrow \Delta}$
is modified  
into the dressed vertex $\bar{\Gamma}_{\gamma N\rightarrow \Delta}$.
The fitting parameters in the analyses of pion photo- and electroproduction
data are the parameters that describe the bare vertex 
$\Gamma_{\gamma N\rightarrow \Delta}$.

In the application of the model, as a first step, 
the amplitude $t_{\pi N}$ 
is obtained from solving of the integral
equations for the $\pi N$ scattering and 
fitting to the $\pi N$ phase shifts.
The subsequent fit to the pion electroproduction
data \cite{Sato2001} allows the separation of the bare and dressed 
contributions to the $\gamma^*p\rightarrow \Delta(1232)P_{33}$ transition.

\subsection{\it Dubna-Mainz-Taipei dynamical model 
\label{sec:dmt}}
The main feature of the DMT model
\cite{Kamalov1999,Kamalov2001} is that the unitarity
is built via direct inclusion of the $\pi N$ final state 
interaction in the $t$-matrix for pion photo- and
electroproduction:
\be 
t_{\gamma \pi} (E)=v_{\gamma\pi} 
+v_{\gamma\pi}G_{\pi N}(E)t_{\pi N},
\label{eq:dmt1}
\ee 
 where $v_{\gamma\pi}$ is the transition potential
for $\gamma^* N\rightarrow \pi N$ and $t_{\pi N}$
is the $\pi N$ scattering $t-$matrix.

For the multipole amplitudes corresponding to
the $\Delta(1232)P_{33}$ resonance, the transition potential
is built from two terms: 
\be 
v_{\gamma \pi} (E)=v^B_{\gamma\pi}(E) 
+v^{\Delta}_{\gamma\pi}(E),
\label{eq:dmt2}
\ee 
where $v^B_{\gamma\pi}$ is the background
potential and $v^{\Delta}_{\gamma\pi}$ corresponds to
the bare $\Delta$ contribution to $\gamma^* N\rightarrow \pi N$.
Conequently, the  $t-$matrix is decomposed in two terms:
\be 
t_{\gamma \pi} (E)=t^B_{\gamma\pi} 
+t^{\Delta}_{\gamma\pi},
\label{eq:dmt3}
\ee 
where
\bea 
&&t^B_{\gamma \pi}=v^B_{\gamma \pi}+
v^B_{\gamma\pi}G_{\pi N}(E)t_{\pi N},
\label{eq:dmt4}\\
&&t^{\Delta}_{\gamma \pi}=v^{\Delta}_{\gamma \pi}+
v^{\Delta}_{\gamma\pi}G_{\pi N}(E)t_{\pi N}.
\label{eq:dmt5}
\eea 
Irrespective of the similarity between  Eqs. (\ref{eq:sl3})
and (\ref{eq:dmt4}), there is a significant difference
in the construction of the background amplitudes
in the SL and DMT models.  In contrast with $t^b_{\gamma \pi}$
in the SL model, the background amplitude (\ref{eq:dmt4})
includes the contributions not only from the nonresonant
mechanisms, but from the full $t_{\pi N}$.

There is also a difference in the construction of the
background contributions to 
$\gamma^* N\rightarrow \pi N$. While in the SL model it is constructed
similar to $\pi N$ scattering using the pseudovector $\pi NN$ coupling,
in the DMT model a mixed type of coupling is used
as in MAID (Eq. \ref{eq:uim1}). In the practical applications
of the DMT model,   
the bare resonance contribution is taken in the Breit-Wigner
form and is parameterized as the resonance contribution
in the UIM (Eq. \ref{eq:uim3}).
The fitting parameters of the model are the bare 
$\hat{A}_{1+},\hat{B}_{1+},\hat{S}_{1+}$
helicity amplitudes for the vertex 
$\Gamma_{\gamma N\rightarrow \Delta}$.
The coupling constants for $\gamma^*\pi\rho$ and $\gamma^*\pi\omega$ vertices, 
as well the parameters that define the phase $\phi(W,Q^2)$
in Eq. (\ref{eq:uim3}),
are adjustable parameters.

\section{Data, Analyses, and Description of the Observables 
\label{sec:data}}

\subsection{\it The $\Delta(1232)P_{33}$ resonance region 
\label{sec:data_delta}}

The electroexcitation of the $\Delta(1232)P_{33}$ resonance
has been studied for more than 50 years, but only in the past
decade have the experimental tools become available
to enable precise measurements in exclusive $\pi$ electroproduction
from protons in a large range of photon virtualities $Q^2$.  
The electroexcitation of the $\Delta(1232)P_{33}$
is dominated by the magnetic-dipole
$\gamma^* N\rightarrow \Delta(1232)P_{33}$ transition
in the entire range $Q^2<8$~GeV$^2$, while 
the electric-quadrupole and scalar-quadrupole amplitudes
remain comparatively much smaller. Precise extraction of the
corresponding ratios
$R_{EM}\equiv ImE^{3/2}_{1+}/ImM^{3/2}_{1+}$ and
$R_{SM}\equiv ImS^{3/2}_{1+}/ImM^{3/2}_{1+}$
at the resonance position
has been one of the main goals of experiments in the
$\Delta(1232)P_{33}$ resonance region.
$R_{EM}$ and $R_{SM}$ are of great interest as their magnitude and
sign are associated with the quadrupole deformation of the nucleon 
and the $\Delta(1232)P_{33}$. 
A thorough discussion of the mechanisms that connect these phenomena
can be found in Refs. \cite{Buchmann1,Buchmann2}.
$R_{EM}$ and $R_{SM}$ are also a measure of the 
$Q^2$ scale where the approach
to the asymptotic domain of QCD may set in.
Earlier experiments at DESY, Bonn, and NINA were
limited to $Q^2 \le 3$~GeV$^2$. The precision and reach
in angular coverage were rather limited and did not allow for  
accurate determination of these quantities as a function of $Q^2$ \cite{Foster}. 
As we will show in section \ref{sec:results}, the new data in the 
$\Delta(1232)P_{33}$ region \cite{Cole,Joo1,Ungaro,Joo2,Biselli,Joo3,
Frolov,Vilano,KELLY1,KELLY2,Stave2006,
Sparveris2007,Stave2008,Pospishil,
Mertz,Kunz,Sparveris2005,Warren}
led to the determination of $R_{EM}$ and $R_{SM}$ 
with high accuracy in the range $Q^2<7$~GeV$^2$.

As the $\Delta(1232)P_{33}$ resonance 
is located at low energies and just above the pion threshold, 
the production cross section is determined 
mainly by contributions from $s$- and $p$-
wave. The corresponding $\pi N$
amplitudes are elastic in this region, and for the $s$- and 
$p$-wave multipole amplitudes
$E^I_{0+}$, $S^I_{0+}$, $M^I_{1-}$, $S^I_{1-}$,
$M^I_{1+}$, $E^I_{1+}$,  $S^I_{1+}$  
($I=\frac{1}{2},\frac{3}{2}$ is the total isospin
in the $s$-channel), 
the Watson theorem \cite{Watson} can be applied:
${\cal M}^I(W,Q^2)=|{\cal M}^I(W,Q^2)|{\rm exp}(i\delta^I(W))$,
where ${\cal M}^I(W,Q^2)$
denotes any of these multipoles, and
$\delta^I(W)$ is the corresponding $\pi N$ phase.
This constrains the energy-dependence of the multipole 
amplitudes in the $\Delta(1232)P_{33}$ region
and reduces the model dependence of the 
$\gamma^* N\rightarrow \Delta(1232)P_{33}$ amplitudes
extracted from the data.

\begin{figure}[ht!]
\begin{center}
\begin{minipage}[t]{14 cm}
\epsfig{file=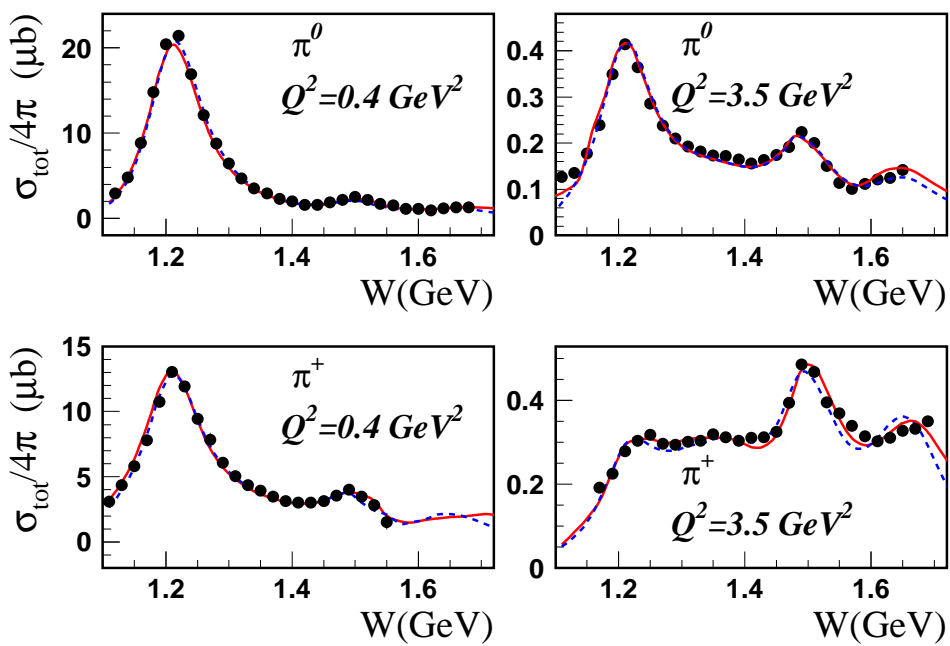,scale=0.75}
\end{minipage}
\begin{minipage}[t]{16.5 cm}
\caption{$W$-dependence of the $\gamma^* p\rightarrow \pi^0 p$
and $\gamma^* p\rightarrow \pi^+ n$ total cross sections at $Q^2=0.4$ 
and $3.5~$GeV$^2$.
Data for $\pi^0 p$ production are from Refs. \cite{Joo1,Ungaro_pr},
and for $\pi^+ n$ from Refs. \cite{Egiyan,Park}. The solid (dashed) 
curves
correspond to the results obtained by
the JLab group \cite{Aznauryan2009}
using the DR (UIM) approach.
\label{sigma}}
\end{minipage}
\end{center}
\end{figure}

The $\Delta(1232)P_{33}$ as an isospin $\frac{3}{2}$ state is 
coupled more strongly to the $\pi^0 p$ final state
than to $\pi^+ n$. 
In addition, the main non-resonant contribution is associated
in the $\Delta(1232)P_{33}$ resonance region
with the multipole amplitude $ReE_{0+}$, which 
at low $Q^2$ is much smaller
in the $\pi^0 p$ channel compared to $\pi^+ n$. 
This has been found in 
the partial-wave analyses of the reactions
$\gamma^* p\rightarrow \pi^0 p$ and
$\gamma^* p\rightarrow \pi^+ n$ and 
in the calculations within DR and
effective Lagrangian approaches
starting from the first investigations  
\cite{Chew,Peccei}.
For this reason, most experiments
studying the electroexcitation of the 
$\Delta(1232)P_{33}$ make use of the process
$ep\rightarrow e \pi^0 p$.  At low $Q^2$, this reaction 
(and in less degree $ep\rightarrow e \pi^+ n$)
is dominated by the $\Delta(1232)P_{33}$
contribution;  however, with increasing $Q^2$, the resonance
structure near $1.5~$GeV and the contribution
of the broad $N(1440)P_{11}$ state become increasingly dominant
in comparison with the $\Delta(1232)P_{33}$.
This is demonstrated in Fig. \ref{sigma}, where the total
photoabsorption cross sections $\gamma^* p\rightarrow \pi^0 p$ and 
$\gamma^* p\rightarrow \pi^+ n$ are shown
at $Q^2=0.4$ and $3.5~$GeV$^2$. 
The dominance of  the $\Delta(1232)P_{33}$
contribution at small $Q^2$ is often used as justification to perform simplified
analyses of the $ep\rightarrow e \pi^0 p$ data
based on the truncated multipole expansion where only terms
that contain $M_{1+}^{3/2}$ linearly or
quadratically are retained. However, 
with increasing $Q^2$ such a truncated multipole analysis is no longer  
justified, and more suitable approaches are needed to get proper results.  

In the following we list experiments for the 
reaction
$ep\rightarrow e\pi^0 p$ in
the $\Delta(1232)P_{33}$ resonance region. These are
experiments at MAMI and MIT/Bates, and also
measurements performed at JLab in Hall A and Hall C.  
The JLab/Hall B measurements using CLAS include
both reactions 
$ep\rightarrow e \pi^0 p$ and $ep\rightarrow e \pi^+ n$,
and the kinematics extends over a wider energy range that includes
the higher mass states 
$N(1440)P_{11}$, $N(1520)D_{13}$, and $N(1535)S_{11}$, 
and partly the third resonance region near 1.68 GeV
(see  Table \ref{tab:data}).
Analysis of these data was made 
by the JLab group 
\cite{Aznauryan2008,Aznauryan2009} and at Mainz \cite{MAID2007,MAID_China}
for both reactions and in both resonance regions combined.
For this reason, we found it expedient to present 
the results related to the CLAS experiments all together
in a separate section \ref{sec:data_hallb}.

\subsubsection{\it Measurements at MAMI 
\label{sec:data_mami}}

\begin{figure}[ht!]
\begin{center}
\begin{minipage}[t]{16.5 cm}
\epsfig{file=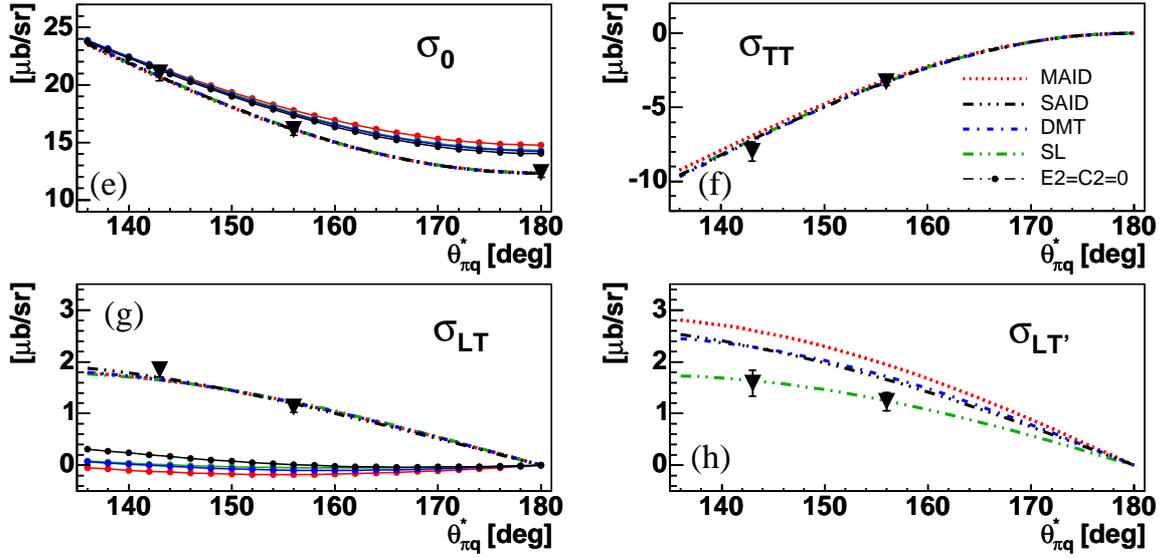,scale=0.6}
\end{minipage}
\begin{minipage}[t]{16.5 cm}
\caption{
In panels (e-h), the structure functions 
$\sigma_0\equiv \sigma_T+\epsilon \sigma_L$,
$\sigma_{LT}$, 
$\sigma_{TT}$, and $\sigma_{LT'}$
are shown that are 
measured in MAMI experiments
\cite{Stave2006,Sparveris2007,Stave2008}
at $W=1.221~$GeV and $Q^2=0.2~$GeV$^2$.
The curves correspond to analyses 
using SAID, MAID, 
and the dynamical models DMT \cite{Kamalov1999} 
and SL \cite{Sato2001}.
The curves with dots (noted by $E2=C2=0$) are the 
model cross sections with $E_{1+}^{3/2}$ and $S_{1+}^{3/2}$
set to zero and fitted $M_{1+}^{3/2}$; they are plotted
only for the sensitive observables, $\sigma_{0}$ and $\sigma_{LT}$.
({\it Source:} From Ref. \cite{Stave2008}.)
\label{mami}}
\end{minipage}
\end{center}
\end{figure}

Experiments in the   
$\Delta(1232)P_{33}$ resonance region carried out at the 
Mainz Microtron (MAMI) include the following measurements:

(i) Differential cross sections and polarized beam 
asymmetries were measured in
$\vec{e}p\rightarrow e\pi^0 p$
for $Q^2=0.06,0.127$,$0.2~$GeV$^2$ 
\cite{Stave2006,Sparveris2007,Stave2008}, and the
structure functions $\sigma_T+\epsilon \sigma_L$,
$\sigma_{LT}$, $\sigma_{TT}$, 
$\sigma_{LT'}$ were determined for polar angles in the range
$\theta^*_{pq} \equiv 180^{\circ}-\theta=120 - 180^{\circ}$
(see Fig. \ref{mami}). 
The data from Refs.
\cite{Stave2006,Sparveris2007,Stave2008} were analyzed in
Ref. \cite{Stave2008} using SAID, MAID, 
and the dynamical models DMT \cite{Kamalov1999}
and SL \cite{Sato2001}.
The results are given as  
average values of those obtained within the different approaches.
They will be shown as MAMI results when presenting the entire set of
results for the $\Delta(1232)P_{33}$. 

(ii) The recoil proton polarization has been measured in 
$\vec{e}p\rightarrow e\pi^0 \vec{p}$ 
for $Q^2=0.121~$GeV$^2$ \cite{Pospishil}. 
The value of $R_{SM}$ was extracted from the data
using MAID. The quoted statistical and systematic uncertainties 
are significantly larger compared to those found in Refs.
\cite{Stave2008,Sparveris2005}; therefore this value of $R_{SM}$
will not be shown while presenting the entire set of results. 

(iii) There are also measurements of 
$\sigma_{LT'}$ \cite{Bartsch}
and $\sigma_{LT}$ \cite{Elsner}; 
however, no results on   the
$\gamma^* N\rightarrow \Delta(1232)P_{33}$ amplitudes
are quoted in these papers. 

\subsubsection{\it MIT/Bates
\label{sec:data_mit}}

\begin{figure}[tb]
\begin{center}
\begin{minipage}[t]{17 cm}
\includegraphics[angle=270,width=6.5in]{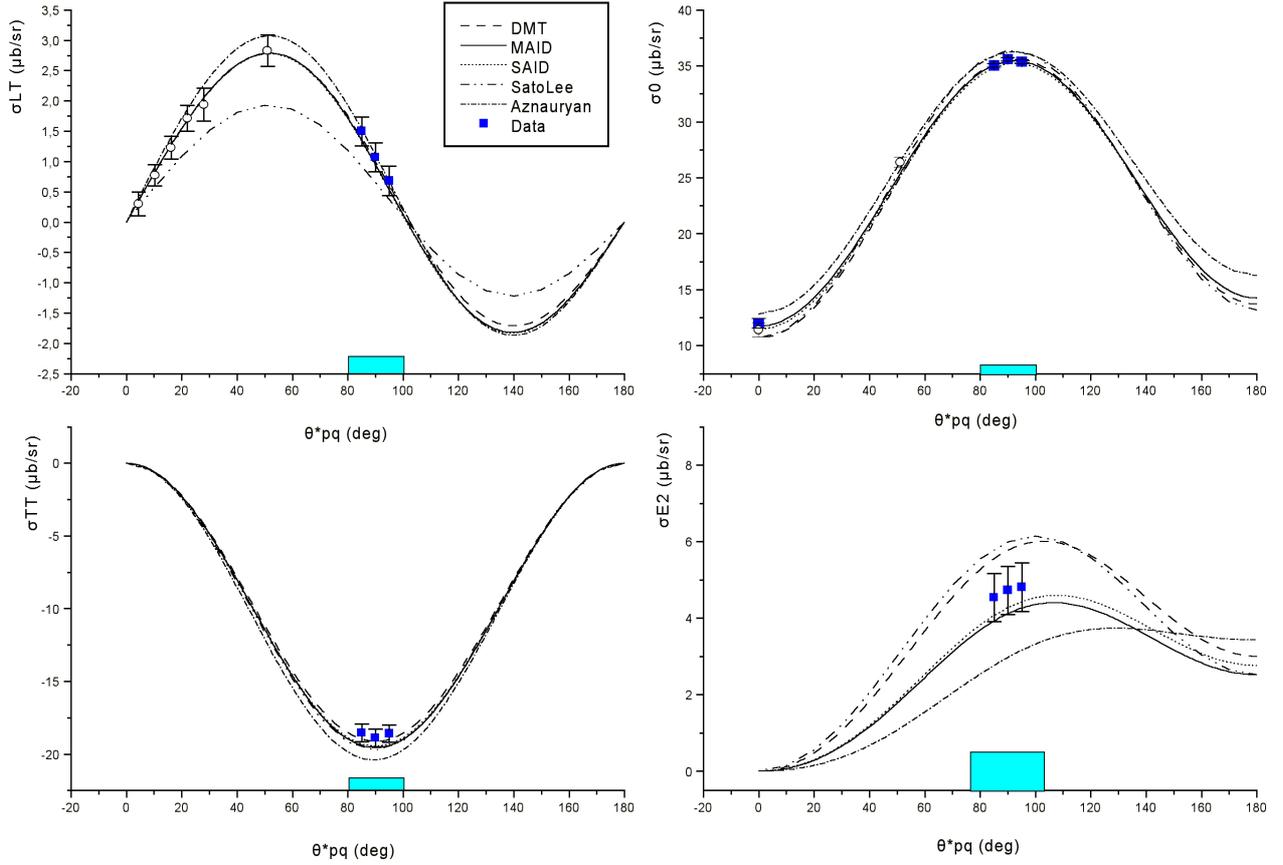}
\end{minipage}
\begin{minipage}[t]{16.5 cm}
\caption{
The structure functions 
$\sigma_0\equiv \sigma_T+\epsilon \sigma_L$,
$\sigma_{LT}$, $\sigma_{TT}$, and 
$\sigma_{E2}\equiv \sigma_0(\theta^*_{pq})+\sigma_{TT}(\theta^*_{pq})
-\sigma_0(\theta^*_{pq}=0^{\circ})$ 
measured in MIT/Bates experiments
\cite{Mertz,Kunz,Sparveris2005}
at $W=1.232~$GeV and $Q^2=0.127~$GeV$^2$.
The open circles are data from 
\cite{Mertz,Kunz} and
the filled squares are from  
\cite{Sparveris2005}.
The curves correspond to model analyses using SAID (dotted), MAID (solid), 
DR \cite{Aznauryan2003} (dashed-dotted),
and dynamical models DMT \cite{Kamalov1999} (dashed)
and SL \cite{Sato2001} (dash-double dotted).
The shaded bands show the estimated systematic uncertainties.
({\it Source:} From Ref. \cite{Sparveris2005}.)  
\label{mit_bates}}
\end{minipage}
\end{center}
\end{figure}

There is a group of three experiments  
\cite{Mertz,Kunz,Sparveris2005} performed
using the MIT/Bates linear accelerator and the out-of-plane 
scattering (OOPS) facility. The $ep\rightarrow e\pi^0 p$
cross section at $Q^2=0.127~$GeV$^2$
was measured for several values of the polar
angle $\theta^*_{pq}$ 
and for azimuthal angles $\phi^*_{pq}\equiv 180^{\circ}+\phi$: 
$\phi^*_{pq}=0^{\circ},180^{\circ}$ \cite{Mertz},
$\phi^*_{pq}=225^{\circ},315^{\circ}$ \cite{Kunz}, and
$\phi^*_{pq}=60^{\circ},90^{\circ},180^{\circ}$ \cite{Sparveris2005}.
The choice of the azimuthal angles allows  separation of 
structure functions $\sigma_T+\epsilon \sigma_L$,
$\sigma_{LT}$, and
$\sigma_{TT}$. They are presented in Fig. \ref{mit_bates} along with
theoretical predictions.
The multipole analysis in Ref. \cite{Sparveris2005}
uses the following approaches: SAID, MAID, DR 
\cite{Aznauryan2003},
and dynamical models DMT 
\cite{Kamalov1999}
and SL \cite{Sato2001}.
The final results that are shown below 
as MIT/Bates results were obtained by
averaging those obtained using these approaches.

There is also a measurement of the induced proton polarization
in $\pi^0$ electroproduction at 
$Q^2=0.126~$GeV$^2$ \cite{Warren}; however,
no result on the electroexcitation of the  
$\Delta(1232)P_{33}$  is presented.

\subsubsection{\it JLab/Hall A
\label{sec:data_halla}}

\begin{figure}[tb]
\begin{center}
\begin{minipage}[t]{17 cm}
\includegraphics[angle=90,width=6.5in]{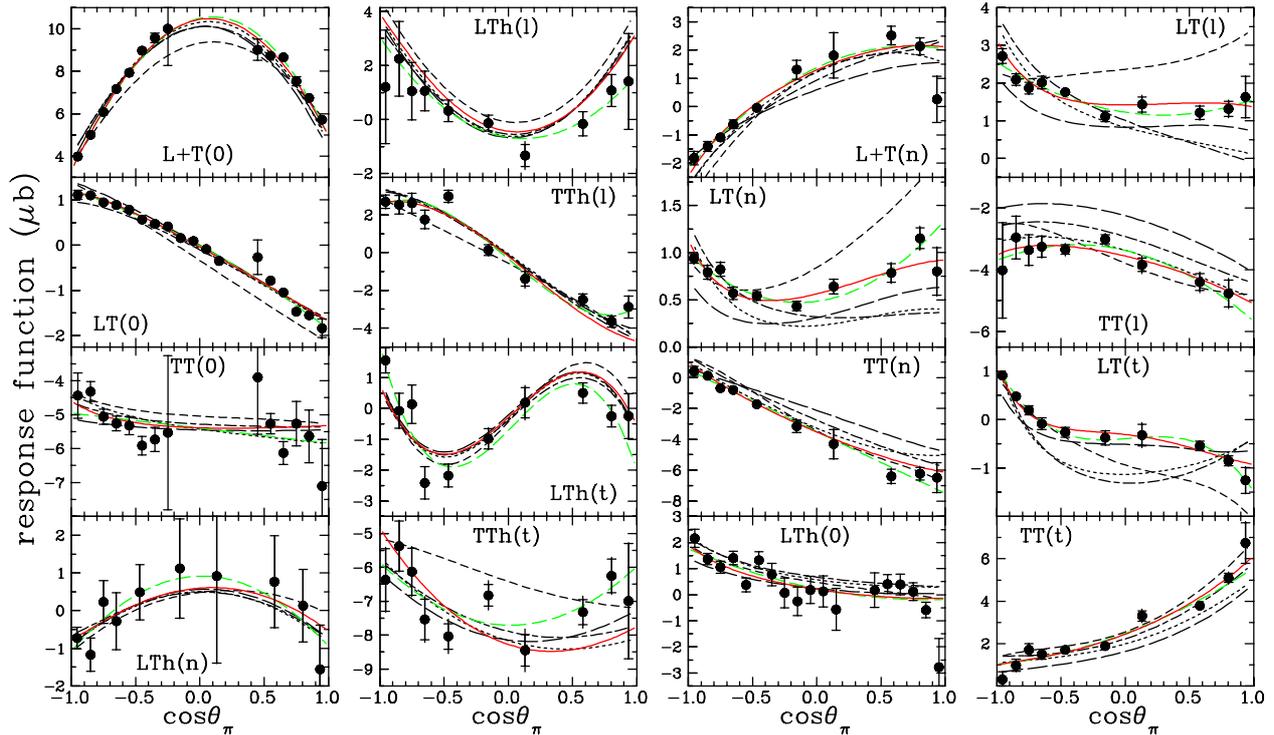}
\end{minipage}
\begin{minipage}[t]{16.5 cm}
\caption{
JLab/Hall A data 
for the $\vec{e}p\rightarrow e\pi^0 \vec{p}$
response functions  
at $W=1.232~$GeV and $Q^2=1.0~$GeV$^2$ \cite{KELLY1,KELLY2}.
Notations refer to transverse (t), normal (n)
and longitudinal (l) components of the proton recoil
polarization.
The curves correspond to the results obtained
using SAID (short-dashed), MAID (dashed-dotted), 
and the dynamical models DMT \cite{Kamalov1999} (dotted)
and SL \cite{Sato2001} (long-dashed/green).
The mid-dashed and solid curves correspond, respectively,
to the Legendre and multipole fits performed by the authors.  
({\it Source:} From Ref. \cite{KELLY1}.)  
\label{kelly}}
\end{minipage}
\end{center}
\end{figure}

The JLab/Hall A experiment \cite{KELLY1,KELLY2}
was performed with a polarized beam and a high resolution magnetic
spectrometer instrumented with a recoil polarimeter to measure the proton
polarization. This setup allowed measurement of 16 response functions at 
$Q^2=1.0~$GeV$^2$ in $\vec{e}p\rightarrow e\pi^0 \vec{p}$.
Twelve of these response functions were measured for the
first time. These data are shown in Fig. \ref{kelly}
at the peak of the $\Delta(1232)P_{33}$ and compared to 
the results of phenomenological analyses.
The experimental information was sufficiently
complete to perform a multipole analysis and to determine 
the ratios $R_{EM}$ and $R_{SM}$ at the resonance mass 
in a model-independent way.
 
\subsubsection{\it JLab/Hall C
\label{sec:data_hallc}}

\begin{figure}[ht!]
\begin{center}
\begin{minipage}[l]{200pt}
\epsfig{file=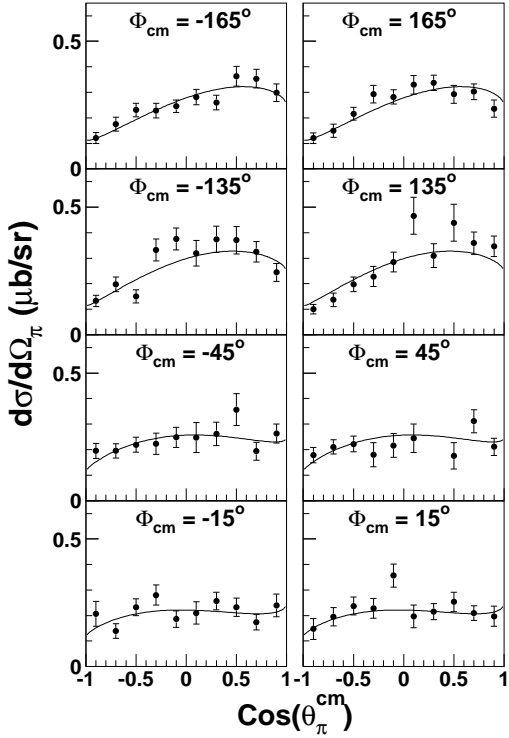,scale=0.55}
\end{minipage}
\hskip 40pt
\begin{minipage}[r]{200pt}
\caption{Examples of angular distributions and the
description of the
JLab/Hall C data
for $\gamma^* p\rightarrow \pi^0 p$
at $W=1.235~$GeV and $Q^2=4.0~$GeV$^2$
from Ref. \cite{Frolov}.
({\it Source:} From Ref. \cite{Frolov}.)  
\label{frolov}}
\end{minipage}
\end{center}
\end{figure}

Two measurements of differential cross sections for 
$ep\rightarrow e\pi^0 p$ in the
$\Delta(1232)P_{33}$ resonance region were performed 
at JLab/Hall C at $Q^2=2.8,~4.2$ \cite{Frolov} and  
$6.4,~7.7~$GeV$^2$ \cite{Vilano}. 
The analysis of the lower $Q^2$ data was
made in Ref. \cite{Frolov} using 
the effective Lagrangian approach 
\cite{David1,David2,David3}.
The data and description
at the resonance position are 
shown in Fig. \ref{frolov}. 

\begin{figure}[ht!]
\begin{center}
\begin{minipage}[t]{16 cm}
\epsfig{file=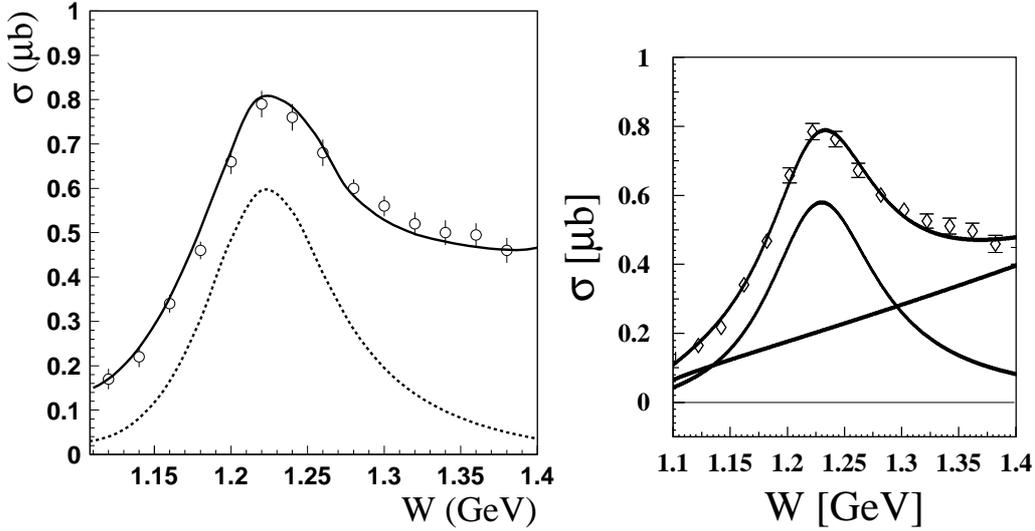,scale=0.4}
\epsfig{file=vil_total.epsi,scale=0.63}
\end{minipage}
\begin{minipage}[t]{16.5 cm}
\caption{Description of 
the $\gamma^* p\rightarrow \pi^0 p$ 
total cross section for
$Q^2\simeq 6.4~$GeV$^2$  \cite{Vilano}. 
Left panel: the description by Aznauryan using UIM; the solid and dashed
curves are the total and resonance contributions, respectively.
Right panel: plot from Ref. \cite{Vilano}, where
the results of the truncated multipole analysis are shown;
the curves represent the total, resonance and background contributions.
\label{vilano}}
\end{minipage}
\end{center}
\end{figure}

The results for $Q^2=6.4,7.7~$GeV$^2$, presented in Ref. \cite{Vilano},
are obtained in two different approaches.
One is based on the truncated multipole expansion, which
as we argued above, cannot be justified
at these large values of $Q^2$. The other analysis performed by Aznauryan
is based on the UIM. The latter approach was used for the analysis 
of all CLAS data in Ref. \cite{Aznauryan2009}, and the corresponding results will be
presented and discussed below. 
The results obtained in the two analyses strongly disagree
with each other both for the magnetic transition form factor and 
the $R_{EM}$ and $R_{SM}$  ratios.
However, in the former case, disagreement for the 
$\gamma^* N\rightarrow \Delta(1232)P_{33}$
magnetic-dipole form factor is caused 
by a numerical mistake in the extraction of this form factor
from experimental data. 
To demonstrate this, we present in Fig. \ref{vilano}
the resonance contributions to the total 
$\gamma^* p\rightarrow \pi^0 p$ 
cross section for
$Q^2\simeq 6.4~$GeV$^2$. 
The resonance contributions
obtained in the two analyses
are practically identical and the magnetic dipole-form factor also should be 
the same.  
The difference in the values of the $\gamma^* N\rightarrow 
\Delta(1232)P_{33}$
magnetic-dipole form factor extracted from these 
contributions is caused by a factor of 2/3 that was missed in the first analysis.
Therefore, when presenting the whole set of
results on the $\Delta(1232)P_{33}$, we will show from
Ref. \cite{Vilano} only those obtained by Aznauryan.

\subsection{\it JLab/Hall B: CLAS data in the $\Delta(1232)P_{33}$, 
$N(1440)P_{11}$, $N(1520)D_{13}$, and $N(1535)S_{11}$
resonance regions in pion electroproduction
\label{sec:data_hallb}}

The CLAS detector at Jefferson Lab
is the first full acceptance instrument
designed for the comprehensive investigation
of exclusive meson production with the goal
to study the excitation of nucleon
resonances in a large kinematics regime with both photon and 
electron beams.
The angular acceptance of CLAS provides nearly $4\pi$
coverage in solid angle. 
In recent years, a variety of measurements
of single pion electroproduction on protons has been performed
at CLAS in a wide range of $Q^2$ from 0.16 to 6 GeV$^2$
\cite{Cole}-\cite{Joo3} (see Table \ref{tab:data}).
The data include nearly 
120,000 points of differential cross sections,
longitudinally polarized
beam asymmetries ($A_{LT'}$), and longitudinal target ($A_{t}$) 
and beam-target ($A_{et}$) asymmetries.
A comprehensive analysis of these data was performed
by the JLab group \cite{
Aznauryan2008,Aznauryan2009} and at Mainz \cite{MAID2007,MAID_China}.
The JLab analysis was performed
using two approaches: DR and UIM.
The analysis at Mainz was based on 
the UIM \cite{Drechsel_UIM} given by its MAID2007  version
\cite{MAID2007}.

The amplitudes for the electroexcitation of the $\Delta(1232)P_{33}$
resonance were determined in the wide range of $Q^2$:
$0.16 \leq Q^2\leq 6~$GeV$^2$.
The extracted amplitudes are in agreement with the 
low $Q^2$ results from
MAMI \cite{Stave2008} and
MIT/Bates \cite{Sparveris2005}, and
with the high $Q^2$ results from
JLab Hall A ($Q^2=1~$GeV$^2$) \cite {KELLY1,KELLY2}
and  Hall C:
$Q^2= 2.8,4.2~$GeV$^2$ \cite{Frolov} and $Q^2= 6.4,7.7~$GeV$^2$ \cite{Vilano}.
The electroexcitation of the
resonances $N(1440)P_{11}$,
$N(1520)D_{13}$, and $N(1535)S_{11}$
was  investigated with high precision in the range
$0.3 \leq Q^2< 4.5~$GeV$^2$. 
For the first time the electrocoupling amplitudes of the Roper resonance 
$\gamma^* p\rightarrow N(1440)P_{11}$ and the longitudinal
amplitudes for the transitions
$\gamma^* p\rightarrow N(1520)D_{13}$ and
$N(1535)S_{11}$
were extracted from data. 

\begin{figure}[ht!]
\begin{center}
\begin{minipage}[t]{13.5 cm}
\epsfig{file=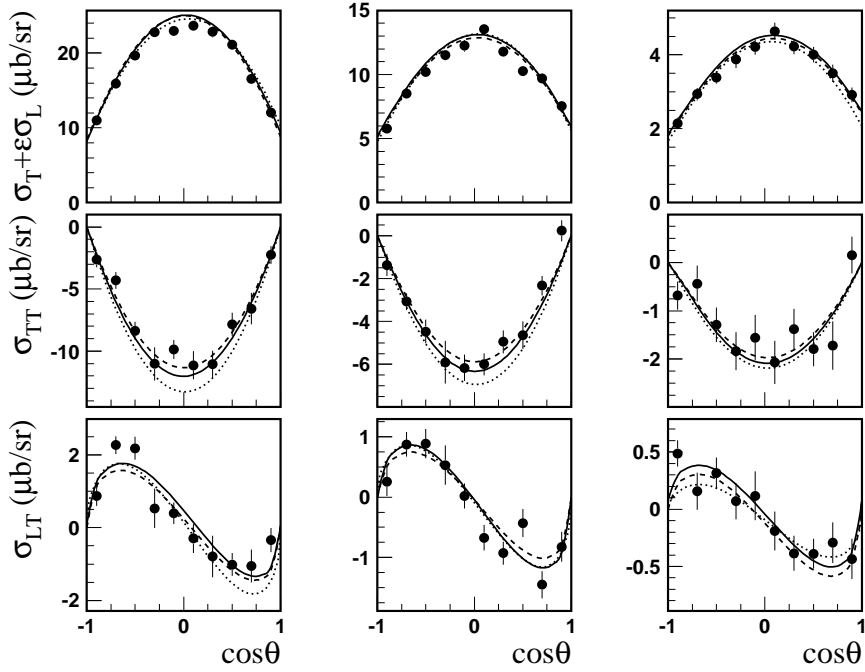,scale=0.65}
\end{minipage}
\begin{minipage}[t]{16.5 cm}
\caption{
JLab \cite{Aznauryan2009} and MAID2007 \cite{MAID2007} 
results for the
$ep\rightarrow ep\pi^0$ structure functions
(in $\mu$b/sr units)
in comparison with experimental data \cite{Joo1} for
$W=1.23$~GeV.
The columns correspond to
$Q^2=0.4,~0.75,~1.45$~GeV$^2$.
The solid (dashed) curves
correspond to the JLab results 
obtained using the DR (UIM) approach.
The dotted curves are from MAID2007.
\label{123_04}}
\end{minipage}
\end{center}
\end{figure}

\begin{figure}[ht!]
\begin{center}
\begin{minipage}[t]{13.5 cm}
\epsfig{file=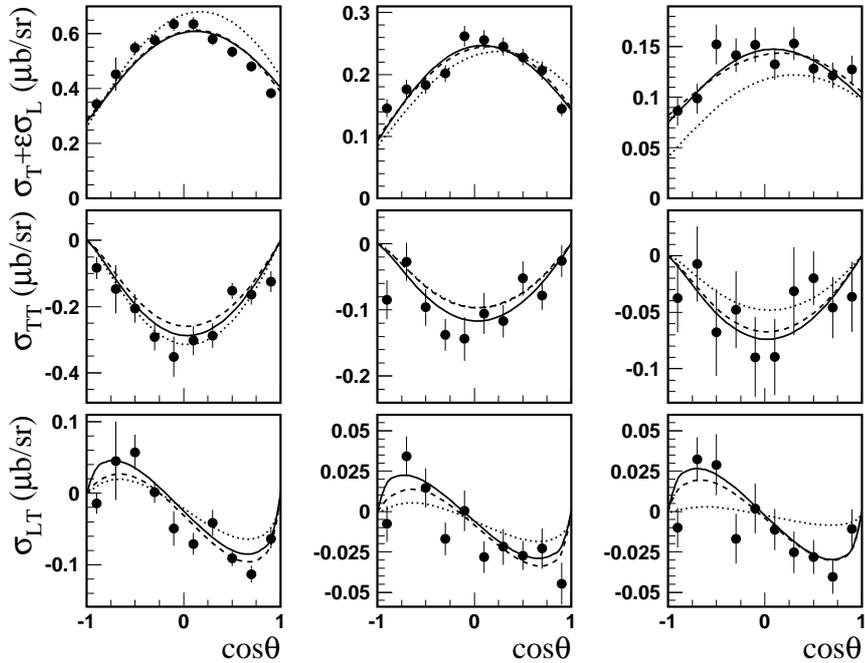,scale=0.65}
\end{minipage}
\begin{minipage}[t]{16.5 cm}
\caption{
JLab \cite{Aznauryan2009} and MAID2007 \cite{MAID2007}
results for the
$ep\rightarrow ep\pi^0$ structure functions
(in $\mu$b/sr units)
in comparison with experimental data \cite{Ungaro} for
$W=1.23~$GeV.
The columns correspond to
$Q^2=3,~4.2,~5~$GeV$^2$.
Other notations are as in Fig. \ref{123_04}.
\label{123_3}}
\end{minipage}
\end{center}
\end{figure}

\begin{figure}[ht!]
\begin{center}
\begin{minipage}[t]{13.5 cm}
\epsfig{file=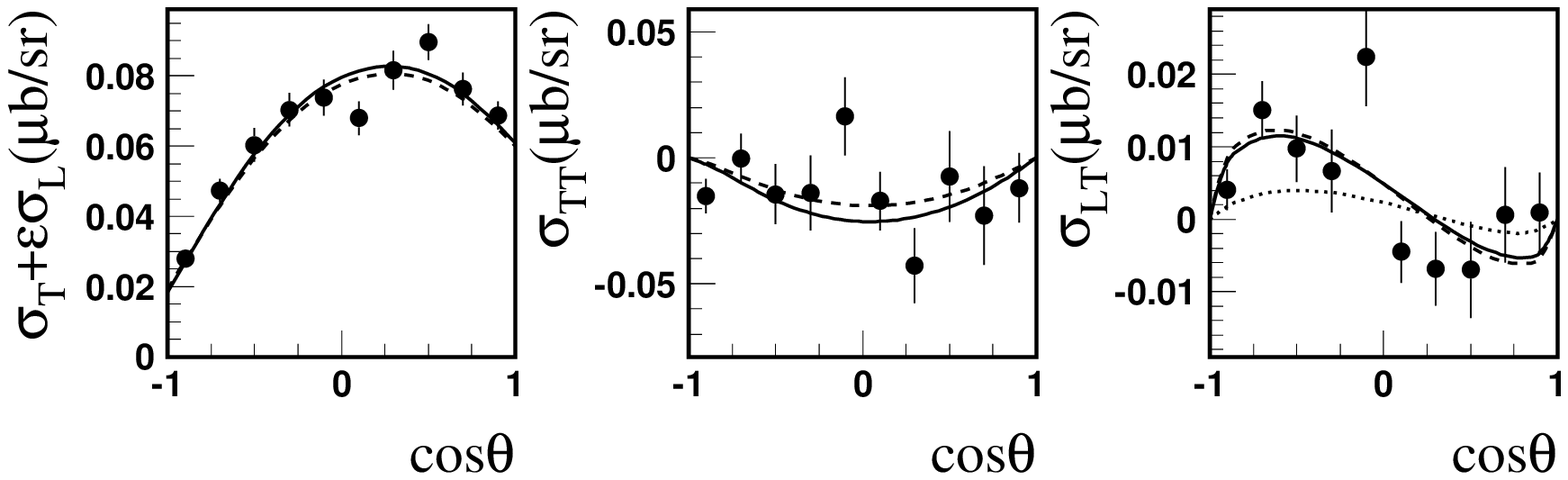,scale=0.7}
\end{minipage}
\begin{minipage}[t]{16.5 cm}
\caption{
JLab \cite{Aznauryan2009} and MAID2007 \cite{MAID2007}
results for the
$ep\rightarrow ep\pi^0$ structure functions
(in $\mu$b/sr units)
in comparison with experimental data \cite{Vilano} for
$W=1.232~$GeV and
$Q^2=6.4~$GeV$^2$.
Other notations are as in Fig. \ref{123_04}.
\label{123_vil}}
\end{minipage}
\end{center}
\end{figure}

The results are presented in section \ref{sec:results}. 
Before discussing the results we address a persistent discrepancy 
between two of the major analysis approaches. 
A consistent picture has 
emerged from the JLab and MAID2007 analyses of the CLAS data for 
the magnetic-dipole $\gamma^* N\rightarrow \Delta(1232)P_{33}$
amplitude and the ratio $R_{EM}$. However, there is significant
difference in the results for $R_{SM}$ at large
$Q^2$ (see below Fig. \ref{delta}). 
According to the results of the JLab group, the magnitude of  $R_{SM}$  
strongly rises at high $Q^2$. 
This behavior of $R_{SM}$  sharply disagrees
with the solution of MAID2007 \cite{MAID2007} which, based on
the same data set, gives an approximately 
$Q^2$-independent behavior of $R_{SM}$ at high $Q^2$. In order to 
resolve the discrepancy we compare the two results in a direct 
comparison with the data on the structure function $\sigma_{LT}$.  
The magnitude of the relevant 
amplitude $S_{1+}^{3/2}$ strongly constrains 
this structure function, whose $\cos{\theta}$
behavior at $W=1.23~$GeV is dominated by the interference of 
$S_{1+}^{3/2}$ with $M_{1+}^{3/2}$:
\begin{equation}
\sigma_{LT}(ep\rightarrow ep\pi^0)\approx
\frac{|\bf{q}|}{K}\frac{Q}{|\bf{k}|}
\sin{\theta}\left[\frac{2}{9}\left(S_{0+}^{1/2}+
2S_{0+}^{3/2}\right)^*M_{1+}^{3/2}+\frac{8}{3}
\cos{\theta}\left(S_{1+}^{3/2}\right)^*M_{1+}^{3/2}\right].
\label{eq:data1}\\
\end{equation}
The comparison is shown in Figs. \ref{123_04} and \ref{123_3}.
For completeness we also show a similar comparison
for the Hall C data \cite{Vilano} at $Q^2=6.4~$GeV$^2$
(Fig. \ref{123_vil}).
At $Q^2=0.4-1.45~$GeV$^2$ (Fig. \ref{123_04}),
both solutions describe the angular behavior
of $\sigma_{LT}$. However, MAID2007 underestimates the 
strong $\cos{\theta}$ dependence of this structure function
with rising $Q^2$, which is a direct consequence of the small
magnitude of $R_{SM}$ in the MAID2007 solution. 

\begin{figure}[ht!]
\begin{center}
\begin{minipage}[t]{12. cm}
\epsfig{file=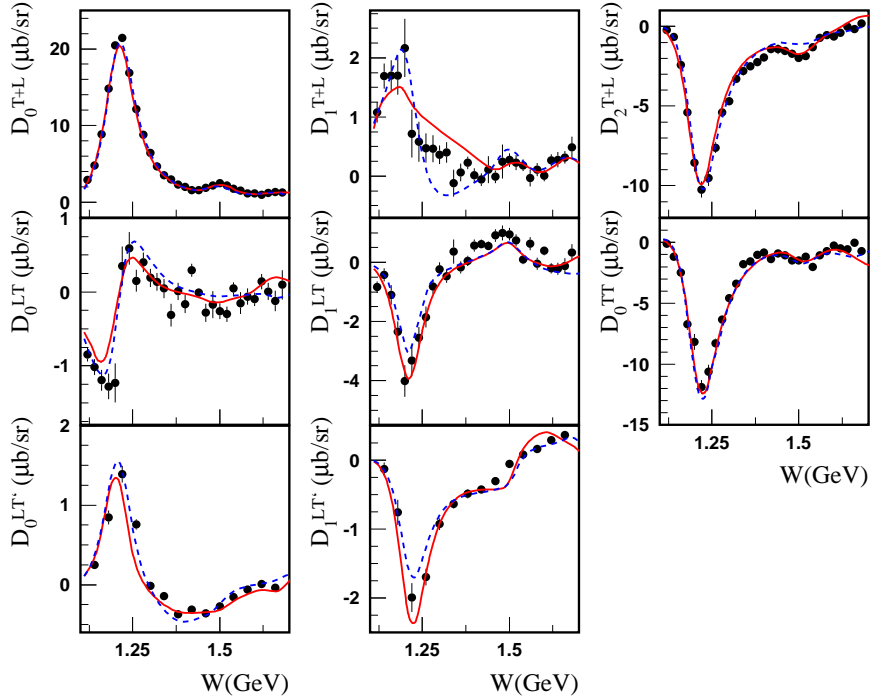,scale=0.7}
\end{minipage}
\begin{minipage}[t]{16.5 cm}
\caption{
The results of the JLab group  \cite{Aznauryan2009} 
for the
Legendre moments of the
$\vec{e}p\rightarrow ep\pi^0$ structure functions
in comparison with experimental data \cite{Joo1} at $Q^2=0.4~$GeV$^2$.
The solid (dashed) curves
correspond to the results obtained using the DR (UIM) approach.
\label{leg_pi0_04}}
\end{minipage}
\end{center}
\end{figure}

\begin{figure}[ht!]
\begin{center}
\begin{minipage}[t]{12. cm}
\epsfig{file=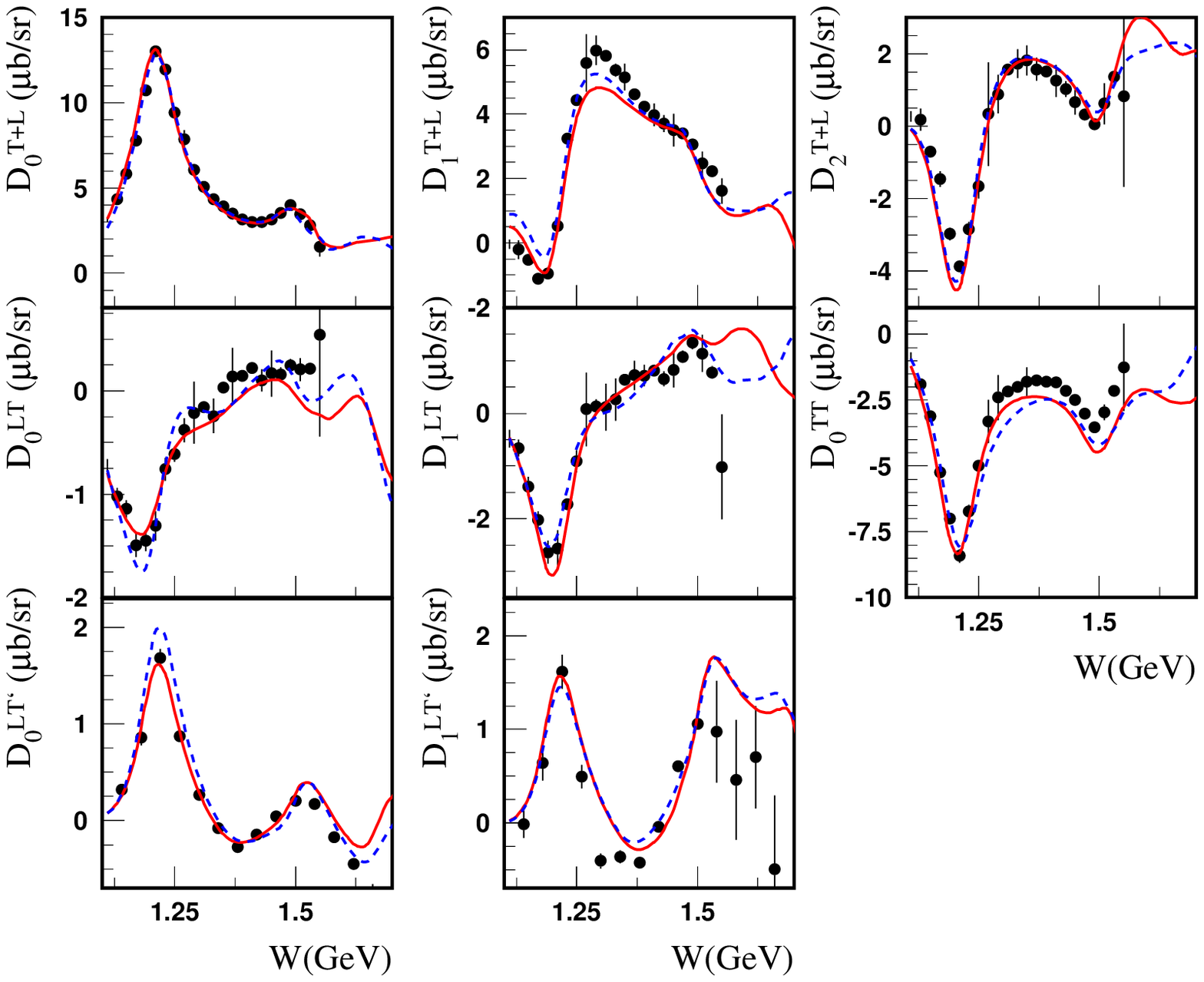,scale=0.7}
\end{minipage}
\begin{minipage}[t]{16.5 cm}
\caption{
The results of the JLab group  \cite{Aznauryan2009} for the
Legendre moments of the
$\vec{e}p\rightarrow en\pi^+$ structure functions
in comparison with experimental data \cite{Egiyan} at
$Q^2=0.4~$GeV$^2$.
Other notations are as in Fig. \ref{leg_pi0_04}.
\label{leg_pip_04}}
\end{minipage}
\end{center}
\end{figure}

\begin{figure}[ht!]
\begin{center}
\begin{minipage}[t]{12. cm}
\epsfig{file=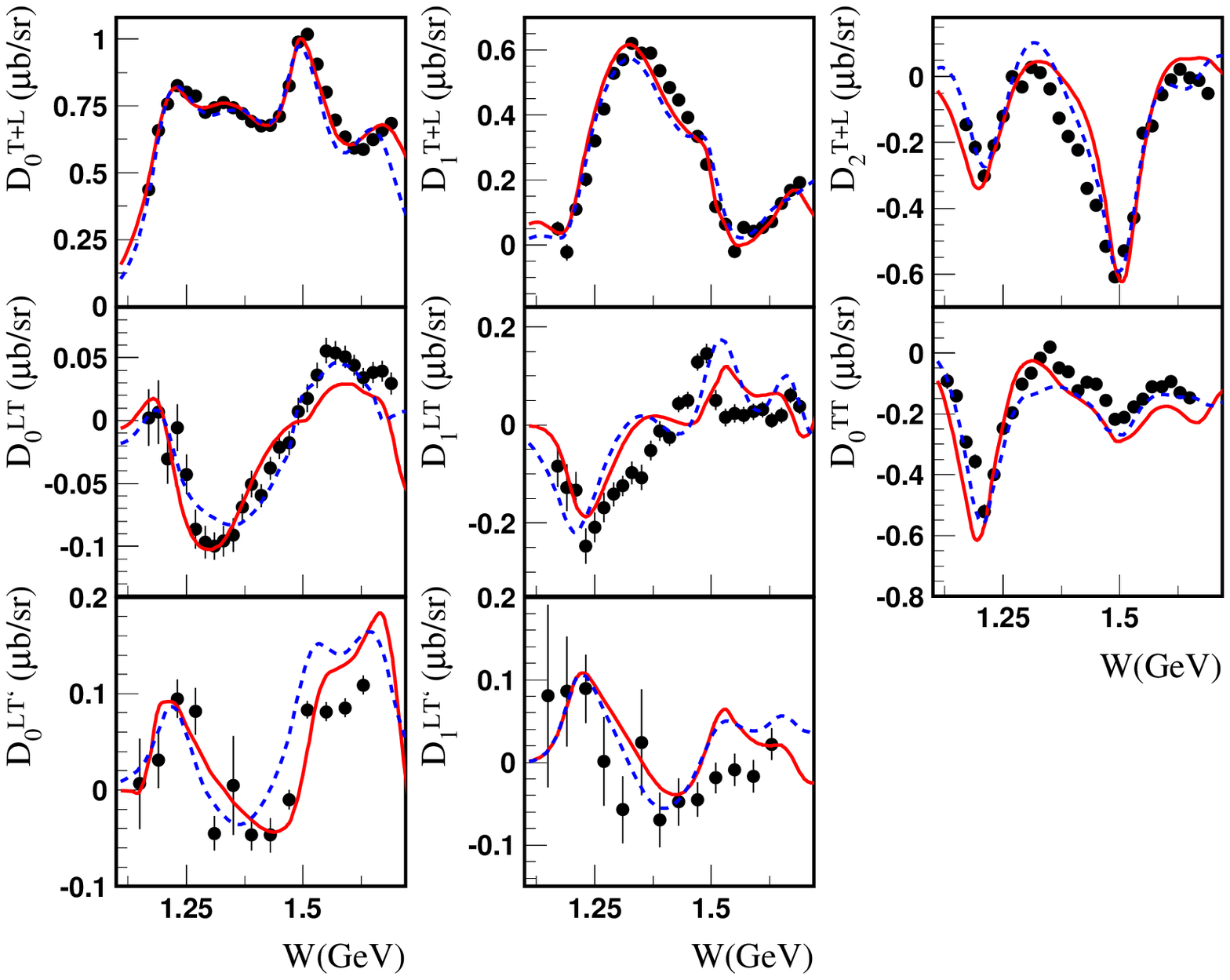,scale=0.7}
\end{minipage}
\begin{minipage}[t]{16.5 cm}
\caption{
The results of the JLab group  \cite{Aznauryan2009} for the
Legendre moments of the
$\vec{e}p\rightarrow en\pi^+$ structure functions
in comparison with experimental data \cite{Park} at $Q^2=2.44~$GeV$^2$.
Other notations are as in Fig. \ref{leg_pi0_04}.
\label{leg_244}}
\end{minipage}
\end{center}
\end{figure}

\begin{figure}[ht!]
\begin{center}
\begin{minipage}[t]{12. cm}
\epsfig{file=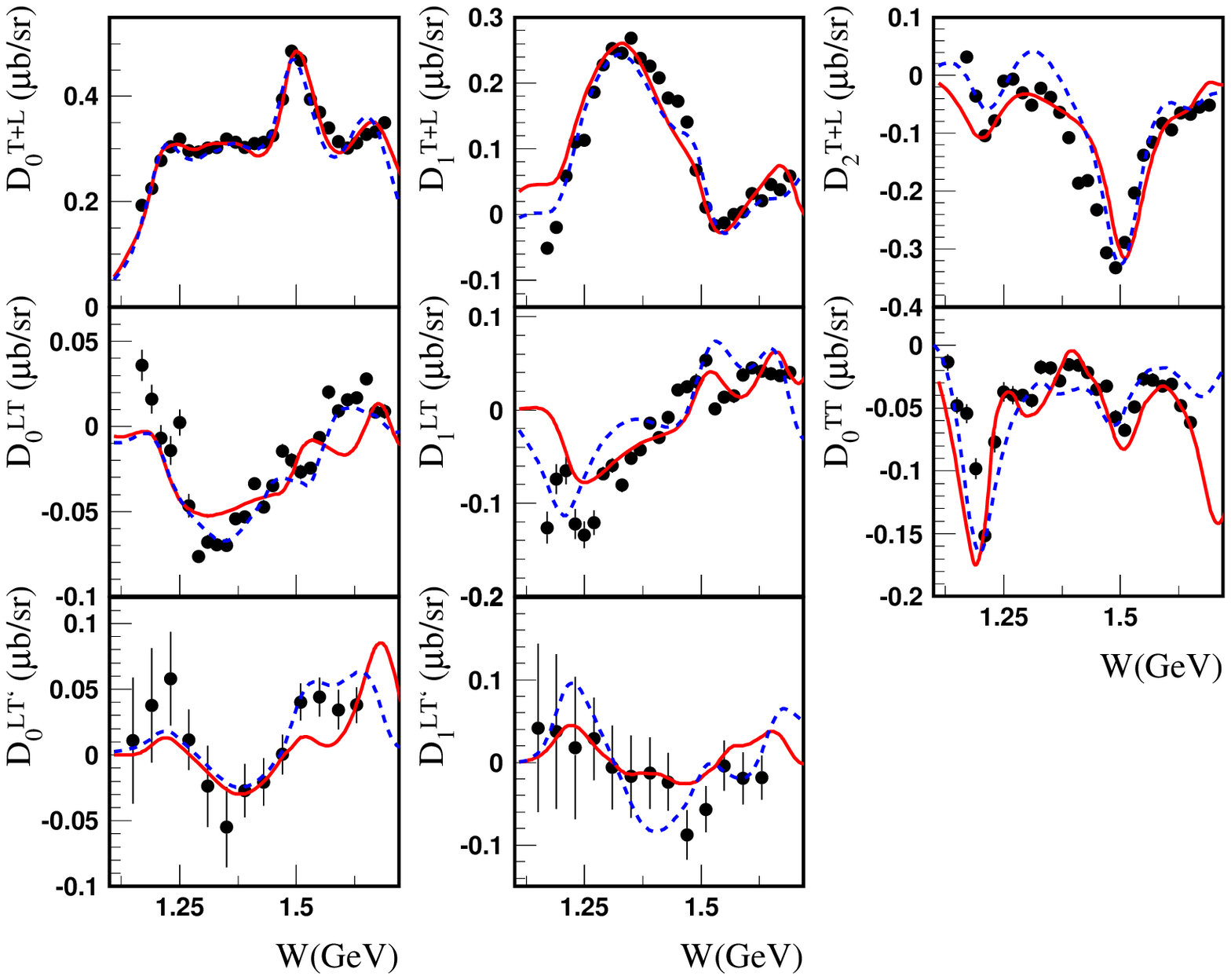,scale=0.7}
\end{minipage}
\begin{minipage}[t]{16.5 cm}
\caption{
The results of the JLab group \cite{Aznauryan2009} for the
Legendre moments of the
$\vec{e}p\rightarrow en\pi^+$ structure functions
in comparison with experimental data \cite{Park} at $Q^2=3.48~$GeV$^2$.
Other notations are as in Fig. \ref{leg_pi0_04}.
\label{leg_348}}
\end{minipage}
\end{center}
\end{figure}

\begin{figure}[tb]
\begin{center}
\begin{minipage}[t]{12. cm}
\epsfig{file=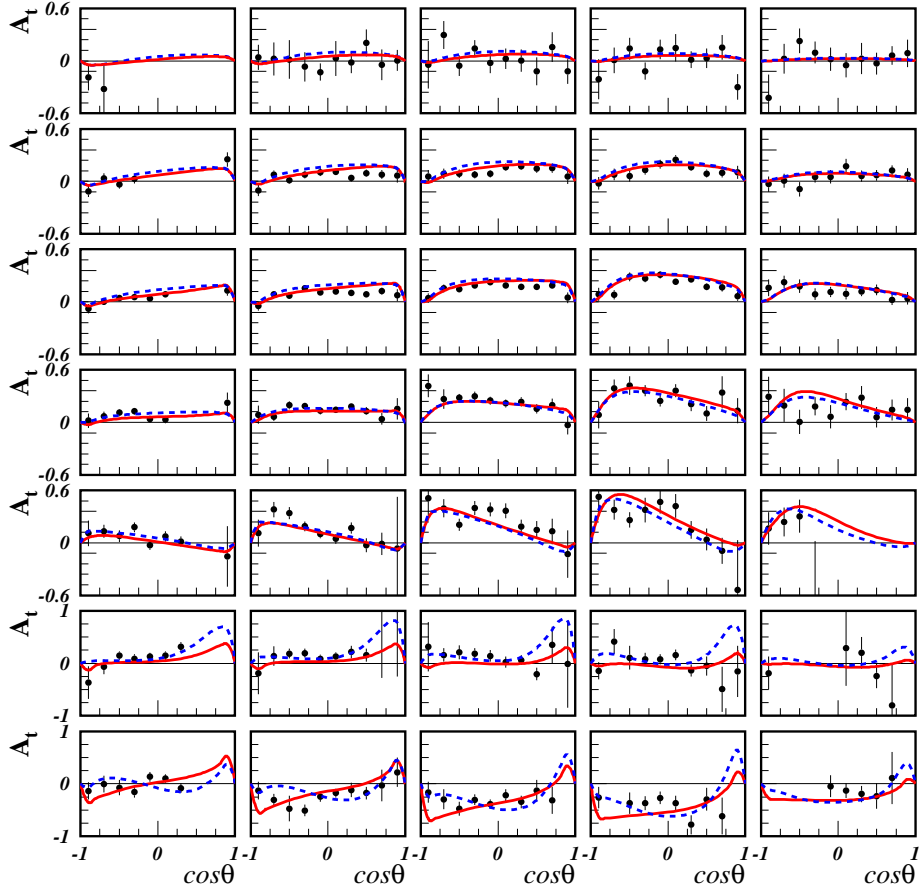,scale=0.7}
\end{minipage}
\begin{minipage}[t]{16.5 cm}
\caption{
The results of the JLab group \cite{Aznauryan2009} for the
the longitudinal target asymmetry $A_t$
in comparison with experimental data at 
$Q^2=0.385~$GeV$^2$ \cite{Biselli}.
The rows correspond to 7 $W$ bins with $W$ mean values of 
1.125, 1.175, 1.225, 1.275, 1.35, 1.45, and 1.55 GeV.
The columns correspond to $\phi$ bins with
$\phi=\pm 72^{\circ},\pm 96^{\circ},\pm 120^{\circ},\pm 144^{\circ},\pm 168^{\circ}$.
The solid circles are the average values of the data for 
positive $\phi$'s and those
at negative  $\phi$'s taken with opposite signs.
Other notations are as in Fig. \ref{leg_pi0_04}.
\label{at_04}}
\end{minipage}
\end{center}
\end{figure}

\begin{figure}[tb]
\begin{center}
\begin{minipage}[t]{12. cm}
\epsfig{file=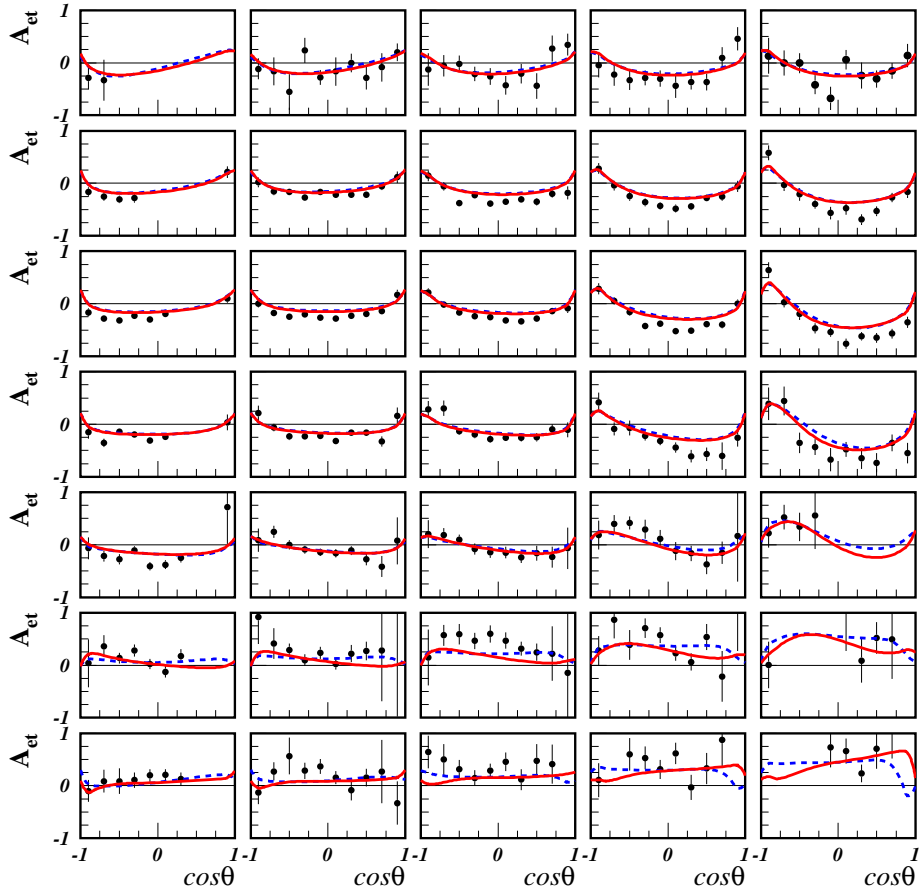,scale=0.7}
\end{minipage}
\begin{minipage}[t]{16.5 cm}
\caption{
The results of the JLab group \cite{Aznauryan2009} for the
beam-target asymmetry $A_{et}$
in comparison with experimental data at
$Q^2=0.385~$GeV$^2$ \cite{Biselli}.
The rows correspond to 7 $W$ bins with $W$ mean values of 
1.125, 1.175,
1.225, 1.275, 1.35, 1.45, and 1.55 GeV.
The columns correspond to $\phi$ bins with
$\phi=\pm 72^{\circ},\pm 96^{\circ},\pm 120^{\circ},\pm 144^{\circ},
\pm 168^{\circ}$.
The average values of the data for positive and negative 
$\phi$'s
are shown by solid circles.
Other notations are as in Fig. \ref{leg_pi0_04}.
\label{aet_04}}
\end{minipage}
\end{center}
\end{figure}

\begin{figure}[tb]
\begin{center}
\begin{minipage}[t]{14. cm}
\epsfig{file=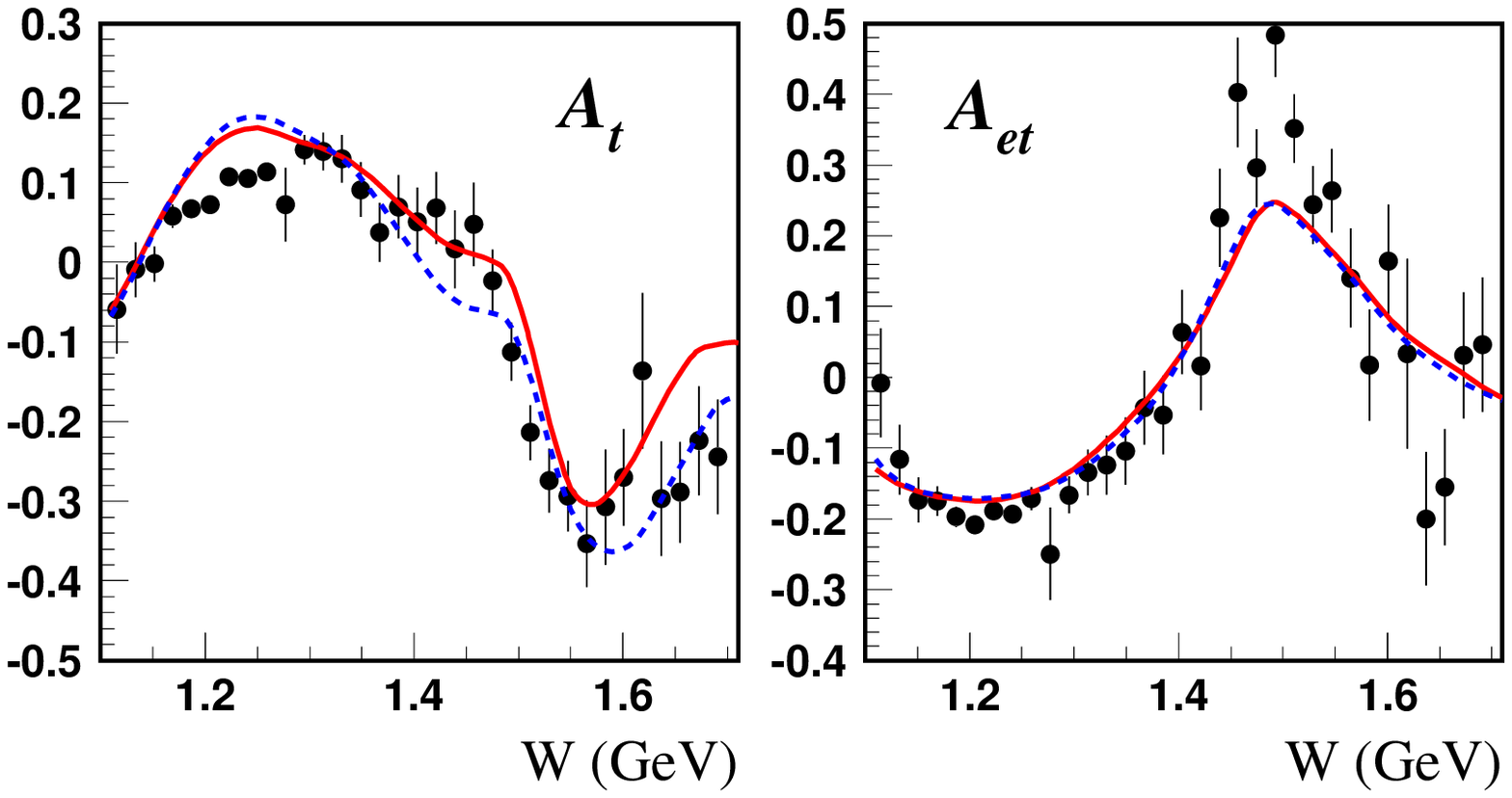,scale=0.8}
\end{minipage}
\begin{minipage}[t]{16.5 cm}
\caption{
$A_{t}$ (left panel) and $A_{et}$ (right panel) as a 
function of the invariant
mass $W$, integrated over the full range in 
$\cos{\theta}$,
$0.252<Q^2<0.611~$GeV$^2$ and $60^{\circ}<\phi<156^{\circ}$.
Experimental data are from Ref. \cite{Biselli}.
Other notations are as in Fig. \ref{leg_pi0_04}.
\label{at}}
\end{minipage}
\end{center}
\end{figure}

Large amount 
of the CLAS data on differential 
cross sections
and longitudinally polarized beam asymmetries 
and their description
are shown in Ref. \cite{Aznauryan2009}
using Legendre moments of the structure functions.
This allows the presentation of the data 
over all energies and angles  
in most complete and compact way. 
The Legendre moments of the structure functions
are defined as the coefficients in the expansion
of structure functions over Legendre polynomials $P_l(\cos{\theta})$:
\begin{eqnarray}
&&\sigma_T(W,\cos {\theta})+
\epsilon \sigma_L(W,\cos{\theta})=
\sum_{l=0}^{n} 
D_l^{T+L}(W)P_l(\cos{\theta}),
\label{eq:data2}\\
&&\sigma_{LT}(W,\cos{\theta})= \sin{\theta}
\sum_{l=0}^{n-1}D_l^{LT}(W)
P_l(\cos{\theta}),
\label{eq:data3}\\
&&\sigma_{LT'}(W,\cos{\theta}) = \sin{\theta} \sum_{l=0}^{n-1}
D_l^{LT'}(W) P_l(\cos{\theta}),
\label{eq:data4}\\
&&\sigma_{TT}(W,\cos{\theta})= \sin^2{\theta}\sum_{l=0}^{n-2}
D_l^{TT}(W)
P_l(\cos{\theta}).
\label{eq:data5}
\end{eqnarray}
Descriptions of the Legendre moments for
the $\vec{e}p\rightarrow ep\pi^0$ and
$\vec{e}p\rightarrow en\pi^+$ structure functions
at low and high $Q^2$
are shown in Figs. \ref{leg_pi0_04}-\ref{leg_348}.

The Legendre moment $D_0^{T+L}$ represents the
$\cos{\theta}$ independent part of
$\sigma_T+\epsilon \sigma_L$,
which is related to the $\gamma^* N\rightarrow \pi N$
total cross section by:
$D_0^{T+ L}=\sigma^{tot}/4\pi$.
The resonant structures
related to the resonances
$\Delta(1232)P_{33}$, $N(1520)D_{13}$, and $N(1535)S_{11}$
are revealed in enhancements in the $W$ dependence of $D_0^{T+L}$.
We observe that with increasing $Q^2$,
the resonant structure near $1.5$~GeV
becomes increasingly dominant in comparison with
the $\Delta(1232)$.
At $Q^2\geq 1.72$~GeV$^2$, there is a shoulder between the $\Delta$
and the structure at $1.5$~GeV, which, 
as is shown in Ref. \cite{Aznauryan2008}, 
is related to the large contribution
of the broad Roper resonance.

The dips in the Legendre moment $D_2^{T+ L}$
are caused by the $\Delta(1232)P_{33}$ resonance
and by the interference of the $N(1520)D_{13}$ and $N(1535)S_{11}$.
They are determined by the following contributions to $D_2^{T}$:
\begin{equation}
D_2^{T}=
-\frac{|\bf{q}|}{K}\left[|M_{1+}|^2+4Re(A_{0+}A^*_{2-})\right].
\label{eq:data6}\\
\end{equation}

The  enhancement in $D_0^{T+ L}$ and the dip in
$D_0^{TT}$ in the $\Delta$ mass region are related mainly 
to the
$M_{1+}^{3/2}$ amplitude
of the
$\gamma^* p \rightarrow ~\Delta(1232)P_{33}$
transition:
\be
D_0^{T+ L}\approx
\frac{8}{9}\frac{|\bf{q}|}{K}|M_{1+}^{3/2}|^2,~~~
D_0^{TT}\approx -\frac{2}{3}
\frac{|\bf{q}|}{K}|M_{1+}^{3/2}|^2.
\label{eq:data7}\\
\ee

The longitudinal target ($A_t$) and beam-target ($A_{et}$)
asymmetries for  $\vec{e}\vec{p}\rightarrow ep\pi^0$
at $Q^2=0.385~$GeV$^2$
\cite{Biselli} 
are shown in Figs. \ref{at_04} and \ref{aet_04}
as a function of $\cos\theta$ for all $W$ and $\phi$ bins.
These observables are defined
in Ref. \cite{Biselli} through the response functions
introduced in Ref. \cite{tiator_def}.
We show also $W$-dependencies of  $A_t$ and $A_{et}$
integrated over the full range in $\cos\theta$, $\phi$, and $Q^2$ (Fig. 
\ref{at}).

\subsection{\it	JLab/Hall B and Hall C data 
on eta electroproduction
in the $N(1535)S_{11}$
resonance region
\label{sec:data_eta}}

\begin{figure}[tb]
\begin{center}
\begin{minipage}[t]{13. cm}
\epsfig{file=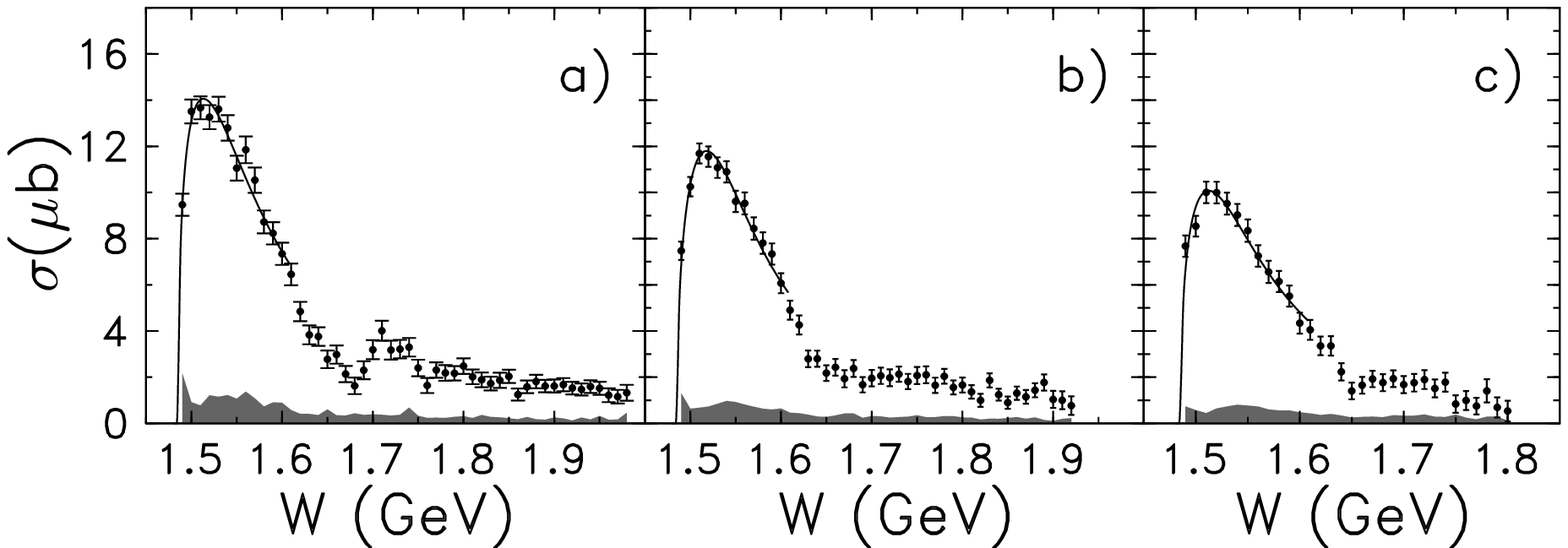,scale=0.6}
\end{minipage}
\begin{minipage}[t]{16.5 cm}
\caption{
The total cross sections for $\eta$ production on protons at 
(a) $Q^2=0.625~$GeV$^2$,
(b) $Q^2=0.875~$GeV$^2$, and
(c) $Q^2=1.125~$GeV$^2$ from Ref. \cite{Thompson}. 
The shaded bands show systematic
uncertainties. The curves correspond to single resonance
Breit-Wigner fits. 
({\it Source:} From Ref. \cite{Thompson}.)
\label{thompson2}}
\end{minipage}
\end{center}
\end{figure}

\begin{figure}[tb]
\begin{center}
\begin{minipage}[t]{12. cm}
\epsfig{file=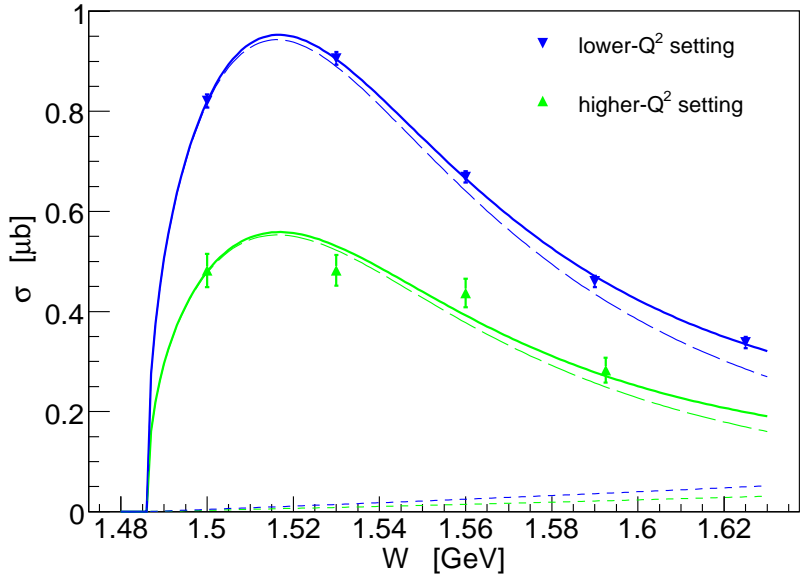,scale=0.55}
\end{minipage}
\begin{minipage}[t]{16.5 cm}
\caption{
Fit to the total cross sections for $\eta$ production at 
$Q^2=5.7$ and
$7~$GeV$^2$ from Ref. \cite{Dalton}.
The solid lines correspond to the sum of Breit-Wigner forms (long dashed lines)
and non-resonant background (short dashed lines). 
({\it Source:} From Ref. \cite{Dalton}.)
\label{dalton}}
\end{minipage}
\end{center}
\end{figure}

There are four JLab measurements of
the differential cross sections in eta electroproduction 
(see Table \ref{tab:data})
 that cover the range $0.13 \leq Q^2 \leq 7~$GeV$^2$.
All experiments include the $N(1535)S_{11}$ resonance     
mass range, and three measurements \cite{Thompson,Denizli,Dalton}
extend to higher energies. The cross section at $W<1.6~$GeV is 
strongly dominated 
by the contribution of the $N(1535)S_{11}$ resonance.
For this reason, the contribution of this resonance
has been found in all experiments using a   
Breit-Wigner form to fit to the total cross section (see Fig. \ref{thompson2}).
Analyses that include phenomenological non-resonant backgrounds
show that this contribution is very small and has little effect on the 
$\gamma^* p\rightarrow N(1535)S_{11}$
amplitudes found from the fit to the total cross section at the
$N(1535)S_{11}$ resonance mass. This can be seen  
from the description of the total cross section data at 
$Q^2=5.7$ and $7~$GeV$^2$ \cite{Dalton} shown in Fig. \ref{dalton} .

\begin{figure}[tb]
\begin{center}
\begin{minipage}[t]{14. cm}
\includegraphics[angle=270,width=5.5in]{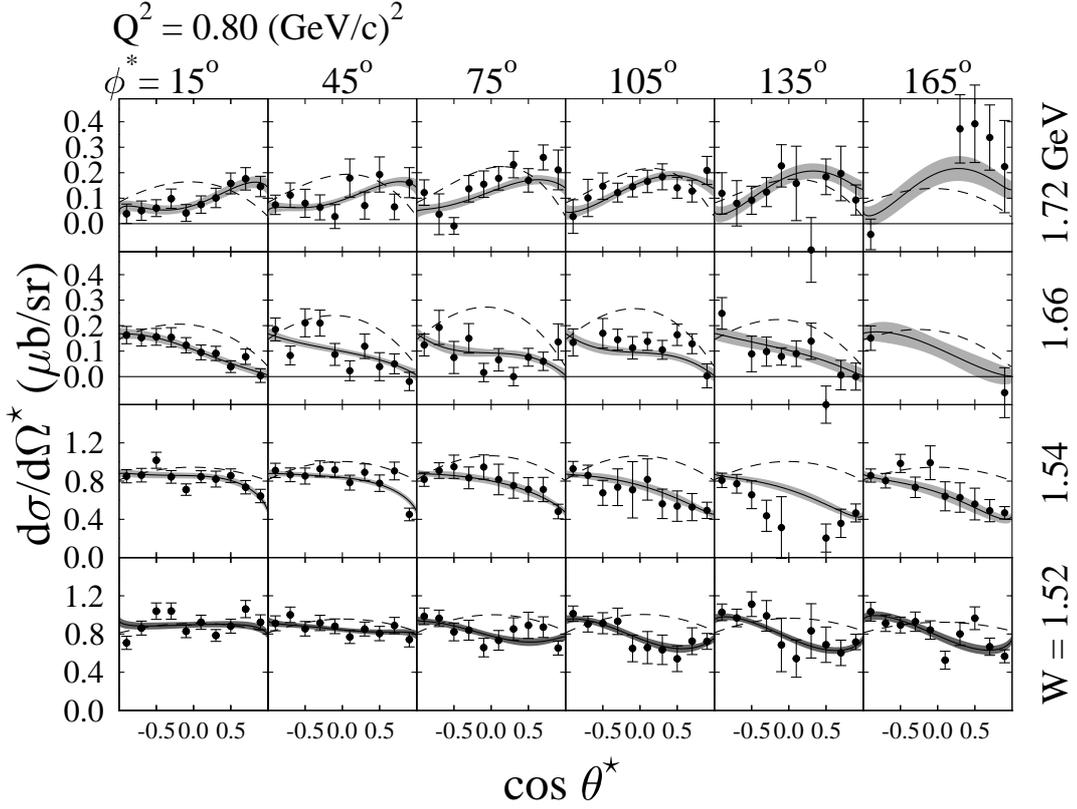}
\end{minipage}
\begin{minipage}[t]{16.5 cm}
\caption{
Angular distribution at different W values 
for $\gamma^* p\rightarrow \eta p$ in the c.m.s. at  
$Q^2=0.8$~GeV$^2$
from Ref. \cite{Denizli}.
The solid lines with error bands are the results of fits to 
separate the structure functions. The dashed lines correspond 
to $\eta$-MAID \cite{etaMAID}.
({\it Source:} From Ref. \cite{Denizli}.)
\label{denizli1}}
\end{minipage}
\end{center}
\end{figure}

\begin{figure}[tb]
\begin{center}
\begin{minipage}[t]{11. cm}
\epsfig{file=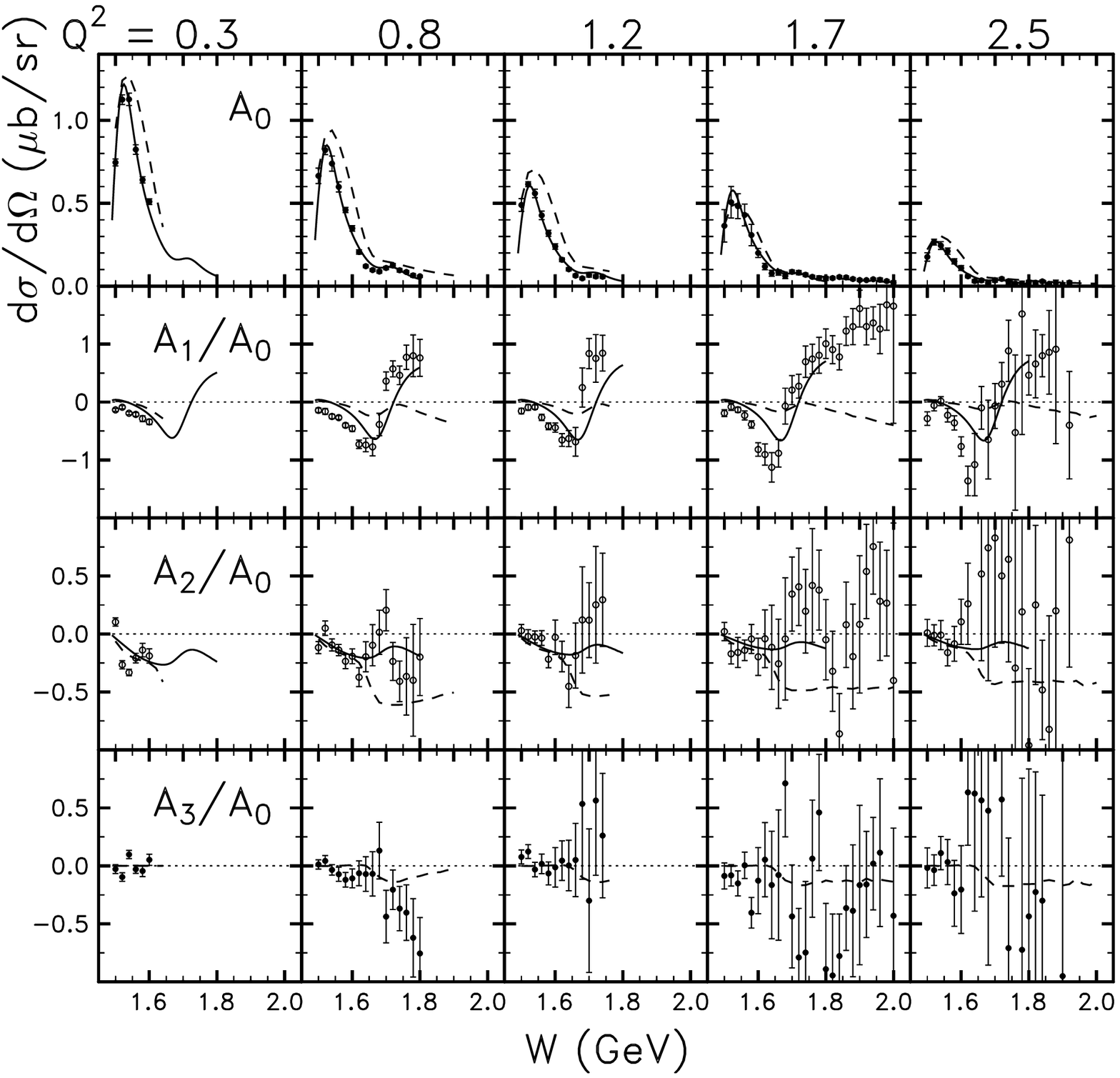,scale=0.55}
\end{minipage}
\begin{minipage}[t]{16.5 cm}
\caption{
The Legendre moments $D_i^{T+L}\equiv A_i$ ( $i=0,1,2,3$)
from Ref. \cite{Denizli}.
The solid lines are from a 4-resonance fit to the Legendre moments
of structure functions; the dashed lines are the $\eta$-MAID predictions 
\cite{etaMAID}.
({\it Source:} From Ref. \cite{Denizli}.)
\label{denizli2}}
\end{minipage}
\end{center}
\end{figure}

The CLAS measurements \cite{Thompson,Denizli} have total angular
coverage: see a sample of the data 
at $Q^2=0.8~$GeV$^2$
from Ref. \cite{Denizli} in Fig. \ref{denizli1}. 
This made it possible to extract the three structure functions:
$\sigma_T+\epsilon \sigma_L$, $\sigma_{TT}$, and $\sigma_{LT}$.
The expansion of the structure functions over the Legendre
moments reveals a sign change of the $D_1^{T+L}$ moment
at $W\sim 1.68$~GeV (see Fig. \ref{denizli2}).
In both publications \cite{Thompson,Denizli}
it is mentioned that this sign change
can be described in a simple isobar model 
that includes the states
$N(1535)S_{11}$, $N(1650)S_{11}$, $N(1520)D_{13}$,
and $N(1710)P_{11}$. With these resonances, it
arises from the interference between 
resonances $N(1535)S_{11}$, $N(1650)S_{11}$
and $N(1710)P_{11}$. However, the sign change in $D_1^{T+L}$ 
can arise from the interference between $N(1535)S_{11}$, $N(1650)S_{11}$
and $N(1720)P_{13}$ too. 
For more definite
conclusions detailed investigations are necessary that need to
include the precise $\eta$ photoproduction results of 
Ref. \cite{McNicoll} and pion
photo- and electroproduction data.

\subsection{\it	JLab data on 
$ep\rightarrow e\pi^-\pi^+ p$
\label{sec:data_2pi}}

The combination of the continuous electron beam and
the CLAS detector enabled the collection of the first precise 
and detailed data sets on two-pion electroproduction.
Nine independent one-fold differential
$\gamma^* p\rightarrow \pi^-\pi^+p$ cross sections,
as well fully integrated cross sections, were measured in the
kinematical areas
presented in Table \ref{tab:2pi}.
Due to the high statistics and the good momentum
resolution of the experiment, the data  
are presented in small $W$ and $Q^2$ bins. 
This made it possible to establish all 
essential mechanisms contributing to $\pi^-\pi^+p$
electroproduction  from their manifestation in the observables. 
The analysis was carried out within the framework of 
the JLAB-Moscow State University (JM) reaction 
model \cite{Mokeev2009}. 

\begin{table}
\begin{center}
\begin{minipage}[t]{16.5 cm}
\caption{Kinematical areas covered by the CLAS
measurements of the $\pi^-\pi^+p$
electroproduction cross sections.}
\label{tab:2pi}
\end{minipage}
\begin{tabular}{|c|c|c|c|}
\hline
$Q^{2}$ coverage, &  $W $coverage, & Bin size over $W/Q^2$, & Data status \\
GeV$^2$  &  GeV  &   GeV/GeV$^2$  &   \\
\hline
0.20-0.60 & 1.30-1.57 & 0.025/0.050  & Completed \cite{Fedotov09}  \\
0.50-1.50 & 1.40-2.10 & 0.025/0.3-0.4 & Completed \cite{Ripani} \\
2.0-5.0 & 1.40-2.00 & 0.025/0.5 & In progress  \\
\hline
\end{tabular}
\end{center}
\end{table}

\begin{figure}[tb]
\begin{center}
\begin{minipage}[t]{13. cm}
\epsfig{file=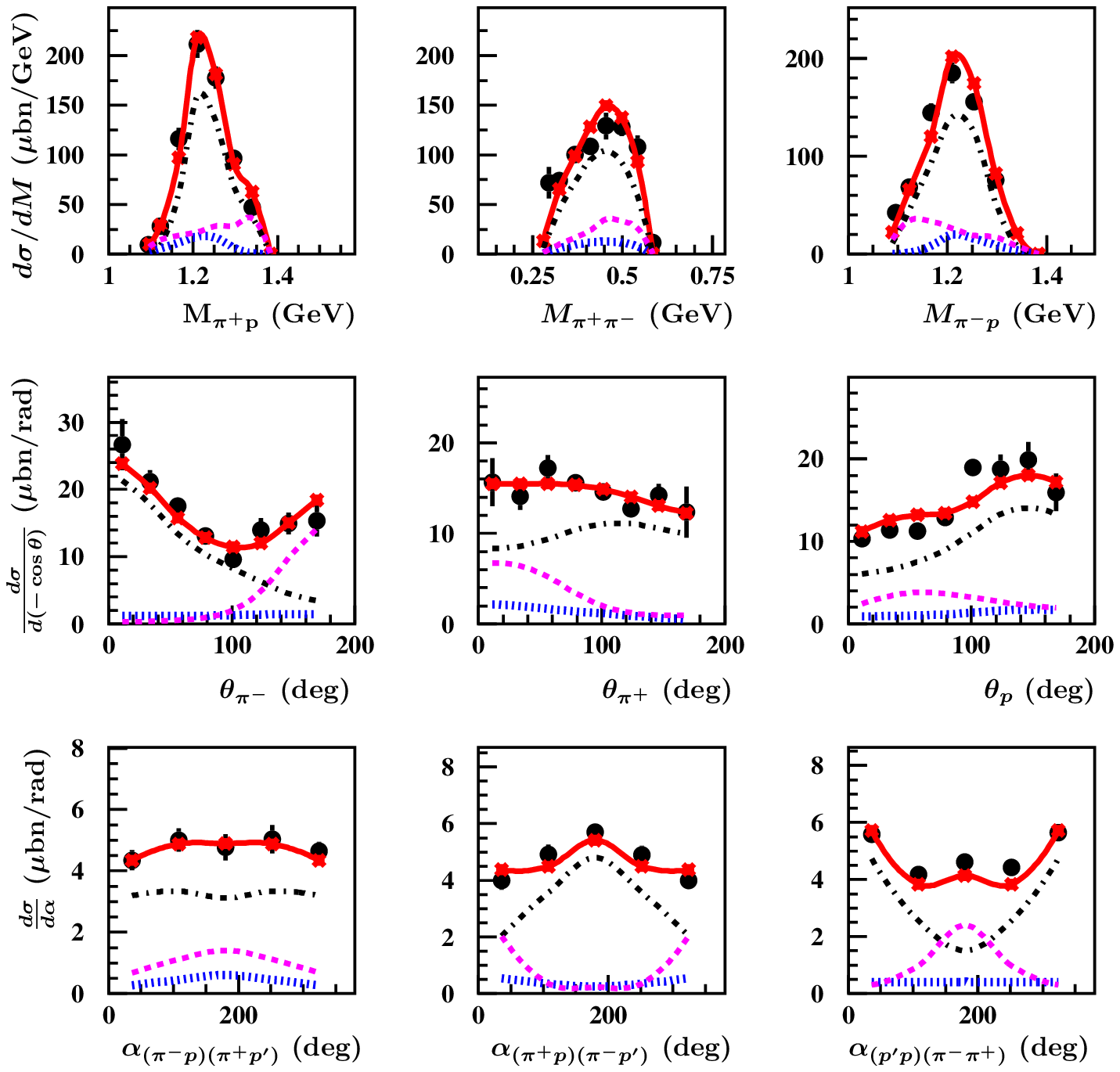,scale=0.9}
\end{minipage}
\begin{minipage}[t]{16.5 cm}
\caption{
Different contributions to the CLAS $\pi^+\pi^-p$
electroproduction  data
\cite{Fedotov09} at $W$= 1.51 GeV and $Q^2=0.43$ GeV$^2$ within the 
JM model \cite{Mokeev2009}. Full calculations are shown by the solid lines, while
the contributions from the $\pi^{-} \Delta^{++}$, $\pi^{+} \Delta^{0}$ isobar
channels and from direct 2$\pi$ production are shown by the dashed-dotted, dotted 
and dashed
lines, respectively.
({\it Source:} From Ref. \cite{Mokeev2009}.)
\label{mokeev}}
\end{minipage}
\end{center}
\end{figure}

The model describes the reaction
$\gamma^*p\rightarrow \pi^-\pi^+p$ through superposition
of the channels 
\bea
&&\gamma^*p\rightarrow\pi^-\Delta^{++} \rightarrow \pi^-\pi^+p, 
\label{eq:mok1}\\
&&\gamma^*p\rightarrow\pi^+\Delta^{0} \rightarrow \pi^+\pi^-p, 
\label{eq:mok2}\\
&&\gamma^*p\rightarrow\rho^0 p \rightarrow \pi^-\pi^+p 
\label{eq:mok3}
\eea
and non-resonant mechanisms.
Quasi two-body channels (\ref{eq:mok1}-\ref{eq:mok3}) 
contain both resonant and non-resonant
parts. The energy-dependence of the resonant parts are described 
by Breit-Wigner forms, taking into account all well-established resonances.
An example of the contribution of different production mechanisms
to the nine independent
one-fold differential
$\gamma^* p\rightarrow \pi^-\pi^+p$ cross sections
  is shown in Fig. \ref{mokeev}.

The analysis of the two-pion data \cite{Ripani} revealed difficulties in 
describing the strong peak structure seen at $W\approx 1.7~$GeV 
using only known resonances. The analysis required a prominent 
contribution of the $N(1720)P_{13}$ state to describe the data. 
This state could be attributed to the known resonance, but with hadronic parameters
that are significantly different from the RPP 
(Review of Particle Physics) values \cite{PDG}. 
Another possibility
consists of the introduction of a new baryon state with the same spin-parity 
as the $N(1720)P_{13}$, but with significantly different strengths of the 
hadronic couplings. We will return to this point in section \ref{sec:new_state}.

\subsection{\it	JLab/Hall B data on 
$K\Lambda$ and $K\Sigma$ electroproduction 
\label{sec:data_strange}}
\begin{figure}[!]
\begin{center}
\begin{minipage}[t]{16.5 cm}
\epsfig{file=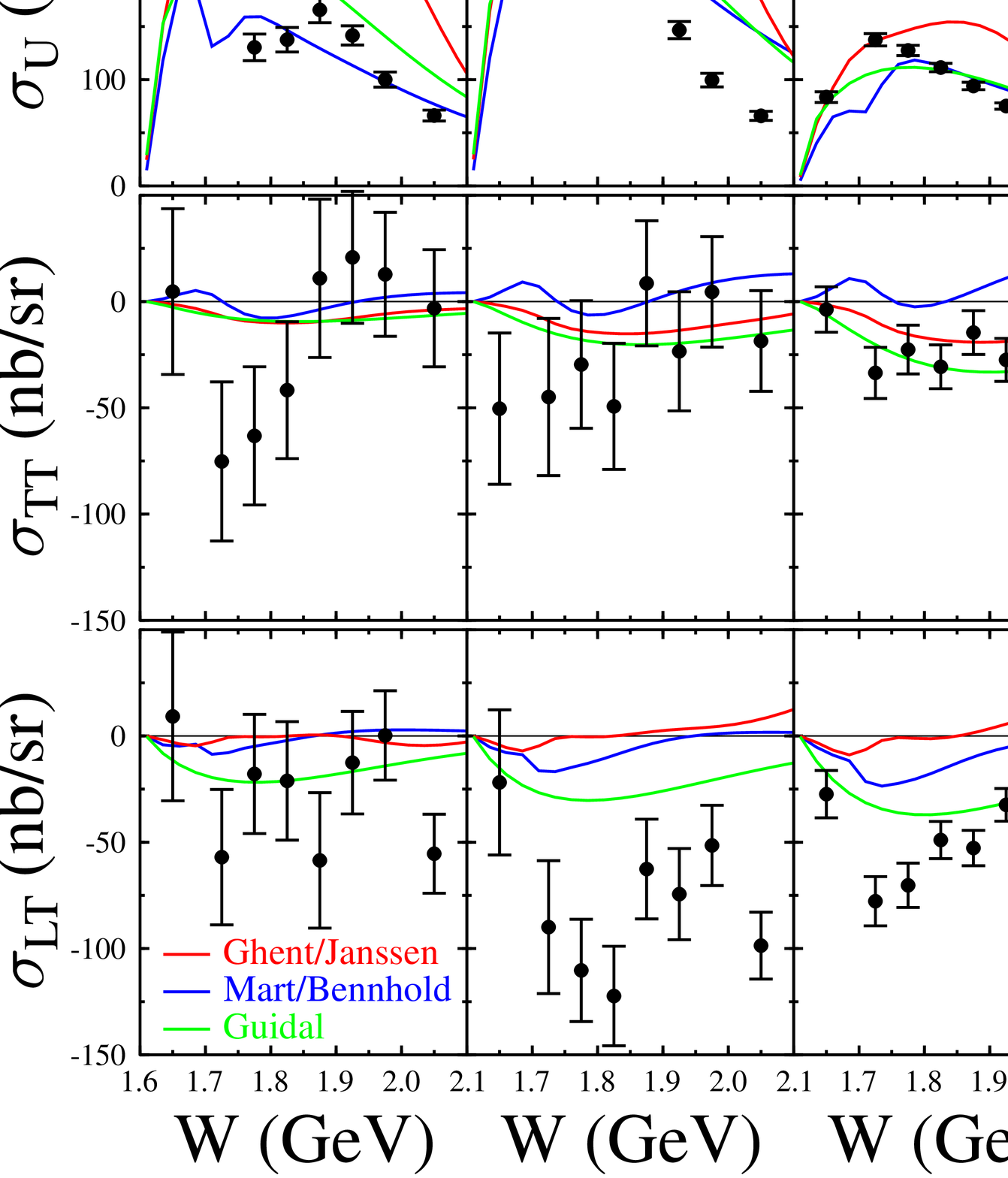,scale=0.3}
\end{minipage}
\begin{minipage}[t]{14.0 cm}
\caption{Separated structure functions for $ep\rightarrow eK^+\Lambda$
at $Q^2=0.65~$GeV$^2$ from CLAS \cite{Ambrozewicz}; 
$\sigma_U\equiv \sigma_T + \epsilon \sigma_L$. The curves
correspond to the results by Mart et al. \cite{Mart},
Guidal et al. \cite{Guidal2000}, and 
Janssen et al. \cite{Janssen}. 
({\it Source:} From Ref. \cite{Ambrozewicz}.)
\label{klambda}}
\end{minipage}
\end{center}
\end{figure}

\begin{figure}[!]
\begin{center}
\begin{minipage}[t]{16.5 cm}
\epsfig{file=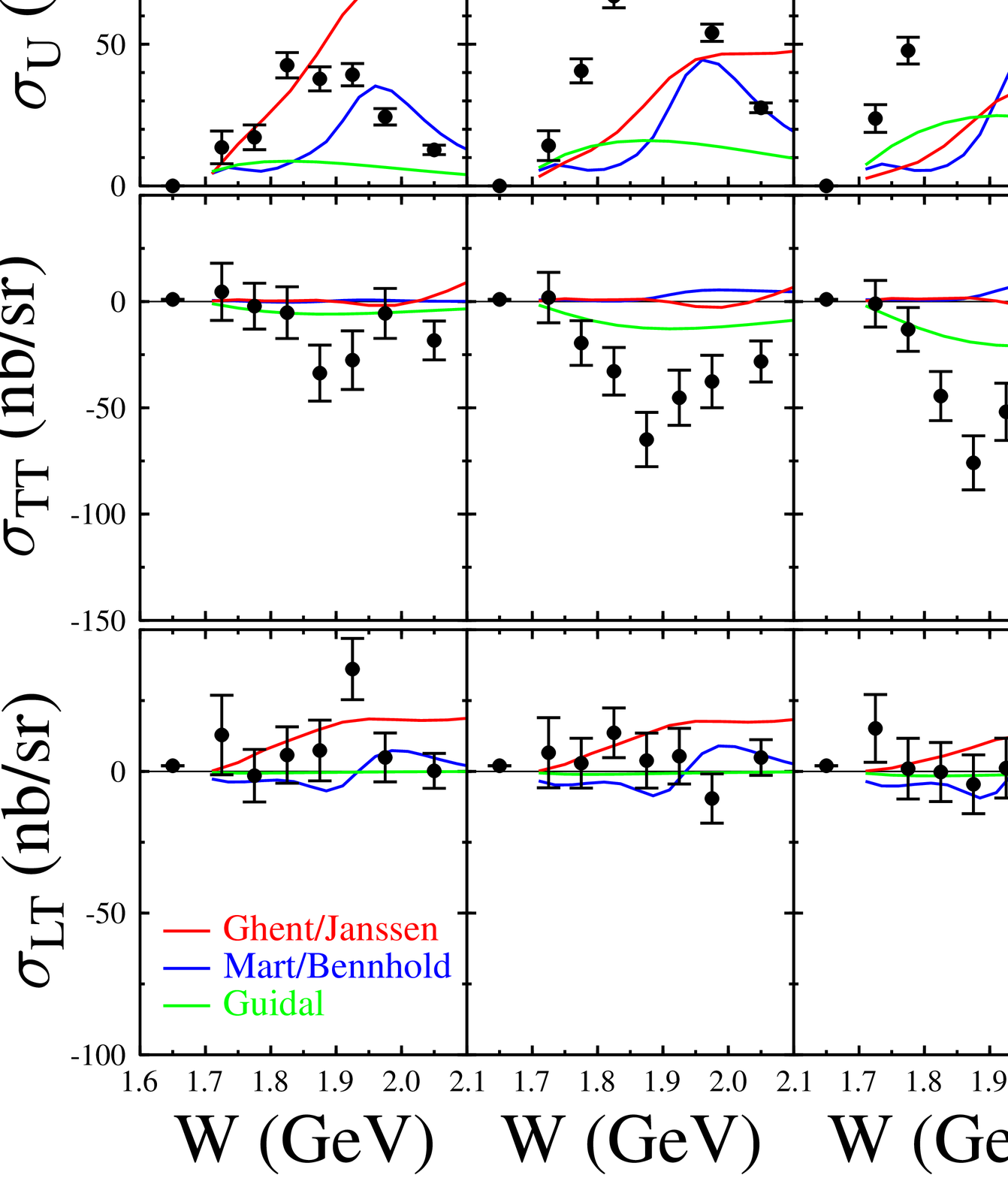,scale=0.3}
\end{minipage}
\begin{minipage}[t]{14.0 cm}
\caption{The same as in Fig. \ref{klambda}
for $ep\rightarrow eK^+\Sigma^0$.
({\it Source:} From Ref. \cite{Ambrozewicz}.)
\label{ksigma}}
\end{minipage}
\end{center}
\end{figure}

 An extensive program of strange-particle electroproduction 
off protons, that includes measurements of a variety of 
polarization observables, 
has been carried out with CLAS at JLab.
The structure functions
$\sigma_T +\epsilon \sigma_L$, $\sigma_{TT}$, and $\sigma_{LT}$
are separated in $ep\rightarrow eK^+\Lambda$ and $K^+\Sigma^0$
(see examples in Figs. \ref{klambda}, \ref{ksigma}).
The polarization measurements include
the transferred polarization in $\vec{e}p\rightarrow eK^+\vec{\Lambda}$
and $\vec{e}p\rightarrow eK^+\vec{\Sigma}^0$ 
\cite{Carman2003,Carman2009},
and the longitudinally polarized beam asymmetry 
in $\vec{e}p\rightarrow eK^+\Lambda$ \cite{Carman2008}.
The data span a range in photon virtuality 
$0.3\leq Q^2\leq 5.4~$GeV$^2$ and energy 
$1.6\leq W\leq 2.6~$GeV.
Exclusive kaon production has much stronger non-resonant 
contributions than pion production and the extraction of 
resonances requires a detailed understanding of the non-resonant 
hadronic couplings, many of which are poorly determined or completely 
unknown.  
While there are indications of $s$-channel resonance behavior in the 
$W$-dependence of the structure function $\sigma_T + \epsilon \sigma_L$, 
the strangeness channels will have to be included in a dynamical 
coupled-channel approach to be fully effective as ingredients in 
the extraction of electrocoupling amplitudes and in the search for 
new excited states. 

\section{Electroexcitation of the 
{\boldmath $\Delta(1232)P_{33}$, $N(1440)P_{11}$,
$N(1520)D_{13}$, and
$N(1535)S_{11}$}
\label{sec:results}}

The new results on 
the electroexcitation of nucleon resonances
in single pion electroproduction 
($ep\rightarrow eN\pi$) have been obtained mostly using the CLAS
detector at JLab/Hall B (Table \ref{tab:data}). Detailed analyses of these data
sets have been performed by two groups: JLab \cite{
Aznauryan2008,Aznauryan2009} and Mainz 
\cite{MAID2007,MAID_China}.
The JLab analysis was carried using two approaches, DR and UIM
\cite{Aznauryan2003,Aznauryan2009},
and the amplitudes of the electroexcitation of
the resonances $\Delta(1232)P_{33}$, $N(1440)P_{11}$,
$N(1520)D_{13}$, and $N(1535)S_{11}$ were obtained
in the range of $Q^2$ covered by the CLAS data.
The goal of this analysis was to determine in detail
the $Q^2$ behavior of the resonance electrocoupling amplitudes or
transition form factors. To achieve this in the least model-dependent way, 
the data were analyzed at each $Q^2$ point separately
without a priori assumptions on the $Q^2$ dependence
of the electrocoupling amplitudes.
Significant effort has been put into accounting 
for model uncertainties and systematic uncertainties of 
the extracted electroexcitation amplitudes. 
Taken into account were uncertainties
in hadronic parameters, such as
masses and widths of resonances,
the amplitudes of higher lying resonances,
the parameters that determine the non-resonant contributions,
as well as the point-to-point systematics of the experimental
data and the overall normalization uncertainties of the
cross sections.
Utilization of two approaches, DR and UIM,
also allowed the estimation of the model dependence of the
results.
The results of the JLab group are presented
with the total model uncertainties that include all uncertainties
listed above.

The analysis of the Mainz group 
includes, in addition to the CLAS data, 
backward $\pi^0$ electroproduction data from Hall A 
at $Q^2= 1$ GeV$^2$ \cite{Laveis}
(Table \ref{tab:data}), and also older data from the SAID database \cite{SAID}
on $ep\rightarrow ep\pi^0,en\pi^+$ for
$W=1.1-2~$GeV and  $Q^2= 0.1- 4.4$ GeV$^2$.
The analysis is based on MAID2007 \cite{MAID2007}.
A global fit of the data simultaneously
at all $Q^2$ has been performed using parameterizations of the
electroexcitation amplitudes as a function of $Q^2$.
In addition, for the $\Delta(1232)P_{33}$, the electroexcitation
amplitudes were constrained  using theoretical predictions
of Ref. \cite{Buchmann}. All amplitudes have been constrained
by the Siegert theorem \cite{Siegert}, which gives relations
between the amplitudes at the unphysical threshold when 
the virtual photon 3-momentum
${\bf{k}}\rightarrow 0$. As the result of the global fit,
parameterizations of the helicity
amplitudes were obtained not only for the resonances
of the first and second resonance regions,
but also for the resonances of the third resonance
region; the latter results will be discussed in section 
\ref{sec:third}. 

The $\Delta(1232)P_{33}$ electroexcitation
amplitudes found in the MAMI 
($Q^2=0.06,0.2~$GeV$^2$)
\cite{Stave2006,Sparveris2007,Stave2008},
MIT/Bates ($Q^2=0.127~$GeV$^2$) \cite{Mertz,Kunz,Sparveris2005}, 
JLab Hall A ($Q^2=1~$GeV$^2$) \cite {KELLY1,KELLY2} 
and  Hall C
($Q^2= 2.8,4.2~$GeV$^2$) \cite{Frolov} experiments
on $ep\rightarrow ep\pi^0$ 
will be presented according to discussions 
in sections \ref{sec:data_mit}-\ref{sec:data_hallc}.
We also present the amplitudes of the transitions
$\gamma^* p\rightarrow N(1535)S_{11}$ and 
$\gamma^* p\rightarrow N(1440)P_{11}$ and 
$N(1520)D_{13}$ found, respectively, from the
$\gamma^* p\rightarrow \eta p$ 
and $\gamma^* p\rightarrow \pi\pi N$ reactions.
The methods used for the extraction of the resonance contributions
in these processes are described in sections \ref{sec:data_eta} and \ref{sec:data_2pi}.

\subsection{\it The 
$\Delta(1232)P_{33}$ resonance
\label{sec:results_delta}}

Historically, the electromagnetic transition amplitudes 
for the $\Delta(1232)P_{33}$ have been presented in terms 
of the  $\gamma^* N\rightarrow \Delta(1232)P_{33}$ 
magnetic-dipole transition form factor
and the ratios $R_{EM}$ and $R_{SM}$. 
For the $\gamma^* N\rightarrow \Delta(1232)P_{33}$
magnetic-dipole transition form factor we use the Ash 
convention \cite{Ash}, which relates $G^*_{M,Ash}(Q^2)$  
to the multipole
amplitude $M_{1+}^{3/2}(Q^2,W)$ at the resonance position
in the following way:
\begin{equation}
G^*_{M,Ash}(Q^2)=\frac{m}{k_r}\sqrt{\frac{8q_r
\Gamma}{3\alpha}}M_{1+}^{3/2}(Q^2,W=M),
\label{eq:res1}
\end{equation}
where $M=1232~$MeV and $\Gamma=118$~MeV are 
the mass and width of the $\Delta(1232)P_{33}$,
$q_r,k_r$ are the pion and virtual photon
three-momenta, respectively, in the c.m.s. of the reaction
$\gamma^* p
\rightarrow p\pi^0$ at the
$\Delta(1232)P_{33}$ resonance position, and $m$ is the nucleon mass.
The Jones-Scadron convention \cite{Scadron}, which is also used, 
is related to $G^*_{M,Ash}$ as:
\begin{equation}
G^*_{M,J-S}(Q^2)=G^*_{M,Ash}(Q^2)\sqrt{
1+\frac{Q^2}{(M+m)^2}}.
\label{eq:res2}
\end{equation}

The results for $G^*_{M,Ash}(Q^2)$, 
$R_{EM}$, and $R_{SM}$
extracted from experiments are shown in Fig. \ref{delta}.
For $G^*_{M,Ash}(Q^2)$, we also include the results of earlier experiments
from NINA \cite{nina} and DESY
\cite{desy_adler,desy_haidan}.  
The earlier results for $R_{EM}$ and $R_{SM}$ had large 
uncertainties and are not shown. 
In the recent experiments, the $Q^2$ range is significantly enlarged,
and accurate results are obtained for 
all quantities $G^*_M(Q^2)$, $R_{EM}$ and $R_{SM}$.

\subsubsection{\it On the  JLab and MAID2007 results
\label{sec:rsm_disagreement}}

The JLab and MAID2007 results for the magnetic-dipole
transition form factor and for $R_{EM}$ are in good agreement
with each other. 
The latter is consistent with a
constant value; its averaged value 
in the range  
$0 < Q^2 < 7$~GeV$^2$
is $R_{EM} = -2.11 \pm 0.06\%$.
However, there are significant
differences in the results for $R_{SM}$, especially at large
$Q^2$. We address this discrepancy as it has led to 
confusion regarding the scale of $Q^2$ where the
asymptotic QCD behavior may set in for the transition
$\gamma^* N\rightarrow \Delta(1232)P_{33}$.

The magnitude of the relevant amplitude $S_{1+}^{3/2}$
can be checked 
using the experimentally determined structure function 
$\sigma_{LT}(ep\rightarrow ep\pi^0)$, 
whose $\cos{\theta}$ behavior at $W=1.232~$GeV
is dominated by the interference of this amplitude
with $M_{1+}^{3/2}$ (see Eq. (\ref{eq:data1})). 
The comparison of the experimental
data for the $ep\rightarrow ep\pi^0$ structure functions                       
with the results of the JLab and MAID2007 solutions is shown
in Figs. \ref{123_04}, \ref{123_3}, and \ref{123_vil}.
At $Q^2=0.4-1.45~$GeV$^2$ (Fig. \ref{123_04}),
the JLab and the MAID2007 solutions describe
equally well the angular behavior of 
$\sigma_{LT}$. However, MAID2007 analysis increasingly
underestimates the strong $\cos{\theta}$ dependence of this
structure function with rising $Q^2$. This is a direct consequence 
of the small magnitude of $R_{SM}$ in the
MAID2007 solution. At $Q^2\geq 3$~GeV$^2$ this is
demonstrated in Figs. \ref{123_3} and \ref{123_vil}. 

\subsubsection{\it CQM and pion-cloud contribution
\label{sec:delta_cloud}}

It is well known that  the prediction of the 
$\gamma N\rightarrow \Delta(1232)P_{33}$
transition magnetic moment
was one of the first successes of the constituent 
quark model \cite{Beg}.
The assumption that the nucleon and the $\Delta(1232)$
consist of three constituent quarks moving non-relativistically 
in an $s$-wave led to the prediction 
\begin{equation}
\mu(\gamma p\rightarrow \Delta)
=\frac{2\sqrt{2}}{3}\mu_p,~~i.e.~~
\mu(\gamma p\rightarrow \Delta)=2.63\frac{e}{2m},
\label{eq:res4}
\end{equation}
where $\mu(\gamma p\rightarrow \Delta)$ is related
to the $\gamma N\rightarrow \Delta(1232)P_{33}$
magnetic-dipole form factor as \cite{Aznauryan1966}: 
\begin{equation}
\mu(\gamma p\rightarrow \Delta)
= \frac{e}{2m}\frac{2M}{M+m}G^*_{M,Ash}(Q^2=0).
\label{eq:res5}
\end{equation}
Another prediction, which is a direct result of  
this assumption, is that the electric-quadrupole and
scalar quadrupole 
$\gamma N\rightarrow \Delta(1232)P_{33}$
transitions are forbidden
\cite{Harari,Becchi_Mor}:
\begin{equation}
R_{EM}(Q^2)=R_{SM}(Q^2)=0. 
\label{eq:res5a}
\end{equation}
This follows from angular momentum conservation:
the corresponding photons carry
total angular momentum $J_{\gamma}=2$ and,
therefore, cannot be absorbed by quarks in an $s$-wave.
Qualitatively these predictions are in good agreement with
experiment and are considered as a success
of the constituent quark model, although the predicted
value of $\mu(\gamma p\rightarrow \Delta)$
(Eq. (\ref{eq:res4})) underestimates the experimental results
\cite{Aznauryan1966,Dalitz}.
With the amplitudes $A_{1/2}$ and $A_{3/2}$ at $Q^2=0$
quoted by RPP \cite{PDG},
one obtains 
\bea
&&G^*_M(0)=3.02\pm 0.03, \\
\label{eq:res8}
&&\mu(\gamma p\rightarrow \Delta)=[3.44\pm 0.03]\frac{e}{2m}.
\label{eq:res9}
\eea
The non-relativistic quark model prediction  
is below the experimental value by about 30\%.
Moreover, experiments also give non-zero values
for $R_{EM}$ and $R_{SM}$.
Despite a large effort to improve the agreement with experiment 
through modifications of the CQM with different schemes
of relativization, configuration mixings in $N$ and $\Delta$,
and the inclusion of quark anomalous magnetic moments, 
no significant progress in the description of the 
$\gamma N\rightarrow \Delta(1232)P_{33}$
transition magnetic moment
was achieved within constituent quark model. 
A satisfactory description of 
$\mu(\gamma p\rightarrow \Delta)$
was achieved in 
the models that include pion-cloud contribution:
the cloudy bag model \cite{Bermuth},
the chiral bag model \cite{Thomas},
and the Lorentz covariant chiral quark model \cite{Faessler1}.
The models of Refs. \cite{Bermuth,Faessler1} gave 
also non-zero values of the ratios $R_{EM}$ and $R_{SM}$, which
are quite close to the experimental data.
The chiral chromodielectric and $\sigma$ models of 
Ref. \cite{Fiolhais}, that include pion-cloud contribution,
give non-zero values of the ratios $R_{EM}$ and $R_{SM}$ too.
In Refs. \cite{Bermuth,Fiolhais}, 
these ratios are determined almost completely by 
the pion-cloud contributions. 
The predictions of Ref. \cite{Fiolhais} extend up to
$Q^2=1~$GeV$^2$ and are shown in Fig. \ref{ratios_new}.

Non-zero values of
the ratios $R_{EM}$ and $R_{SM}$
are obtained also in the constituent quark model with
the inclusion of two-body exchange currents 
that may be associated with the cloud
of quark-antiquark pairs
\cite{Buchmann,Buchmann3,Buchmann4,Buchmann5}.
The obtained ratio $R_{EM}(0)$ receives sizeable
contribution from exchange currents and is quite close to
the experimental value.
The prediction for $R_{SM}$ is:
\be 
R_{SM}(Q^2)=\frac{mk_r}{2Q^2}\frac{G_C^n(Q^2)}{G_M^n(Q^2)},
\label{eq:buch}
\ee 
where $G_C^n(Q^2)$ and $G_M^n(Q^2)$ are the neutron charge and
magnetic form factors. The results that follow from
Eq. (\ref{eq:buch}) for two different parameterizations of
the neutron charge form factor are shown in Fig. \ref{ratios_new}.
It can be seen that the approach describes the sign and 
order of magnitude of the ratio $R_{SM}$ extracted
from experimental data.

\begin{figure}[tb]
\begin{center}
\begin{minipage}[t]{18 cm}
\epsfig{file=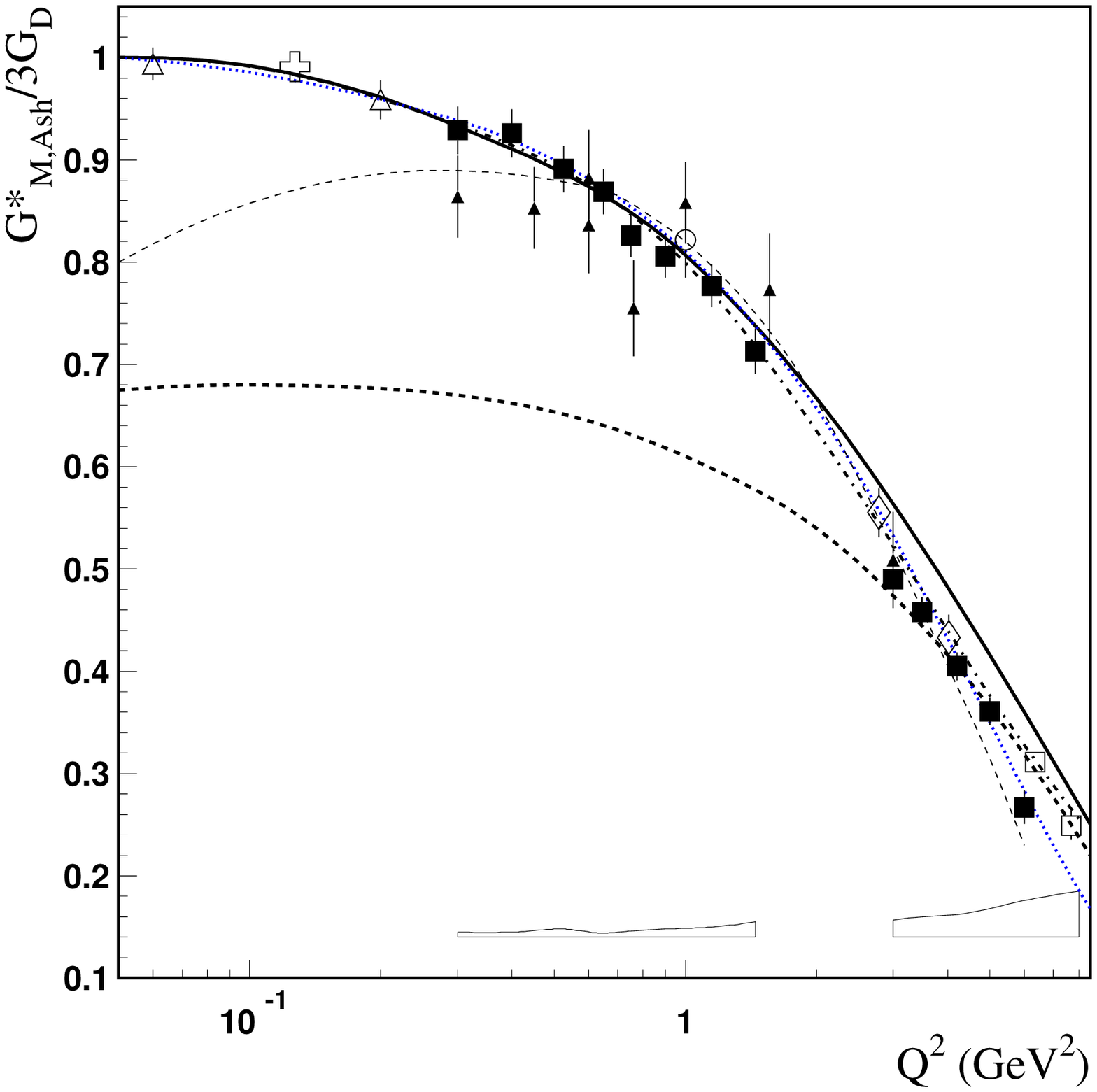,scale=0.5}
\epsfig{file=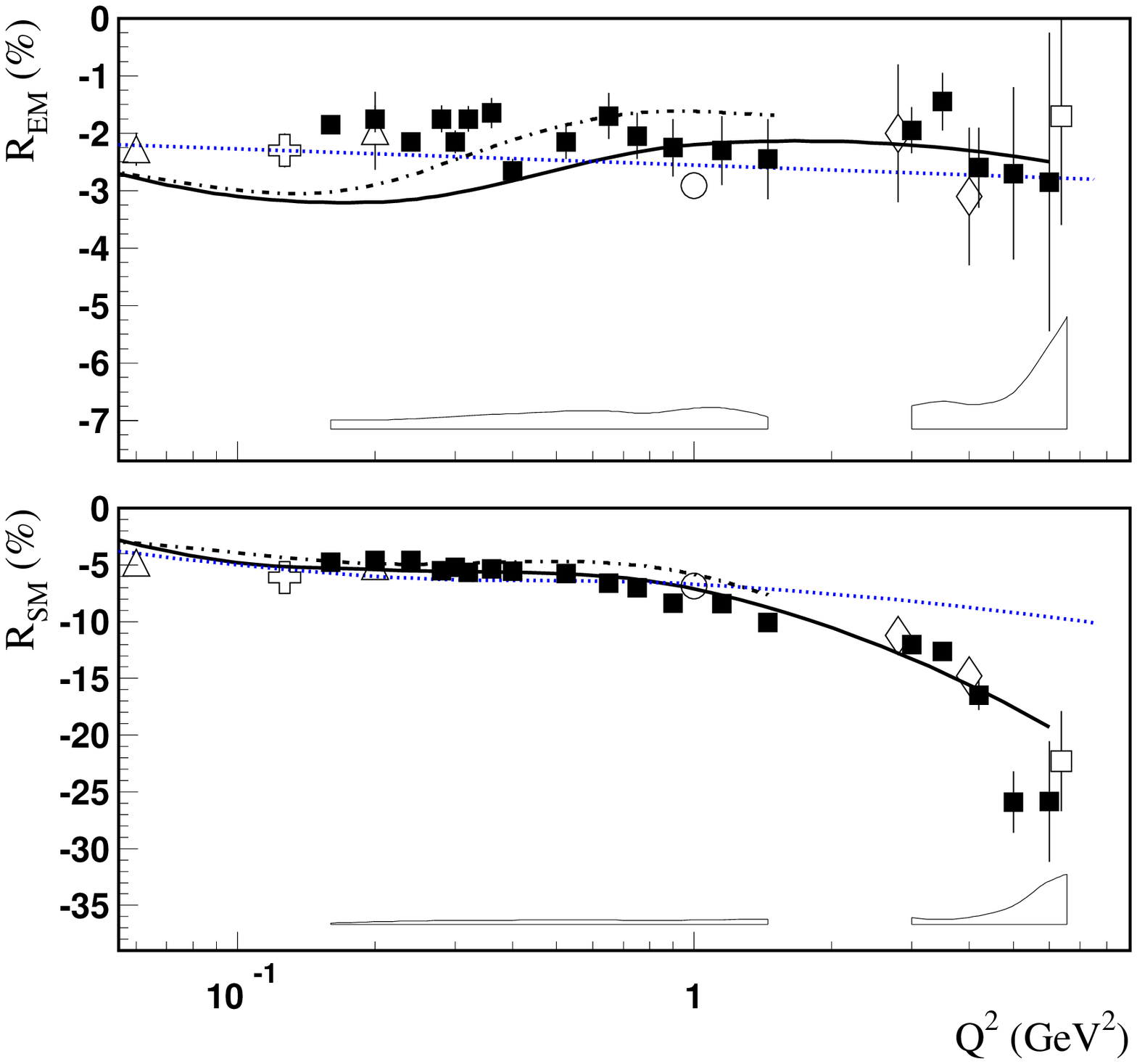,scale=0.53}
\end{minipage}
\begin{minipage}[t]{18.5 cm}
\caption{
Left panel: the form factor $G^*_{M,Ash}(Q^2)$
for the  $\gamma^* p \rightarrow~\Delta(1232)P_{33}$
transition relative to $3G_D(Q^2)$;
$G_D(Q^2)
=1/(1+\frac{Q^2}{0.71\rm{GeV}^2})^2$.
Right panel:
the ratios $R_{EM},~R_{SM}$.
The full boxes correspond
to CLAS data; they are extracted
in the analysis of the JLab group \cite{Aznauryan2009}.
The bands show the model uncertainties obtained
in this analysis.
The results of the global analysis of the Mainz group using
MAID2007 \cite{MAID2007}
are shown by the dotted curves.
The results from other experiments are: open triangles,
MAMI \cite{Stave2006,Sparveris2007,Stave2008};
open crosses, 
MIT/Bates \cite{Mertz,Kunz,Sparveris2005}; open rhombuses,
JLab/Hall C from Ref. \cite{Frolov};
open boxes, JLab/Hall C from Ref. \cite{Vilano};
and open circles,
JLab/Hall A \cite {KELLY1,KELLY2}.
The results of old experiments 
from NINA
\cite{nina} and DESY
\cite{desy_adler,desy_haidan}
are shown by full triangles.
The solid and dashed curves correspond to
the `dressed' and `bare' contributions from Ref. \cite{Sato2001};
for $R_{EM}$ and $R_{SM}$, only the `dressed'
contributions are shown; the
`bare' contributions are close to zero.
The dashed-dotted and thin dashed curves are
the predictions obtained
in the large-$N_c$ limit of QCD in Refs. \cite{Pascalutsa1,Pascalutsa2}
and \cite{Grigoryan}, respectively.
\label{delta}}
\end{minipage}
\end{center}
\end{figure}

\begin{figure}[tb]
\begin{center}
\begin{minipage}[l]{200pt}
\epsfig{file=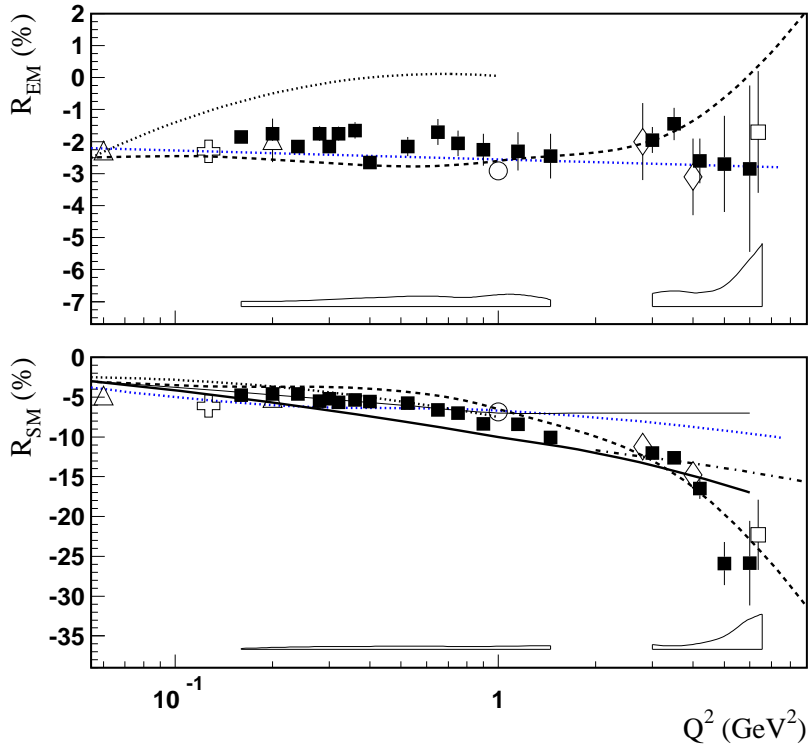,scale=0.6}
\end{minipage}
\hskip 130pt
\begin{minipage}[r]{150pt}
\caption{
The ratios $R_{EM}$ and $R_{SM}$
for the  $\gamma^* p \rightarrow~\Delta(1232)P_{33}$
transition.
The legend is partly as for Fig. \ref{delta}.
The dotted curves are the predictions of the chiral chromodielectric 
model \cite{Fiolhais}.
The solid curves are the predictions of Ref. \cite{Buchmann5} 
obtained within 
constituent quark model with
two-body exchange currents; see Eq. (\ref{eq:buch}).
The dashed curves are obtained
in Ref. \cite{Vereshkov} under assumption of early pQCD scaling
for the ratios of the  $\gamma^* p \rightarrow~\Delta(1232)P_{33}$
transition form factors. 
The dashed-dotted curve for $R_{SM}$ is the prediction of pQCD with
logarithmic corrections \cite{Idilbi}; see Eq. (\ref{eq:cor}). 
\label{ratios_new}}
\end{minipage}
\end{center}
\end{figure}

\begin{figure}[tb]
\begin{center}
\begin{minipage}[l]{200pt}
\epsfig{file=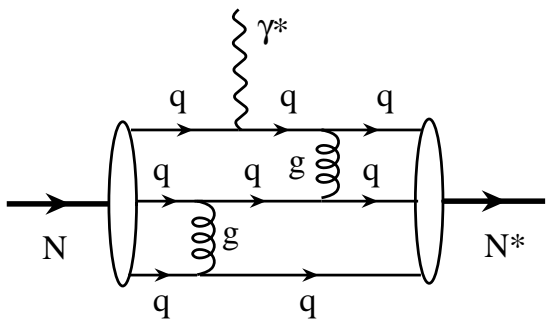,scale=0.7}
\end{minipage}
\hskip 70pt
\begin{minipage}[r]{200pt}
\caption{
One of the diagrams corresponding to the hard
mechanism for the transition 
$\gamma N\rightarrow N^*$ in the pQCD asymptotics.
\label{pQCD}}
\end{minipage}
\end{center}
\end{figure}

\begin{figure}[tb]
\begin{center}
\begin{minipage}[t]{18 cm}
\epsfig{file=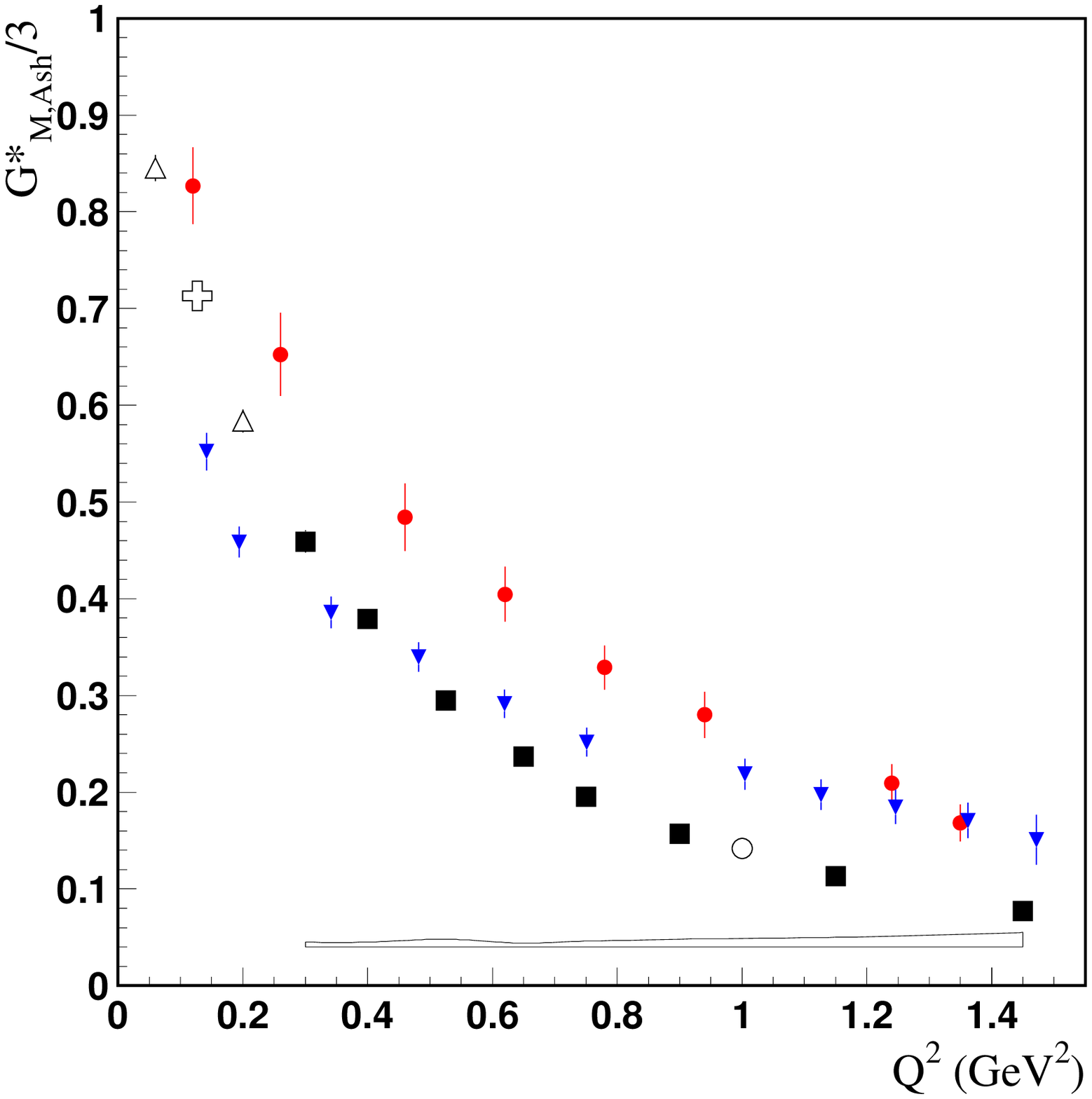,scale=0.5}
\epsfig{file=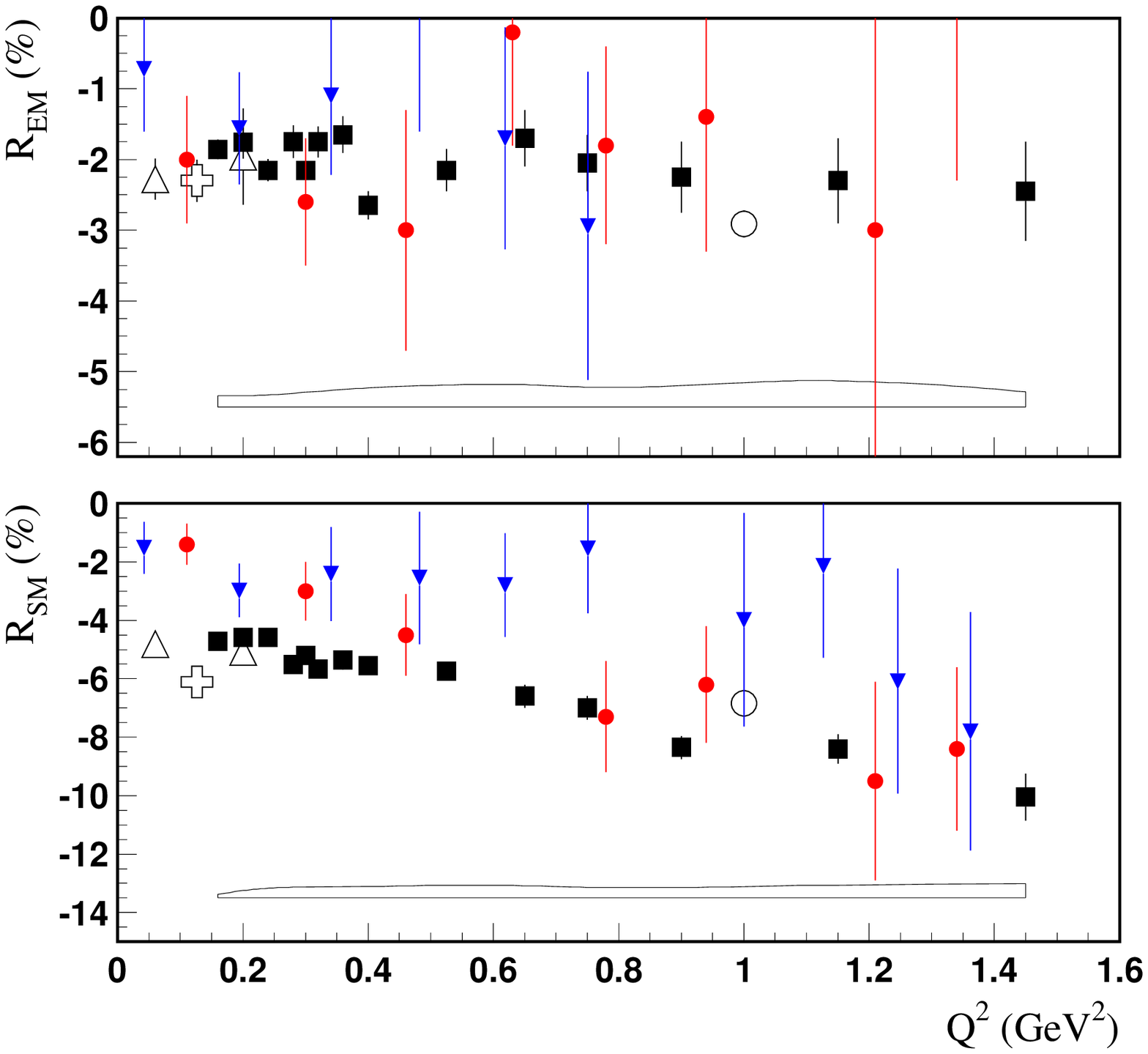,scale=0.53}
\end{minipage}
\begin{minipage}[t]{16.5 cm}
\caption{
Lattice QCD results for the 
$\gamma N\rightarrow \Delta(1232)P_{33}$
transition 
in comparison with experimental data.
Left panel: form factor $G^*_{M,Ash}(Q^2)$;
right panel: ratios $R_{EM}$ 
and $R_{SM}$.
The filled circles show the results of quenched calculations 
\cite{Alexandrou2005} extrapolated
linearly to the physical pion mass, and the filled triangles
correspond to results with dynamical quarks \cite{Alexandrou}.
Other notations as in Fig. \ref{delta}.
\label{ratios_lattice}}
\end{minipage}
\end{center}
\end{figure}

The contributions of the quark core and meson cloud to 
$\gamma N\rightarrow \Delta(1232)P_{33}$
can be separated within dynamical reaction models.
One may expect that the quark core contribution
can be identified in these models with the
`bare' resonance contribution, while the
meson-cloud effects correspond to the $t$-channel
meson exchanges followed by the $\pi N$ rescattering.
The total $\gamma N\rightarrow \Delta(1232)P_{33}$
amplitude corresponds to the `dressed' resonance.
Fig. \ref{delta} shows the `bare' and `dressed' resonance
contributions to  $G^*_{M,Ash}(Q^2)$,
$R_{EM}$, and $R_{SM}$ obtained in the SL dynamical
model \cite{Sato2001}
through the description
of the experimental data on $ep\rightarrow ep\pi^0$
at $Q^2 < 4$~GeV$^2$.
The corresponding results of the DMT dynamical model 
\cite{Kamalov1999,Kamalov2001} 
are very similar to those obtained by SL.
The pion-cloud contribution is significant and accounts
for more than 30\% of $G^*_{M,Ash}(Q^2)$ at the photon
point and remains sizable even at large $Q^2$. 
As in the models \cite{Bermuth,Fiolhais},
the $R_{EM}$ and $R_{SM}$ ratios of the dynamical
models \cite{Kamalov1999,Kamalov2001,Sato2001} are almost exclusively 
determined by the meson-cloud contributions. 

\subsubsection{\it Perturbative QCD asymptotic limits
\label{sec:delta_pqcd}}

The asymptotic limit of pQCD puts 
restrictions on the $Q^2$ behavior of the 
$\gamma^* N\rightarrow N^*$ transition
amplitudes that follow from hadron helicity
conservation \cite{Brodsky1} and dimensional
counting rules \cite{Matveev,Brodsky2,Brodsky3,Brodsky4,Efremov}.
These restrictions are specific to 
the hard scattering mechanism when
large transferred momentum
is shared among the three quarks 
through two hard gluon exchanges
(Fig. \ref{pQCD}).
Below we discuss in more detail the predictions for the helicity
amplitudes $A_{1/2}$, $A_{3/2}$, and $S_{1/2}$. Following 
the discussion in Ref. \cite{Carlson}, we consider
the $\gamma^* N\rightarrow N^*$ transition in the Breit system,
where
\be 
{\bf p}=-{\bf p^*}; 
{\bf k}=-2{\bf p}.
\label{eq:res22}
\ee 

From relativistic invariance it follows that the matrix elements
entering Eqs. (\ref{eq:def44}-\ref{eq:def46}) can be written
in the Breit frame in the following way:
\be 
<N^*,S_z^*|\epsilon^{(\lambda_{\gamma})}_{\mu}J_{em}^{\mu}|N,S_z>
=<\lambda_{N^*}|\epsilon^{(\lambda_{\gamma})}_{\mu}J_{em}^{\mu}
|\lambda_N>_{Breit~frame},
\label{eq:ress22}
\ee 
where $\lambda_N=-S_z$ and $\lambda_{N^*}=S_z^*$.

In the limit $Q^2\rightarrow \infty$, 
we have $|{\bf p}|\simeq |{\bf p^*}|\sim \frac{Q}{2}$;
therefore, the $N$ and $N^*$ masses, and also the quark
masses and transverse momenta, can be neglected compared 
to their longitudinal momenta. In this case,
the gluon and photon interaction vertices with
quarks ($\gamma_{\mu}$) preserve the quark helicity,
and every quark that requires its helicity to be flipped
introduces an additional factor $\sim 1/Q$.
For the matrix elements that enter 
Eqs. (\ref{eq:def44},\ref{eq:def45},\ref{eq:def46}) we have,
respectively, 
$\lambda_{N^*}-\lambda_{N}=0,2,1$. Therefore:
\be
\tilde{A}^N_{3/2}/\tilde{A}^N_{1/2}\sim 1/Q^2, 
~~~~~\tilde{A}^N_{1/2}/\tilde{S}^N_{1/2}\sim const. 
\label{eq:resss22}
\ee 
Here, we took into account the factor 
$\frac{|{\bf k}|}{Q}\approx \frac{Q}{2m}$ 
in the definition (\ref{eq:def46}) for $\tilde{S}^N_{1/2}$.

Using the more detailed evaluation of the asymptotic
$Q^2$ behavior of the helicity amplitudes in the Breit
frame in Ref. \cite{Carlson},
we get the following results:
\be
\tilde{A}^N_{1/2}\sim 1/Q^3, 
~~~~~\tilde{A}^N_{3/2}\sim 1/Q^5, 
~~~~~\tilde{S}^N_{1/2}\sim 1/Q^3.
\label{eq:ress23}
\ee 
Furthermore, taking into account the definitions (\ref{eq:def25}), 
(\ref{eq:def38}-\ref{eq:def40}), and (\ref{eq:def48})
of section \ref{sec:definitions}, we obtain the asymptotic predictions
for $R_{EM}$ and $R_{SM}$:
\be 
R_{EM}\rightarrow 100\%,
~~~~R_{SM}\rightarrow const,~~~~Q^2\rightarrow \infty.
\label{eq:res23}
\ee 
The study of $R_{EM}(Q^2)$ and $R_{SM}(Q^2)$
should give clear understanding of the $Q^2$ range where
the pQCD regime for the 
$\gamma N\rightarrow \Delta(1232)P_{33}$
transition may set in.

The results for  $R_{EM}(Q^2)$ and $R_{SM}(Q^2)$ extracted from experimental
data show that $R_{EM}$ remains negative, small,
and nearly constant in the entire range 
$0<Q^2<7$~GeV$^2$; $R_{SM}$ remains negative, but its magnitude
strongly rises at high $Q^2$. Consequently, there is no indication that 
asymptotic pQCD is applicable in the range  
$0<Q^2<7~$GeV$^2$ covered by experiment. 
The very small value of $R_{EM}$ at the highest 
$Q^2$ values indicates that there is not even a trend towards that limit. 

It should be mentioned however, that the asymptotic behaviour
given by Eqs. (\ref{eq:resss22},\ref{eq:ress23},\ref{eq:res23}) 
corresponds to the inner part
of the hard scattering diagram of Fig. \ref{pQCD}, i.e.
to the part related to the quarks only.
Higher order corrections and convolution with the soft
$N$ and $N^*$ distribution amplitudes  may introduce
logarithmic corrections. The convolution
is needed also
to calculate numerical coefficients in 
Eqs. (\ref{eq:resss22},\ref{eq:ress23},\ref{eq:res23}).
In Ref. \cite{Carlson}, for example, it is shown that
the asymptotic form of nucleon wave function
suggested in Ref. 
\cite{Chernyak}
results in significant destructive interference between
contributions related to different parts of this wave
function. This can strongly suppress the 
$\gamma N\rightarrow \Delta(1232)P_{33}$ amplitude
in the pQCD limit. 

For the ratio $R_{SM}$, logarithmic corrections 
caused by
the orbital motion of the constituents in the nucleon
and the  $\Delta(1232)P_{33}$
are found in Ref. \cite{Idilbi}: 
\be 
R_{SM}=c\frac{|{\bf k}|}{Q^2}ln^2(Q^2/\Lambda^2),
\label{eq:cor}
\ee 
where $c$ is an unknown factor, and $\Lambda=0.2\div 0.4$ is
a soft scale related to the size of hadrons.
Corresponding results with $\Lambda=0.2$ that gives
maximal slope for $R_{SM}$ are shown in Fig. \ref{ratios_new}.
The coefficient $c$ has been found from the data at 
$Q^2=2.8\div 4.2~$GeV$^2$. It can be seen that logarithmic
corrections are not sufficient to describe strongly rising
magnitude of $R_{SM}$ found in experiment. 

To conclude this section, we would like to mention
the results obtained in Ref. \cite{Vereshkov}.
The investigation exploits the observation of Refs. \cite{Belitsky,Karmanov}
that the experimental data on the ratio of the nucleon
form factors $F_1(Q^2)$ and $F_2(Q^2)$ 
\cite{Jones,Gayou} 
exhibit the pQCD scaling behaviour starting
with small $Q^2=0.5-1~$GeV$^2$, while the form factors
themselves do not.
Good description of this ratio has been obtained
due to the logarithmic corrections found in Ref. \cite{Belitsky}.
On this basis in Ref. \cite{Vereshkov}, 
a phenomenological multi-parameter approach
was proposed to describe the ratios 
$R_{EM}$ and $R_{SM}$. With the parameters
found from the fit to the data
at $Q^2<7~$GeV$^2$, the extension was made
to higher $Q^2$. According to the results
shown in Fig. \ref{ratios_new}, 
$R_{SM}$ will continue to grow in 
magnitude,
while $R_{EM}$ will cross zero around
$Q^2=5$GeV$^2$ and become positive. 
While the Jlab data disfavor
this prediction for $R_{EM}$, a zero-crossing at somewhat 
higher $Q^2$
values is not excluded
taking into account statistical
and systematic uncertainties of the $R_{EM}$ values
extracted from experiment. 
These results can be checked
in future JLab experiments at 12 GeV. 

\subsubsection{\it Large $N_c$ limit and GPDs
\label{sec:delta_nc}}

It is well known that in the limit of a large number of colors, $N_c$, 
many results of the $SU(6)$ symmetric quark model can be obtained without
making model assumptions. For example, in that limit
we have: ${\mu_p}/{\mu_n}=-\frac{3}{2}$,
i.e. the same result as in the non-relativistic quark model.
The prediction for the $\gamma N\rightarrow \Delta(1232)P_{33}$
magnetic moment is \cite{Jenkins1}:
\begin{equation}
\mu(\gamma p\rightarrow \Delta)
=\frac{1}{\sqrt{2}}(\mu_p-\mu_n).
\label{eq:res15}
\end{equation}
This gives $\mu(\gamma p\rightarrow 
\Delta)=3.23\frac{e}{2m}$, which is 
very close to the experimental value 
(Eq. \ref{eq:res9}).
The predictions for $R_{EM}(0)$ and $R_{SM}(0)$  
in the large $N_c$ limit with $M=m$ are 
$R_{EM}(0)=R_{SM}(0)=0$
\cite{Pascalutsa2,Grigoryan,Jenkins2},
i.e. the same as in the non-relativistic quark model (Eq. \ref{eq:res5a}).
With empirical masses ($M-m\neq 0$), there are
two kinds of predictions for $R_{EM}(0)$ and $R_{SM}(0)$:
\begin{equation}
R_{EM}(0)=R_{SM}(0)
=\frac{1}{12}\left(\frac{m}{M}\right)^{3/2}(M^2-m^2)\frac{r_n}{\kappa_V}
\label{eq:res16}
\end{equation}
and
\begin{equation}
R_{EM}(0)=R_{SM}(0)
=-\frac{M-m}{3M+m},
\label{eq:res16b}
\end{equation}
where $r_n$ is the neutron charge radius     and
$\kappa_V=\frac{2m}{e}(\mu_p-\mu_n)$.
Eqs. (\ref{eq:res16}) are derived in Ref. \cite{Pascalutsa2}
using the relation between the $N\rightarrow \Delta$ quadrupole moment
and $r_n$ found in the large $N_c$ limit in Ref. \cite{Jenkins2}.
Eqs. (\ref{eq:res16b}) are obtained in Ref. \cite{Grigoryan}
in the approach based on the conjunction of large $N_c$ QCD
with the idea of holography. 

Numerically 
the relations (\ref{eq:res16}) and (\ref{eq:res16b}) give,
respectively, 
\begin{equation}
R_{EM}(0)=R_{SM}(0)=-2.77\%
\label{eq:res17}
\end{equation}
and
\begin{equation}
R_{EM}(0)=R_{SM}(0)=-6.3\%.
\label{eq:res17a}
\end{equation}
Therefore,  with empirical masses $M$ and $m$,
the ratios $R_{EM}(0)$ and 
$R_{SM}(0)$ gain correct signs in both approaches 
\cite{Pascalutsa2} and \cite{Grigoryan};
with this
the predictions for $R_{EM}$ (\ref{eq:res16},\ref{eq:res17}) 
and $R_{SM}$ (\ref{eq:res16b},\ref{eq:res17a}) are close 
to the experimental values:
\bea
&&R_{EM}=-(2.5\pm 0.5)\%~~(Q^2=0)~~\cite{PDG},
\label{eq:res13}\\
&&R_{SM}=-(4.81\pm 0.27\pm 0.26)\%~~(Q^2=0.06~\rm{GeV}^2)
~~\cite{Stave2008}.
\label{eq:res14}
\eea

In Ref. \cite{Pascalutsa2}, 
the relations (\ref{eq:res16}) are extended to finite values of $Q^2$
via parameterizations:
\bea
&&R_{EM}=
-\left(\frac{m}{M}\right)^{3/2}\frac{M^2-m^2}{2Q^2}
\frac{G_{En}(Q^2)}{F_{2p}(Q^2)-F_{2n}(Q^2)},
\label{eq:res18}\\
&&R_{SM}=
-\left(\frac{m}{M}\right)^{3/2}
\frac{\sqrt{[(M+m)^2+Q^2][(M-m)^2+Q^2]}}{2Q^2}
\frac{G_{En}(Q^2)}{F_{2p}(Q^2)-F_{2n}(Q^2)},
\label{eq:res19}
\eea
where $F_{2N}(Q^2)$ is the nucleon Pauli form factor 
and $G_{En}(Q^2)$ is the neutron electric form factor. We note that 
in this parameterization $R_{EM}$ and $R_{SM}$ have the same 
dependence on the nucleon form factors. Hence, the double ratio 
$R_{SM} / R_{EM}$ 
depends only on kinematical quantities. 
The relations (\ref{eq:res18},\ref{eq:res19})
give a reasonably good description
of the experimental data for $Q^2<1.5~$GeV$^2$. The 
corresponding results are shown in Fig. \ref{delta}.

The generalization of the predictions (\ref{eq:res16b})
to $Q^2\neq 0$ is obtained in Ref. \cite{Grigoryan} in the leading 
order over $1/N_C$:
\bea
&&R_{EM}=
-\frac{M^2-m^2-Q^2}{(3M+m)(M+m)+Q^2},
\label{eq:res18a}\\
&&R_{SM}=
-\frac{\sqrt{4M^2Q^2+(M^2-m^2-Q^2)^2}}
{(3M+m)(M+m)+Q^2}.
\label{eq:res19a}
\eea
Both ratios have the correct pQCD behavior. With this it is remarkable
that at $Q^2\rightarrow\infty$, $R_{SM}\rightarrow -100\%$.
Such asymptotic value of $R_{SM}$ might be expected
taking into account 
the rapidly rising magnitude and negative value of $R_{SM}$ found 
experimentally (Fig. \ref{delta}).
In Fig. \ref{delta}, we show also the prediction
 for the magnetic-dipole
$\gamma p\rightarrow \Delta$
transition obtained in the same Ref. \cite{Grigoryan}.
The description at $Q^2> 0.5$~GeV$^2$ is quite
satisfactory. The disagreement at small $Q^2$ may be attributed
to the lack of higher $1/N_c$ corrections in the calculations.  

The prediction (\ref{eq:res15}) for the magnetic-dipole 
$\gamma p\rightarrow \Delta$
transition was extended in Ref. \cite{Pascalutsa1}
to finite values of $Q^2$. This was made by utilizing the GPD formalism,
a recently developed framework that
utilizes hard exclusive processes  to
determine spin-dependent and spin-independent leading twist
amplitudes which characterize the 3-dimensional internal structure of the
nucleon and of the transitions like
$N\rightarrow \Delta$ (see, for example, reviews \cite{Goeke,GPD}). 
The relation between the $N\rightarrow \Delta$ GPD  
$H_M$ and the nucleon isovector GPD $E^u-E^d$ 
in the large $N_c$ limit 
is derived in Refs. \cite{Goeke,Frankfurt}:
\begin{equation}
H_M(x,\xi,Q^2)=2\frac{G_M^*(0)}{\kappa_V}[E^u(x,\xi,Q^2)-E^d(x,\xi,Q^2)].
\label{eq:res20}
\end{equation}

\noindent Using this relation in  Ref. \cite{Pascalutsa1},
the following sum rule was found:
\bea
G^*_{M,J-S}(Q^2)&&=\frac{G_M^*(0)}{\kappa_V}\int^{+1}_{-1}
dx[E^u(x,\xi,Q^2)-E^d(x,\xi,Q^2)]
\nonumber \\
&&=\frac{G_M^*(0)}{\kappa_V}[F_{2p}(Q^2)-F_{2n}(Q^2)].
\label{eq:res21}
\eea
Furthermore, with the Regge parameterization of GPDs 
suggested in Ref. \cite{Guidal}, a good description of 
the magnetic-dipole $\gamma p\rightarrow \Delta$
transition was obtained up to $Q^2=8~$GeV$^2$
\cite{Pascalutsa1}. This result is presented 
in Fig. \ref{delta}.

\subsubsection{\it Lattice QCD
\label{sec:delta_lattice}}

The first lattice QCD study of the 
$\gamma p\rightarrow \Delta$
transition was carried out in Ref. \cite{lattice_1}
in the quenched approximation and with limited
statistics. Negative values were
obtained 
for both $R_{EM}$ and $R_{SM}$;
however, because of large statistical uncertainties, a zero value
could not be excluded for either of these ratios.
Increased statistics, and a number of improvements
introduced later, allowed the evaluation of the   
$\gamma p\rightarrow \Delta$
form factors with better accuracy \cite{
Alexandrou,AlexBeij,Alexandrou2005,Alexandrou2007}.
The results obtained in Ref. \cite{Alexandrou2005}  
in the quenched approximation and linearly
extrapolated to the physical pion mass are shown in 
Fig. \ref{ratios_lattice}. A non-zero value
with the correct sign was obtained
for $R_{EM}$. $R_{SM}$ is in good agreement
with the experimental results in the entire $Q^2$ range
covered by the calculations, with the exception of small
$Q^2$. At $Q^2<0.4$~GeV$^2$, the lattice results are 
negative but smaller in magnitude than the experimental values.
The first calculations with dynamical quarks were performed in Ref. 
\cite{Alexandrou}. 
The results that correspond to the lowest pion
mass $m_{\pi}=350$~MeV are shown in 
Fig. \ref{ratios_lattice}.
For the ratios $R_{EM}$ and $R_{SM}$,
they confirm non-zero and negative values obtained
in the quenched approximation.
For $R_{SM}$ at small $Q^2$, the unquenched
calculations show a slightly decreasing gap between the lattice
results and experiment.

For the magnetic-dipole form factor, unquenched calculations
give better agreement with experiment. However,
a complete description is not achieved in spite of
a quite small pion mass. In Ref. 
\cite{Alexandrou}, the conclusion is made that
the most likely source for this disagreement is the fact
that pion masses are still too large.  

\subsection{\it The Roper resonance:
$N(1440)P_{11}$
\label{sec:results_roper}}

\begin{figure}[tb]
\begin{center}
\begin{minipage}[t]{14.5 cm}
\epsfig{file=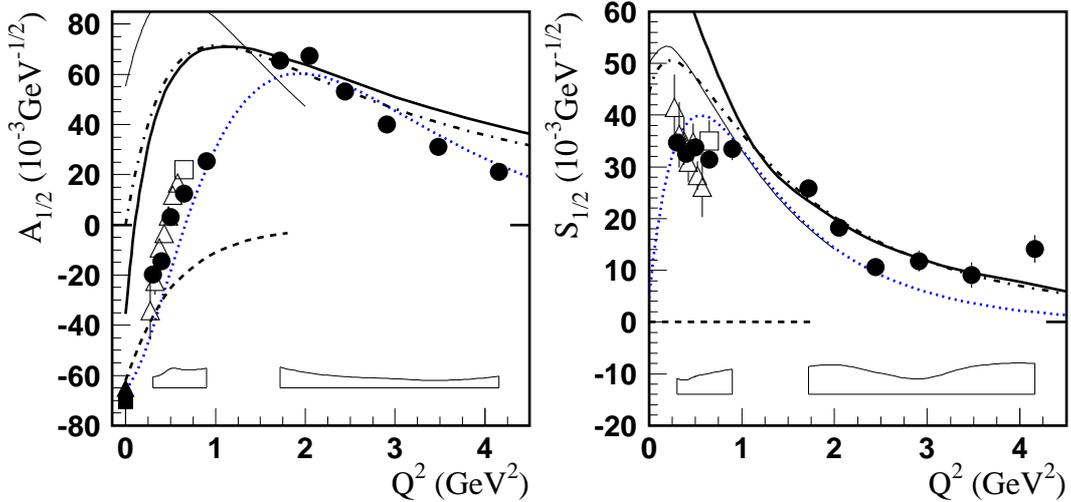,scale=0.8}
\end{minipage}
\begin{minipage}[t]{16.5 cm}
\caption{
Helicity amplitudes
for the  $\gamma^* p \rightarrow N(1440)P_{11}$
transition.
The full circles correspond
to CLAS data on pion electroproduction; they are extracted
in the analysis of the JLab group
\cite{Aznauryan2008,Aznauryan2009}.
The bands show the model uncertainties obtained
in this analysis.
The results of the global analysis by the Mainz group
using
MAID2007 \cite{MAID2007}
are shown by the dotted curves.
The full box at $Q^2=0$ is
the amplitude extracted from CLAS $\pi$ photoproduction
data \cite{Dugger_pip}.
All these results correspond to
$M=1440~$MeV, $\Gamma_{tot}=350~$MeV, and $\beta_{\pi N}=0.6$.
The full triangle at $Q^2=0$ is
the RPP estimate \cite{PDG}.
The open boxes are  the results of the combined analysis
of CLAS single $\pi$ and 2$\pi$ electroproduction data
\cite{Azn065}.
The open triangles correspond to the amplitudes  extracted from
CLAS 2$\pi$ electroproduction data using $\beta_{2\pi N}=0.4$
\cite{Mokeev2009}.
The solid and dashed-dotted  curves are the results obtained,
respectively,  in the
LF relativistic quark model \cite{Aznauryan_Roper}
and the covariant spectator quark model \cite{Ramalho},
assuming that $N(1440)P_{11}$
is a first radial excitation of the $3q$ ground state.
The thin solid curves are non-relativistic quark model
predictions from Ref. \cite{Li2}, taken with the correct sign.
The dashed curves are
obtained assuming that $N(1440)P_{11}$
is a gluonic baryon excitation (q$^3$G hybrid state) \cite{Li2}.
\label{p11}}
\end{minipage}
\end{center}
\end{figure}

\begin{figure}[tb]
\begin{center}
\begin{minipage}[t]{14.5 cm}
\epsfig{file=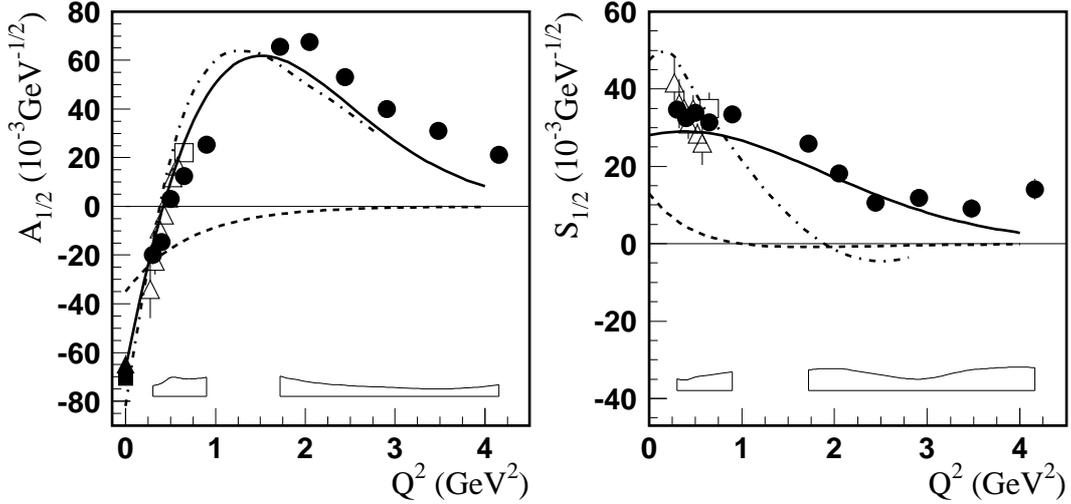,scale=0.8}
\end{minipage}
\begin{minipage}[t]{16.5 cm}
\caption{
The preditions for
the  $\gamma^* p \rightarrow N(1440)P_{11}$
helicity amplitudes obtained by combining
a quark core taken as the first
radial excitation of the $3q$ ground state
with a meson cloud.
The solid curves are
from Ref. \cite{Faessler2011} where the meson-cloud
contribution corresponds to the $N+\sigma$ content
of the Roper resonance; this contribution is shown separately
by the dashed curves.
The dashed-dotted curves are
from Ref. \cite{Cano1998}.
Notations for the amplitudes extracted from the experimental data
are as in Fig. \ref{p11}.
\label{p11_faes}}
\end{minipage}
\end{center}
\end{figure}

The results
for the  $\gamma^* p \rightarrow N(1440)P_{11}$
helicity amplitudes
are presented in Fig. \ref{p11}.
The CLAS measurements allowed for the first time 
the determination of the electroexcitation
amplitudes for this resonance at $Q^2>0$
in the range  $Q^2 < 4.5~$GeV$^2$.
The amplitudes are
extracted from CLAS $\pi$ and 2$\pi$ electroproduction data,
both of which are sensitive to the
$N(1440)P_{11}$ contribution owing to the large branching
ratios of this resonance to the $\pi N$ and $2\pi N$ channels.
The amplitudes extracted from the two
channels are in good agreement with each other, thus
confirming the reliability of the obtained results,
as the non-resonant contributions and resonance
decay mechanisms are different
in these reactions. Within uncertainties,
the results of the
JLab and Mainz
analyses of pion electroproduction data
are also in good agreement with
each other.

\subsubsection{\it The Roper resonance: a predominantly first
radial excitation of the $3q$ ground state
\label{sec:roper_radial}}

The so-called Roper resonance, $N(1440)P_{11}$, is the lowest
excited state of the nucleon. In the CQM, the simplest and most
natural assumption is that this resonance is the first
radial excitation of the three quark ($3q$) ground state
and belongs to the multiplet $[56, 0^+]_r$.
However, the first calculations within the non-relativistic CQM with
the oscillator potential failed to reproduce the mass and
width of the resonance (see Ref.
\cite{Capstick_Isgur} and numerous references therein).
Moreover, the oscillator potential led to the wrong mass
ordering between the Roper resonance
and the resonances of
the multiplet $[70,1^-]$, in particular, the $N(1535)S_{11}$.
It was realized later \cite{Capstick} that
the sign of the $\gamma^* p \rightarrow~N(1440)P_{11}$
transition amplitude at the photon point is also not
consistent with the non-relativistic CQM.
These discrepancies made the state a problematic object
for the CQM. 

Further investigations provided better results.
Estimations of the mass in the
relativized quark model with more realistic potential
motivated by QCD significantly reduced the difference
of the computed and empirical mass from initially $\sim 600~$MeV to
$\sim 100~$MeV \cite{Capstick_Isgur,Richard}.
The quark models, where the quarks
were assumed to interact by Goldstone boson
exchanges, gave
the correct mass ordering between the
$N(1440)P_{11}$ and $N(1535)S_{11}$ \cite{Glozman}.
The description of the width was also improved.
In the quark model that incorporates some of the features
expected from QCD,
the gap between the empirical width and theoretical calculations
was significantly reduced \cite{Koniuk_Isgur};
a good description of the width was obtained
within the pair-creation ${}^3P_0$ model
\cite{Capstick_Roberts,Gavela1980,Cano1996}.
The LF relativistic quark model
gave the correct sign of the
$\gamma^* p \rightarrow~N(1440)P_{11}$ amplitude
at the photon point \cite{Capstick}.

In spite of the positive developments in the CQM, the Roper
resonance continued to be considered as a `puzzle'
and gave rise to the attempts to describe this
resonance in alternative approaches or by complementing
the $3q$ state with $qqqq\bar{q}$ components.
The electroexcitation amplitudes of the Roper resonance
were evaluated assuming this resonance has a gluonic component
in its wave function, i.e.
it is  
a hybrid state $q^3G$ \cite{Li1,Li2}.
Such an approach was
motivated by the fact that in the bag model, 
the lightest hybrid state has quantum numbers of
the Roper resonance, and its mass can be $< 1.5~$GeV \cite{Close}.
The helicity amplitudes $A_{1/2}(Q^2)$ and $S_{1/2}(Q^2)$
predicted in the hybrid model were consistent with the early
data that showed the rapid disappearance of $A_{1/2}$
with $Q^2$ and $S_{1/2}(Q^2)\sim 0$.
Also suggested were alternative descriptions of the $N(1440)P_{11}$
as a $N\sigma$ molecule \cite{Oset_1983,Dillig}
or a dynamically generated resonance \cite{Krehl1}.
The empirical width of the $N(1440)P_{11}$ was described
within the non-relativistic CQM
by complementing the $3q$ state with a
$30\%$ $qqqq\bar{q}$ component \cite{Riska}.

The CLAS measurements
made possible,
for the first time,
the determination of the electroexcitation
amplitudes of the Roper resonance on the proton
in a wide range of $Q^2$. These results
are crucial to get a better 
understanding of the nature of this state.
Fig. \ref{p11} shows that
the transverse helicity amplitude $A_{1/2}$
of the  $\gamma^* p \rightarrow N(1440)P_{11}$
transition extracted
from CLAS data exhibits a very specific behavior.
This amplitude,
being large and negative at $Q^2=0$, changes sign
in the range $0.4<Q^2<0.65~$GeV$^2$,
and becomes large and positive
at $Q^2>1.5~$GeV$^2$. With
increasing $Q^2$, $A_{1/2}$ drops smoothly in magnitude.
Such a behaviour is qualitatively reproduced by the LF
relativistic quark models
\cite{Capstick,Aznauryan_Roper,Weber,Simula,Bruno}
assuming
that the Roper resonance $N(1440)P_{11}$ is the first
radial excitation of the $3q$ ground state. The results
obtained
in these models are presented and reviewed in Ref.
\cite{Aznauryan_Roper}.
In Fig. \ref{p11}, we show the predictions
\cite{Aznauryan_Roper}, which extend over the full 
$Q^2$ range investigated in experiment.
Under the assumption that the Roper
resonance is a first radial excitation
of the $3q$ ground state,
the LF relativistic quark models
also describe the sign of the
longitudinal
$\gamma^* p \rightarrow N(1440)P_{11}$ amplitude.
Under the same assumption, a
good description of both amplitudes  at high $Q^2$
is obtained
within the covariant
quark model of Ref. \cite{Ramalho}.

The LF relativistic quark models
\cite{Capstick,Aznauryan_Roper,Weber,Simula,Bruno}
and the approach of Ref. \cite{Ramalho} fail, 
however, to describe
the data at small $Q^2$.
This can
have a natural explanation in the meson-cloud contribution.
Indeed, the description
at small $Q^2$ is significantly improved \cite{Cano1998,Faessler2011}
when a quark core, taken as the first
radial excitation of the $3q$ ground state,
is combined with a meson cloud. The last contribution
is constructed 
in the approaches \cite{Cano1998,Faessler2011}
in different ways.
In Ref. \cite{Cano1998}, the meson-cloud contribution is found in
the pair-creation ${}^3P_0$ model, while
in Ref. \cite{Faessler2011} it corresponds to
the $N+\sigma$ contribution. It should be mentioned that the
strong $N+\sigma$ content is found in the Roper
resonance via estimations
in a gluon exchange model \cite{Dillig},
and also in the analysis of data
on pion- and photon-induced reactions
\cite{Sarantsev}.
A large meson-cloud contribution to $A_{1/2}$
is expected also according to the
coupled-channel analysis of pion photoproduction
data \cite{Sato2008}.

An important aspect of the transition amplitudes is their signs.
As we discussed in section \ref{sec:amplitudes}, the experimental
results on the $\gamma^* N \rightarrow~N^*$ amplitudes, extracted
from the contribution of diagram (d) in Fig.~\ref{fig2} to
$\gamma^* N \rightarrow~N\pi$, contain the sign of the
hadronic $\pi N N^*$ vertex.
The sign of the $N(1440)P_{11}\rightarrow \pi N$
vertex was first found in Ref. \cite{Capstick} using the ${}^3 P_0$
model, and confirmed by a computation using the LF relativistic
quark model \cite{Aznauryan_Roper}.
A comparison between the empirical amplitudes and the quark model predictions
is given in Figs. \ref{p11} and \ref{p11_faes} taking into account this sign.

Here we briefly comment on the sign of the
$\gamma^* p \rightarrow~N(1440)P_{11}$
amplitudes in the non-relativistic quark models.
Traditionally, the sign of $A_{1/2}(Q^2=0)$ has been fixed
to the experimentally determined sign of that amplitude
at $Q^2=0$. In this case, the non-relativistic quark model
predictions for both amplitudes $A_{1/2}$ and $S_{1/2}$
have negative signs. As examples, we refer to the
predictions from Ref. \cite{Li2}.
The negative signs of $A_{1/2}$ and $S_{1/2}$
are in contradiction to the experimental data:
for $S_{1/2}$ at all $Q^2$, and for $A_{1/2}$
at $Q^2 > 0.5$~GeV$^2$. They also contradict the LF relativistic
quark model predictions.
However, if we take into account the sign
of the $N(1440)P_{11}\rightarrow \pi N$ vertex
found in Refs. \cite{Capstick,Aznauryan_Roper},
the non-relativistic quark model predictions for the
$\gamma^* p \rightarrow~N(1440)P_{11}$ amplitudes
get signs opposite to those found in the traditional way,
and the aforementioned disagreement disappears
(see the results from Ref.
\cite{Li2} presented in Fig. \ref{p11}
with the correct signs).

Taken together, the arguments presented above 
provide strong evidence in favor of the
Roper resonance  $N(1440)P_{11}$ as predominantly the first
radial excitation of the $3q$ ground state.

In Fig. \ref{p11}, we present the predictions
obtained assuming that the Roper resonance is a hybrid state \cite{Li1,Li2}.
They definitely contradict the amplitudes extracted from
experimental data. It is important to mention that although the suppression
of the longitudinal amplitude $S_{1/2}$ is obtained
in Refs. \cite{Li1,Li2} up
to $O(v^2/c^2)$, it has a physical origin,
which makes this result practically independent of
relativistic effects. The presentation of the
$N(1440)P_{11}$ as a $q^3G$ hybrid state is definitely ruled out.

\subsubsection{\it Light-front holographic QCD
\label{sec:holographic}}
Convincing arguments in favor of the point of view
that the Roper resonance is a first radial excitation
of the nucleon
have been obtained very recently in the light-front
holographic QCD  
\cite{Teramond1,Teramond2}. This approach is built on
the correspondence between semiclassical QCD
quantized on the light-front and a dual gravity model
in anti-de Sitter (AdS) space providing an approximation
to QCD in its strongly coupled regime. 
The arguments are based on the description of the mass
of the Roper resonance
and the $F_1(Q^2)\equiv Q^2G_1(Q^2)$ form factor
for the  $\gamma^* p \rightarrow~N(1440)P_{11}$ transition
(see definitions (\ref{eq:def53},\ref{eq:def54})).

Using Schr\"odinger and Dirac equations 
with a linear confining potential and fixing
the overall energy scale to be identical
for mesons and baryons, the masses of  mesons and baryons
on the Regge trajectories that correspond to
$\pi,\rho,\omega$, and positive parity baryon states are derived.
These masses are consistent with the empirically observed
masses. The Roper resonance $N(1440)P_{11}$ 
and the $N(1710)P_{11}$ are well accounted in this
scheme as the first and second radial states
of the nucleon family,
likewise the $\Delta(1600)P_{33}$ corresponds to
the first radial excitation of the $\Delta$ family.  

\begin{figure}[tb]
\begin{center}
\begin{minipage}[t]{15.5 cm}
\begin{center}
\epsfig{file=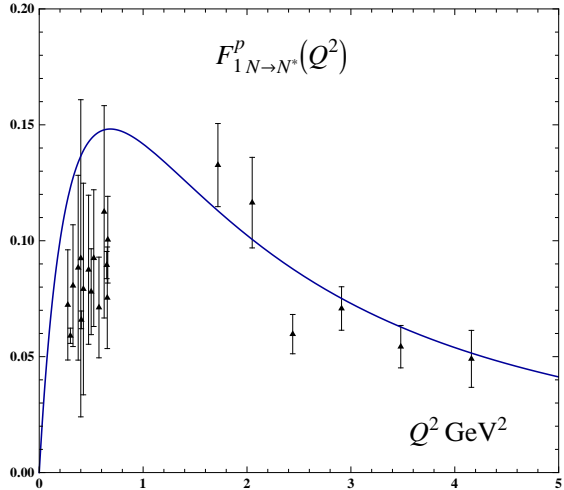,scale=0.8}
\end{center}
\end{minipage}
\begin{minipage}[t]{16.5 cm}
\caption{
Predictions of light-front holographic QCD (Eq. (\ref{eq:hologr}))
for the $F_1(Q^2)$ form factor of 
the  $\gamma^* p \rightarrow~N(1440)P_{11}$ transition
from Ref. \cite{Teramond2}.
Experimental data correspond to CLAS $\pi$
and $2\pi$ electroproduction data \cite{Aznauryan2009,Azn065,Mokeev2009}.
{\it Source:} From Ref. \cite{Teramond2}.
\label{teram}}
\end{minipage}
\end{center}
\end{figure}

The $F_1(Q^2)$ form factor is predicted without
free parameters as:
\begin{equation}
F_1(Q^2)=\frac{\sqrt{2}}{3}
\frac
{\frac{Q^2}{M_{\rho}^2}}
{\left(1+\frac{Q^2}{M_{\rho}^2}\right)
\left(1+\frac{Q^2}{M_{\rho'}^2}\right)
\left(1+\frac{Q^2}{M_{\rho''}^2}\right)},
\label{eq:hologr} 
\end{equation}
where $M_{\rho}$, $M_{\rho'}$, and $M_{\rho''}$
are masses of the $\rho$ meson and its first two radial states.
The results are shown in Fig. \ref{teram} and are in good agreement
with experimental data.

\begin{figure}[tb]
\begin{center}
\begin{minipage}[t]{14.5 cm}
\epsfig{file=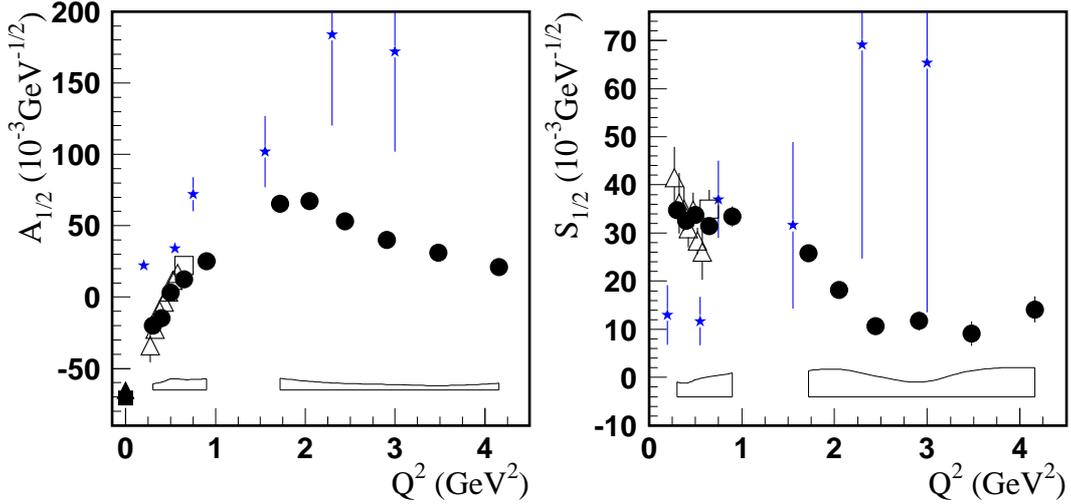,scale=0.8}
\end{minipage}
\begin{minipage}[t]{16.5 cm}
\caption{
Lattice QCD results (stars) for
the  $\gamma^* p \rightarrow~N(1440)P_{11}$
helicity amplitudes from Ref. \cite{Lin}.
Notations for the amplitudes extracted from experimental data
are as in Fig. \ref{p11}.
\label{p11_lat}}
\end{minipage}
\end{center}
\end{figure}

\subsubsection{\it Lattice QCD
\label{sec:roper_lattice}}

The electromagnetic transitions from the nucleon's 
ground state to the excited states of the nucleon have not been studied
in lattice QCD until very recently. The first attempts to evaluate 
the $\gamma^* p \rightarrow N(1440)P_{11}$
transition in lattice QCD were made in Ref. \cite{Lin}.
The calculations used the quenched approximation 
and a large pion mass of $\sim 720$~MeV. The Roper resonance 
is taken as the lowest excited state 
with the same quantum numbers and quark content as the 
nucleon. The results are given in terms
of the form factors $F^*_{1,2}(Q^2)$ that are related
to the form factors $G_{1,2}(Q^2)$ defined in Sec. \ref{sec:formfactors}
(Eqs. (\ref{eq:def53},\ref{eq:def54})) by
\be
F^*_{1}(Q^2)=Q^2 G_1(Q^2),
~~~~F^*_{2}(Q^2)=-\frac{M^2-m^2}{2} G_2(Q^2).
\label{eq:res25} 
\ee
The lattice QCD results \cite{Lin} are shown in Fig. \ref{p11_lat}
in terms of the $\gamma^* p \rightarrow~N(1440)P_{11}$
helicity amplitudes. The signs of the amplitudes and orders 
of magnitudes are correctly predicted. 
There are obvious 
discrepancies at the quantitative level, especially at low $Q^2$. 
These are not unexpected given the large pion mass and the use 
of the quenched approximation. As we discussed in the previous 
section, the pion-cloud contributions are expected to be large at small
$Q^2$. 

\subsection{\it The
$N(1535)S_{11}$ resonance
\label{sec:results_s11}}

\subsubsection{\it JLab data on $ep\rightarrow e\pi N,e\eta p$
and determination
of the branching ratios 
$N(1535)S_{11}\rightarrow \pi N,~\eta N$
\label{sec:s11_branching}}
The $N(1535)S_{11}$ resonance has large couplings
to both the $\pi N$ and $\eta N$ channels 
and has been extensively studied
in $\pi$ and $\eta$ electroproduction
off protons. 
In the early generations of electron beam accelerators
the electroexcitation of the $N(1535)S_{11}$ was studied
up to $Q^2=3~$GeV$^2$ 
(see reviews \cite{PDG82,Foster}). A falloff of the transverse 
$\gamma^* p \rightarrow~N(1535)S_{11}$ amplitude with $Q^2$ 
was observed that was deemed unusually slow in comparison 
to other resonances. However, the experimental information was 
scarce, and the data did not allow for definite results on the 
longitudinal amplitude. We should also mention that modern quark 
models explain the observed $Q^2$ dependence at least qualitatively.

\begin{figure}[tb]
\begin{center}
\begin{minipage}[t]{14.5 cm}
\epsfig{file=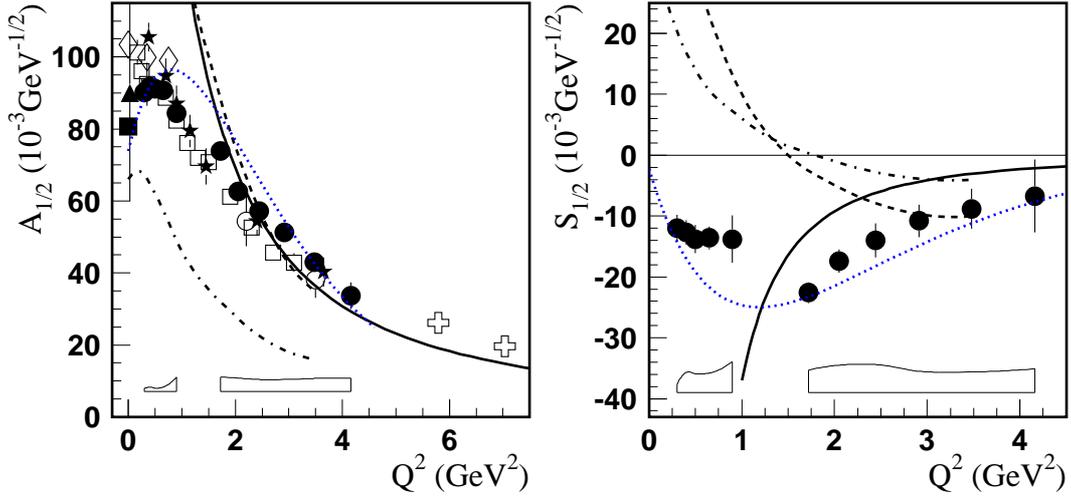,scale=0.8}
\end{minipage}
\begin{minipage}[t]{16.5 cm}
\caption{
Helicity amplitudes
for $\gamma^* p \rightarrow~N(1535)S_{11}$.
The legend for the amplitudes extracted from $\pi$ electroproduction 
data is as for Fig. \ref{p11}.
The amplitudes extracted from $\eta$ electroproduction
data are: 
the stars \cite{Thompson},
the open boxes \cite{Denizli}, 
the open circles \cite{Armstrong},
the crosses \cite{Dalton}, and
the rhombuses \cite{Aznauryan2005,Aznauryaneta}.
All amplitudes, except the RPP values, correspond to
$M=1535~$MeV, $\Gamma_{tot}=150~$MeV, and to
the branching ratios from Eqs. (\ref{eq:res27},\ref{eq:res28}): 
$\beta_{\pi N}=0.485$ and
$\beta_{\eta N}=0.46$. 
The dashed
and dashed-dotted curves
show predictions of
LF relativistic quark models \cite{Capstick}
and \cite{Simula1}, respectively.
The solid curves
show the central values of the amplitudes found within
light-cone sum rules using lattice results for the light-cone
distribution amplitudes of the $N(1535)S_{11}$ resonance \cite{Braun}.
\label{s11}}
\end{minipage}
\end{center}
\end{figure}

Starting in 1999, rich information on the 
$N(1535)S_{11}$ has been obtained
at JLab in a wide range of $Q^2$ up to $4.5$ and $7$~GeV$^2$, 
respectively, in $\pi$ and $\eta$ 
electroproduction off protons (see Fig. \ref{s11}). 
Accurate results were obtained in both reactions 
for the transverse amplitude $A_{1/2}$;
they show a consistent $Q^2$ slope 
and confirm the slow fall off with $Q^2$ observed in 
earlier experiments.

In $\eta$ electroproduction, the contribution of the
$N(1535)S_{11}$ is 
dominant at $W<1.6~$GeV
and is extracted from the data in a nearly model-independent way.
However, in the recent high statistics measurements of this channel, 
no explicit separation of the longitudinal and transverse terms that
contribute to the cross section has been undertaken.  
The analyses assume that the longitudinal contribution to the total 
cross section is small enough to have a negligible effect on the
extraction of the $A_{1/2}$ amplitude.  This assumption 
is motivated by results of the early experiments \cite{Foster}. 

In $\pi$ electroproduction, the longitudinal
amplitude can be revealed due to an interference with other
contributions, resonant and non-resonant, which is absent 
in $\eta$ production. Due to new $n\pi^+$ data, for the first time
definite results were obtained for this amplitude
\cite{Aznauryan2009,MAID2007,MAID_China}.
Estimations based on these results confirm 
that the inclusion of the longitudinal component
would have a slight effect on the  
amplitude $A_{1/2}$ extracted from $\eta$ electroproduction data
\cite{Thompson,Denizli,Armstrong}.
For example, the contribution of the longitudinal amplitude
found in $\pi$ electroproduction
to the total cross sections shown in Fig. \ref{thompson2}
at the $N(1535)S_{11}$ resonance position 
is within shown statistical and systematic errors.
This means that the amplitude $A_{1/2}$
estimated at the resonance position would 
decrease in magnitude  
within half of its error. 
However, for
accurate conclusions the analyses of
complete sets of data over all angles and energies
are required.

Numerical comparison of the  results extracted from
the  $\pi$ and $\eta$ photo- and electroproduction data
depends on the relation between the branching ratios
to the $\pi N$ and $\eta N$ channels.
Consequently, it contains an arbitrariness connected
with the uncertainties of these branching ratios:
$\beta_{\pi N}=0.35\div 0.55$, $\beta_{\eta N}=0.45\div 0.6$ \cite{PDG}.

The accurate results on the transverse amplitudes $A_{1/2}$ found from
the new $\pi$ and $\eta$ data were used in Ref. \cite{Aznauryan2009} to specify
the relation between $\beta_{\pi N}$ and $\beta_{\eta N}$. 
From the fit to these amplitudes at $0\leq Q^2 \leq 4.5$~GeV$^2$,
it was found
\begin{equation}
\frac{\beta_{\eta N}}{ \beta_{\pi N}}=0.95\pm 0.03.
\label{eq:res26} 
\end{equation}
Furthermore, taking into account the branching
ratio to the $\pi\pi N$ channel, 
$\beta_{\pi\pi N}=0.01\div 0.1$ \cite{PDG},
which accounts
practically for all channels different
from $\pi N$ and $\eta N$,
more accurate values of the branching
ratios $\beta_{\pi N}$ and $\beta_{\eta N}$
were obtained:
\begin{eqnarray}
&&\beta_{\pi N}=0.485\pm 0.008\pm 0.023, 
\label{eq:res27} 
\\
&&\beta_{\eta N}=0.460\pm 0.008\pm 0.022.
\label{eq:res28} 
\end{eqnarray}
The first uncertainty corresponds to the fit uncertainty in Eq. (\ref{eq:res26}),
the second one accounts for the uncertainty of
$\beta_{\pi\pi N}$. 
The results shown in Fig. \ref{s11} correspond to
$\beta_{\pi N}=0.485,~\beta_{\eta N}=0.46$.

\subsubsection{\it Model predictions and the sign of the $S_{1/2}$
amplitude for $\gamma^* p\rightarrow N(1535)S_{11}$
\label{sec:s11_sign}}

Constituent quark models give the correct sign for the transverse
 amplitude $A_{1/2}$ taking into account the sign
of the $N(1535)S_{11}\rightarrow \pi N$ vertex
found in the LF relativistic quark model \cite{Aznauryan_1985}.
However, there is a large diversity among
predictions of different models, which is clearly seen
in the results obtained from the LF relativistic quark models
of Refs. \cite{Capstick} and \cite{Simula1} (see Fig. \ref{s11}).
This diversity prevents us from drawing quantitative
conclusions on the quark core and meson-cloud contributions
by utilization of the empirical values of $A_{1/2}$.

The CQM predictions for the longitudinal amplitude $S_{1/2}$
are positive at $Q^2<1$~GeV$^2$, both in the relativistic
and non-relativistic approaches. This is in contradiction
to the amplitudes extracted from experiment.
The only possibility to reach agreement with the empirical
values would be large and negative contributions produced
by the meson cloud or by additional $qqqq\bar{q}$ components
to the $N(1535)S_{11}$ wave function.
Such information is currently not available, and definite conclusions
will have to wait until results
of analyses within dynamical coupled-channel approaches become
available. 
Concerning a possible 5-quark component
it was found  in Ref. \cite{An} that the most likely lowest energy configuration
in the $N(1535)S_{11}$ is given by
a $qqqs\bar{s}$ component. This can solve, in principle, the problem
of mass ordering between the $N(1440)P_{11}$ and $N(1535)S_{11}$,
and may also explain the large branching ratio of the $N(1535)S_{11}$
to the $\eta N$ channel and
recently observed large couplings
to the $\phi N$ and $K\Lambda$ channels \cite{Xie,Liu}.
However, from the results of Ref. \cite{An}, one can conclude
that the 5-quark contribution
to both amplitudes $A_{1/2}$ and $S_{1/2}$ is small, and would
also not resolve the disagreement between experiment and models
regarding the $S_{1/2}$ amplitude.

It is remarkable that the signs of both amplitudes
$A_{1/2}$ and $S_{1/2}$ are described within the approach
of Ref. \cite{Braun}. This light-cone sum rule approach may be
considered a tool to derive the $\gamma^* p\rightarrow N(1535)S_{11}$
amplitudes from the first principles of QCD. The
$N(1535)S_{11}$ light-cone distribution amplitudes
are found through lattice calculations and are used to compute
the $\gamma^* p\rightarrow N(1535)S_{11}$ transition amplitudes
by utilization of light-cone sum rules. At $Q^2>2$~GeV$^2$,
where the approach may be applicable,
there is good quantitative agreement between
the predictions and the amplitudes extracted from experiment.

\subsection{\it The
$N(1520)D_{13}$ resonance
\label{sec:results_d13}}

\begin{figure}[tb]
\begin{center}
\begin{minipage}[t]{16.2 cm}
\epsfig{file=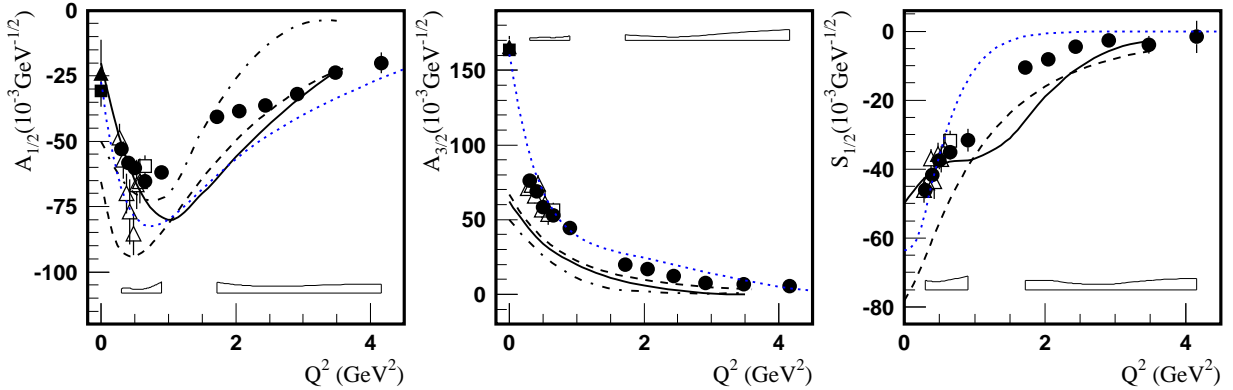,scale=0.95}
\end{minipage}
\begin{minipage}[t]{16.5 cm}
\caption{
Helicity amplitudes
for the  $\gamma^* p \rightarrow N(1520)D_{13}$
transition.
The legend for the amplitudes extracted from $\pi$ and $2\pi$ electroproduction 
data is as for Fig. \ref{p11}.
All amplitudes, except the RPP values, correspond to
$M=1520~$MeV, $\Gamma_{tot}=112~$MeV, 
$\beta_{\pi N}=0.6$, and
$\beta_{2\pi N}=0.4$. 
The solid, dashed and dashed-dotted curves are, respectively,
the predictions of
the quark models \cite{Warns},
\cite{Santopinto},
and \cite{Merten}.
\label{d13}}
\end{minipage}
\end{center}
\end{figure}

Results
for the  $\gamma^* p \rightarrow~N(1520)D_{13}$
helicity amplitudes
are presented in Fig. \ref{d13}.
The amplitudes are extracted from CLAS data on $\pi$ and 2$\pi$ electroproduction,
which are the main channels for the investigation of the $N(1520)D_{13}$.
The precise new data enabled the determination of the 
$\gamma^* p \rightarrow N(1520)D_{13}$ transition in a wider $Q^2$ range, and
with much higher accuracy for the transverse amplitudes 
compared to earlier experiments.
The sensitivity of the earlier data
to the $S_{1/2}$ amplitude was very limited.
The CLAS data allowed this amplitude to be determined with good precision.

\begin{figure}[tb]
\begin{center}
\begin{minipage}[l]{200pt}
\epsfig{file=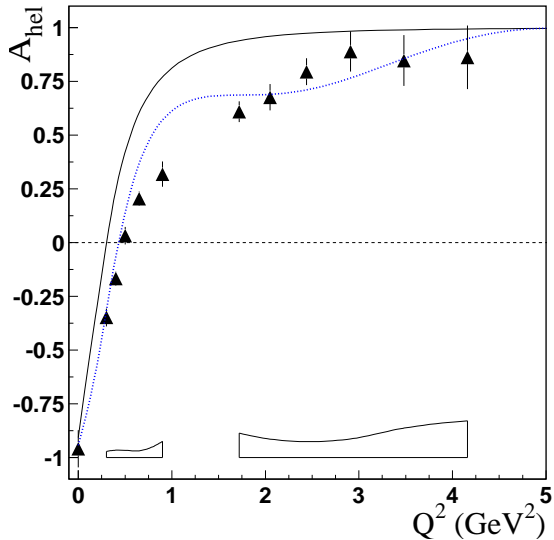,scale=0.4}
\end{minipage}
\hskip 70pt
\begin{minipage}[r]{200pt}
\caption{
The helicity asymmetry
$A_{hel}\equiv (A^2_{1/2}-A^2_{3/2})/(A^2_{1/2}+A^2_{3/2})$
for the  $\gamma^* p \rightarrow N(1520)D_{13}$
transition.
Triangles show the results
of the JLab analysis of CLAS 
$\pi$ electroproduction data \cite{Aznauryan2009}.
The bands indicate the model uncertainties corresponding to these results.
The result of the global MAID2007 fit \cite{MAID2007}
is shown by the dotted curve. The solid curve is the prediction of
the quark model with the harmonic oscillator potential \cite{Koniuk_Isgur}.
\label{d13_asym}}
\end{minipage}
\end{center}
\end{figure}

The signs of all three helicity amplitudes are described by quark models 
taking into account
the signs of both vertices $\gamma^* p \rightarrow~N(1520)D_{13}$
and $N(1520)D_{13}\rightarrow \pi N$,
with the last one taken from the 
LF relativistic quark model \cite{Aznauryan_1985}. 
The shapes of the amplitudes are also reproduced.
However, there is a significant shortfall in the quark models 
with regard to the $A_{3/2}$ amplitude at $Q^2<2$~GeV$^2$, 
which again may hint
at large meson-cloud contributions.
A coupled-channel analysis of pion photoproduction data
indeed shows large meson-cloud contributions to $A_{3/2}$  
\cite{Sato2008}, which could explain this discrepancy.
The data show a clear dominance of the $A_{3/2}$ amplitude at 
the photon point. 
With increasing $Q^2$, 
this amplitude drops rapidly, 
and its magnitude is smaller than the magnitude of 
$A_{1/2}$ at $Q^2 > 0.6$~GeV$^2$. In fact, $A_{1/2}$ dominates 
the resonance strength at high $Q^2$.   
This is demonstrated in Fig. \ref{d13_asym} in terms of the 
helicity asymmetry. The ``helicity switch" was predicted in 
the nonrelativistic quark model with a harmonic oscillator 
potential \cite{Close1}. It is interesting that in spite of possible 
large meson-cloud contributions, the empirical
amplitudes reveal a behavior that is qualitatively consistent with 
the prediction of the naive constituent quark model. 

\subsection{\it Helicity amplitudes and the pQCD asymptotic behavior
\label{sec:pqcd}}

\begin{figure}[tb]
\begin{center}
\begin{minipage}[l]{200pt}
\epsfig{file=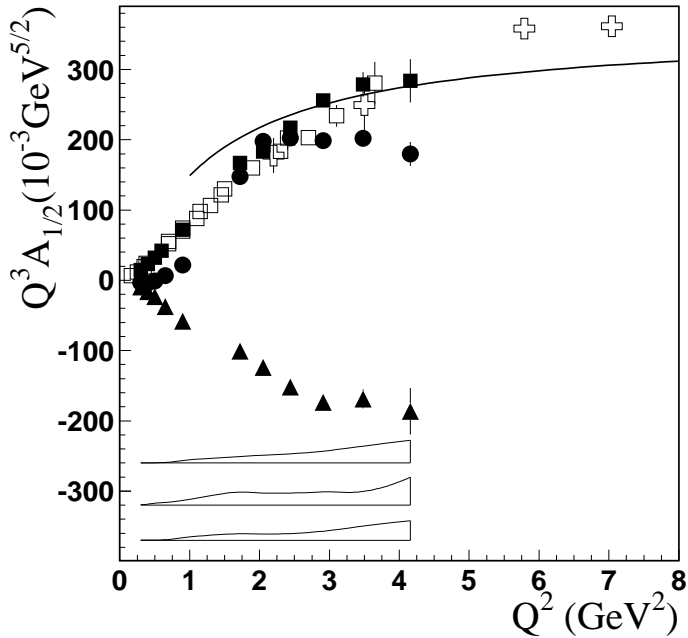,scale=0.5}
\end{minipage}
\hskip 70pt
\begin{minipage}[r]{200pt}
\caption{
The helicity amplitudes $A_{1/2}$
for the  $\gamma^* p \rightarrow 
N(1440)P_{11}$,
$N(1520)D_{13}$, 
and $N(1535)S_{11}$
transitions, multiplied by $Q^3$.
The results obtained from the CLAS data on
pion electroproduction off protons
by the JLab group \cite{Aznauryan2009}
are shown by the full circles ($N(1440)P_{11}$),
the full triangles ($N(1520)D_{13}$),
and the full boxes ($N(1535)S_{11}$).
The upper, middle, and lower bands correspond
to systematic uncertainties of these results for the $N(1535)S_{11}$,
$N(1440)P_{11}$, and $N(1520)D_{13}$, respectively.
The open boxes and crosses are the results for the $N(1535)S_{11}$ 
obtained at JLab
in $\eta$ electroproduction, respectively,
in Hall B \cite{Thompson,Denizli} and Hall C
\cite{Armstrong,Dalton}.
The solid curve corresponds
to the amplitude  $A_{1/2}$
for the  $\gamma^* p \rightarrow~
N(1535)S_{11}$ transition
found within
light-cone sum rules \cite{Braun}.
\label{asymptotics}}
\end{minipage}
\end{center}
\end{figure}

\begin{figure}[tb]
\begin{center}
\begin{minipage}[l]{200pt}
\epsfig{file=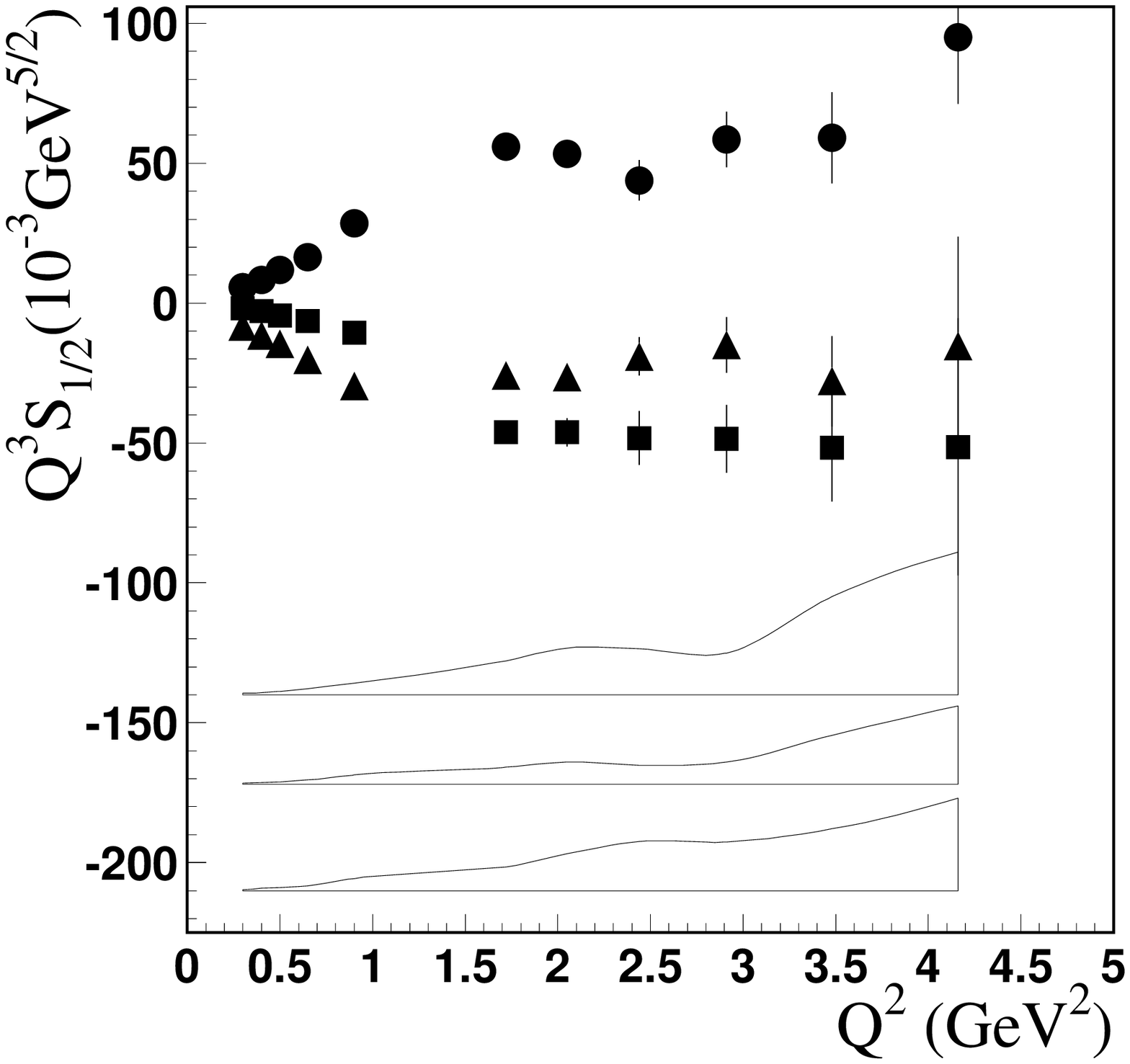,scale=0.5}
\end{minipage}
\hskip 70pt
\begin{minipage}[r]{200pt}
\caption{
The helicity amplitudes $S_{1/2}$
for the  $\gamma^* p \rightarrow 
N(1440)P_{11}$,
$N(1520)D_{13}$, 
and $N(1535)S_{11}$
transitions, multiplied by $Q^3$.
The results are 
obtained from the CLAS data on
pion electroproduction off protons
by the JLab group \cite{Aznauryan2009}:
the solid circles - $N(1440)P_{11}$,
the solid trangles - $N(1520)D_{13}$,
and the solid boxes - $N(1535)S_{11}$.
The upper, middle, and lower bands correspond
to systematic uncertainties of these results for the 
$N(1440)P_{11}$, $N(1520)D_{13}$, and $N(1535)S_{11}$, respectively.
\label{asymp_s}}
\end{minipage}
\end{center}
\end{figure}

Using the helicity amplitudes for the three excited  states  
$N(1440)P_{11}$,
$N(1535)S_{11}$, and $N(1520)D_{13}$,
one can check the $1/Q^3$ scaling prediction of pQCD 
for the $A_{1/2}$ and $S_{1/2}$ amplitudes.
As we discussed in Sec. \ref{sec:delta_pqcd}, such scaling
is expected in the asymptotic limit of pQCD (see Eqs. (\ref{eq:ress23})).
Figs. \ref{asymptotics} and \ref{asymp_s}
show the helicity amplitudes $A_{1/2}$ and $S_{1/2}$ multiplied by $Q^3$.
We note that starting with $3~$GeV$^2$ for
$A_{1/2}$ and $1.5-2~$GeV$^2$ for
$S_{1/2}$, the empirical amplitudes show a $Q^2$ trend close to 
the expected $1/Q^3$ dependence, although measurements at 
higher $Q^2$ are needed for more definite conclusions.

\subsection{\it Empirical transverse charge densities in the 
$\gamma^* p \rightarrow \Delta(1232)P_{33}$, $N(1440)P_{11}$,
$N(1535)S_{11}$, and $N(1520)D_{13}$ 
transitions
\label{sec:densities}}

The precise results on the amplitudes of the resonance transitions
$\gamma^* p \rightarrow~\Delta(1232)P_{33}$, $N(1440)P_{11}$,
$N(1535)S_{11}$, and $N(1520)D_{13}$ provide the basis needed 
to study the spatial characteristics of excited nucleon states. 
A proper density interpretation of the empirical amplitudes 
is obtained in Refs. 
\cite{Carlson_Van,Tiator_Van,Tiator_Van1}
in a frame where the baryons have a large
momentum-component along the $z$-axis chosen along
the direction of $\bf{P}$  
($P=\frac{p^* +p}{2}$) and
the virtual photon four-momentum has $k^+=0$ ($k^+\equiv k^0+k^z$). 
In the $xy$-plane, the virtual photon momentum has
a transverse component $\bf{k}_{\perp}$:
$k^2=-|\bf{k}_{\perp}|^2=-Q^2$.
The transition charge densities are defined by the 
Fourier transfrom 
\be
\rho_{0(T)}^{NN^*}({\bf{b}})=~
\int \frac{d^2{\bf{k}}_{\perp}}{(2\pi)^2}
e^{-i{\bf{k}}_{\perp}{\bf{b}}}\frac{1}{2P^+}
<p^*,\lambda({\bf{S}}^*_{\perp})
|J_{em}^{+}|p,\lambda({\bf{S}}_{\perp})>,
\label{eq:tcd1}
\ee
where the electromagnetic current $J_{em}$ is related
to the $\gamma^* N \rightarrow~N^*$ transition form factors
by Eqs. (\ref{eq:def53},\ref{eq:def57}), the 2-dimensional
vector $\bf{b}$ denotes the position in the $xy$-plane;
$\rho_{0}$ is the transition charge
density for unpolarized 
$N$ and $N^*$, 
$\lambda$ denotes their helicities;
$\rho_{T}$ is transition charge density for
transversely polarized 
$N$ and $N^*$ along the directions $\bf{S}_{\perp}$
and $\bf{S}^*_{\perp}$, respectively. 

\begin{figure}[ht!]
\begin{center}
\begin{minipage}[t]{16.0 cm}
\epsfig{file=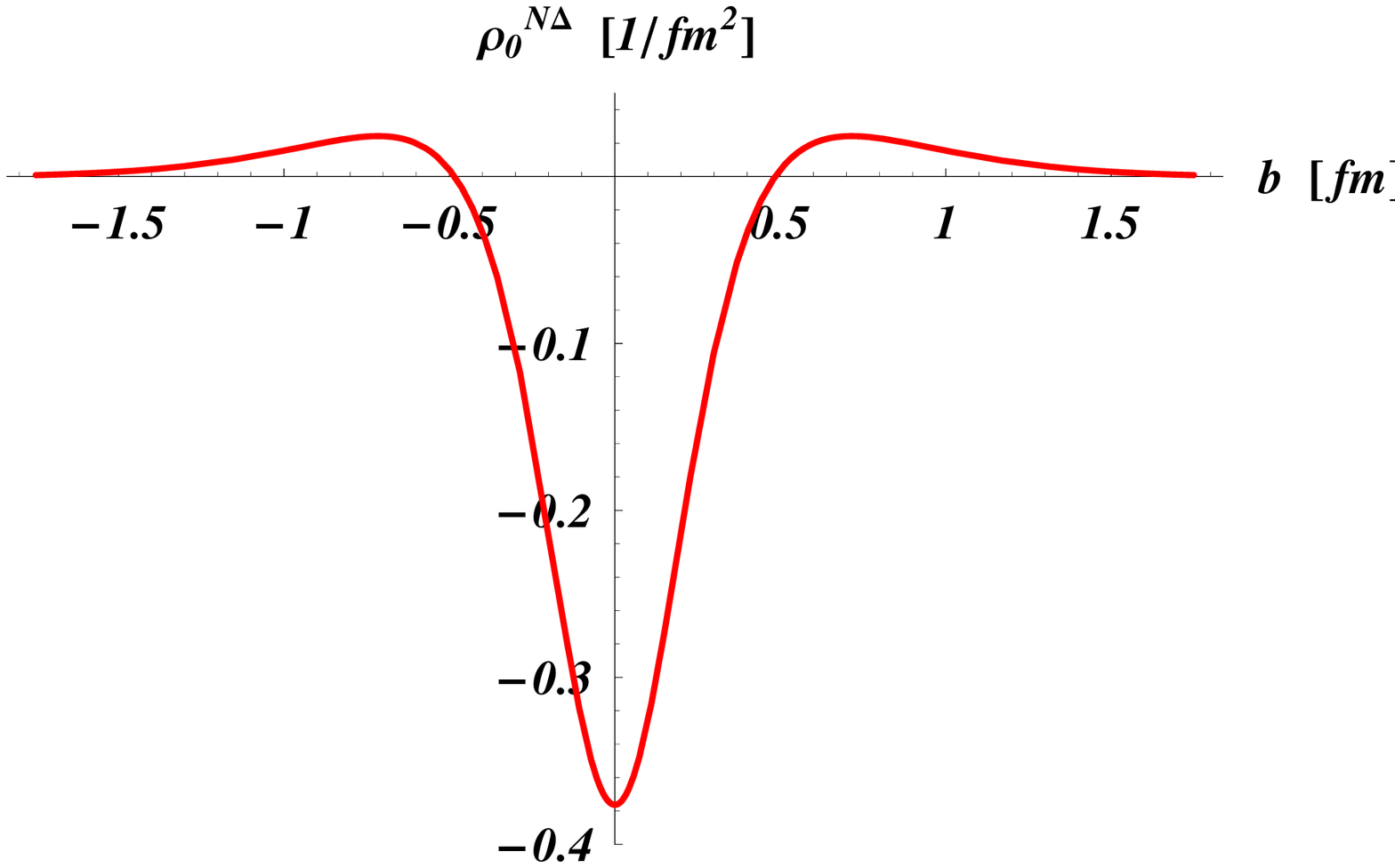,scale=0.29}
\epsfig{file=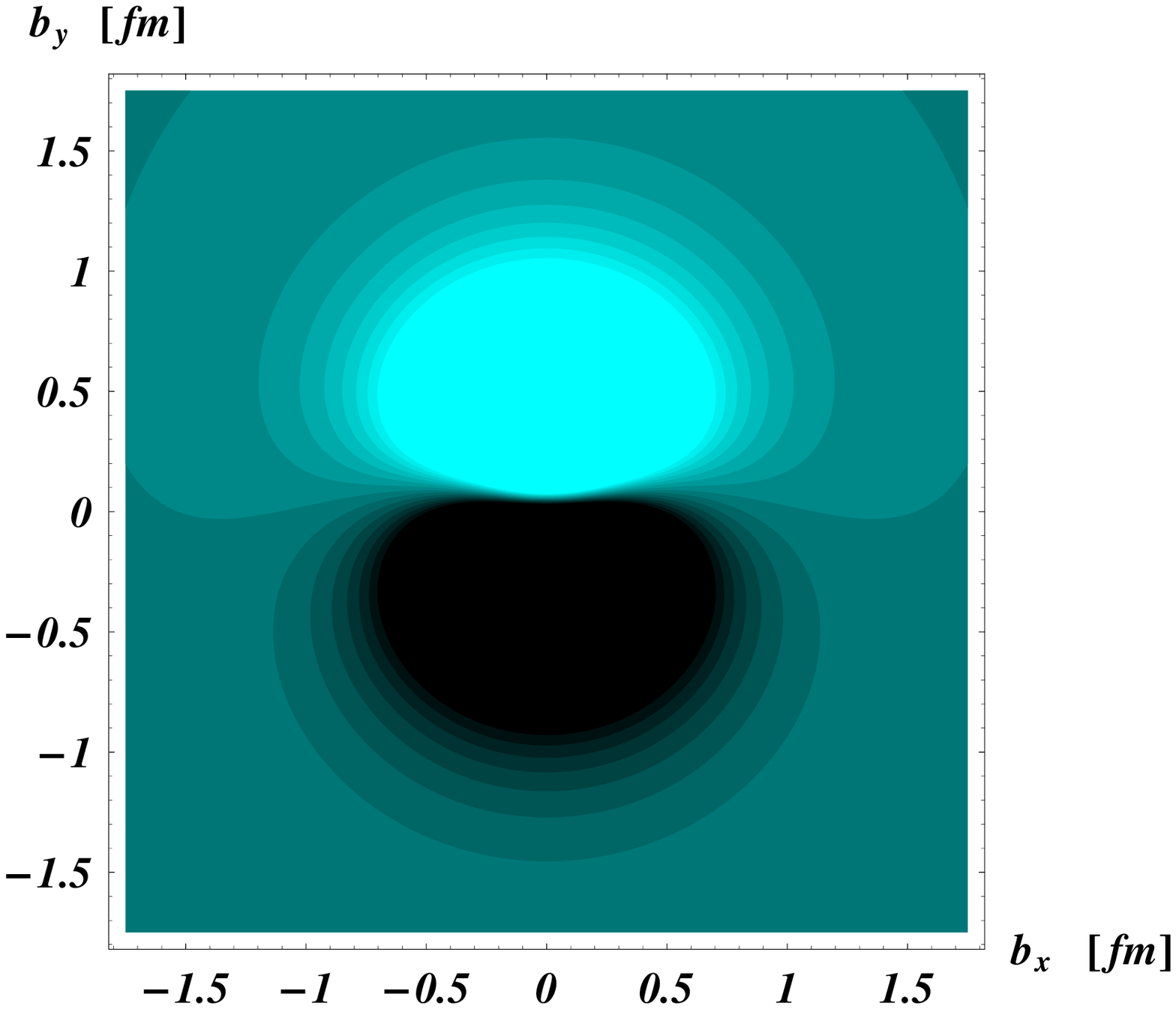,scale=0.29}
\epsfig{file=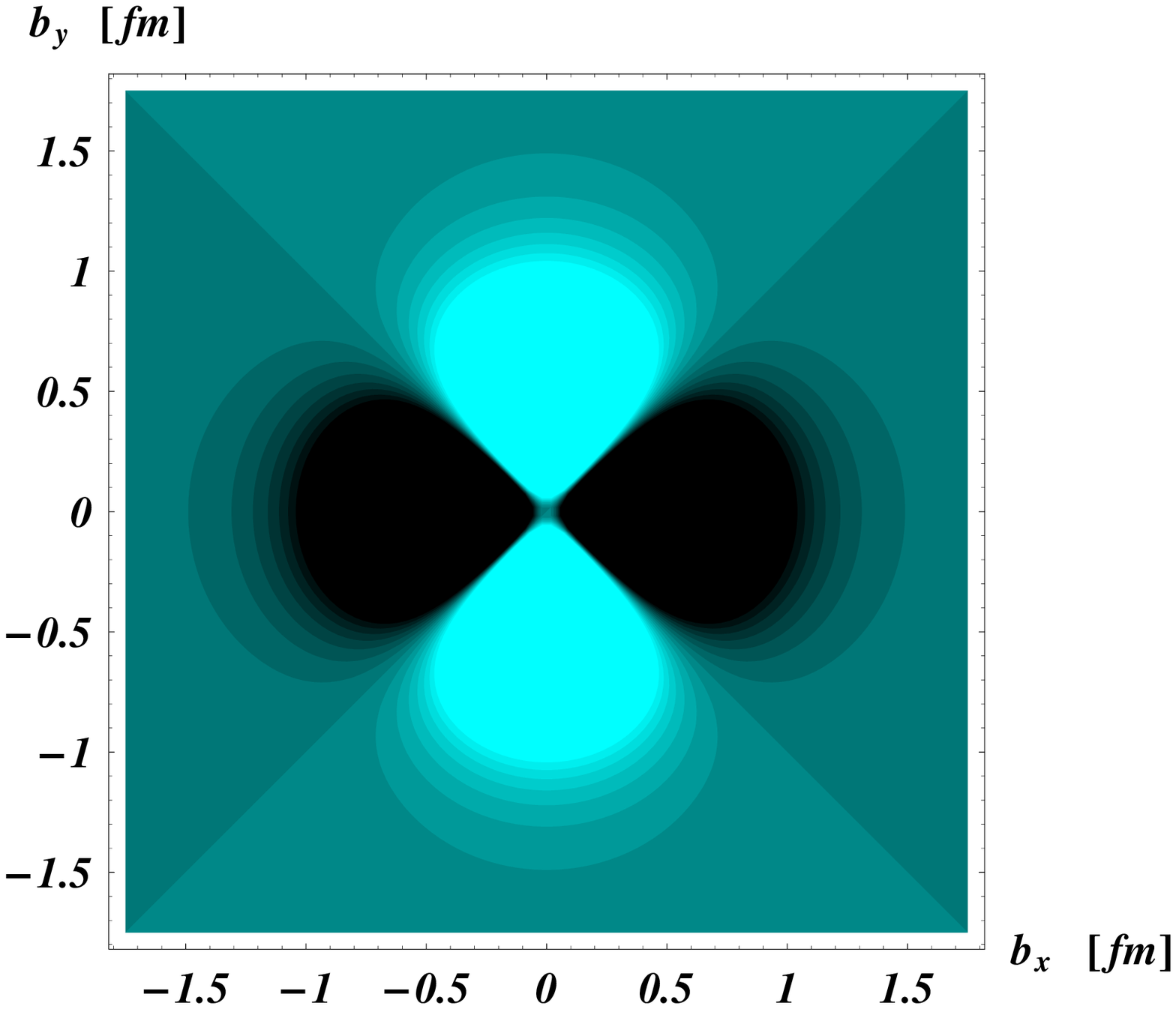,scale=0.29}
\end{minipage}
\begin{minipage}[t]{16.5 cm}
\caption{
Quark transverse charge density for the
$\gamma^* p \rightarrow \Delta(1232)P_{33}$
transition \cite{Carlson_Van}. 
Left panel corresponds to the unpolarized $N$ and $\Delta$.
Middle panel shows the density for the $N$ and $\Delta$
polarized along the $x$-axis. Right panel presents
the quadrupole contribution in this case.
The light (dark) areas are dominated by 
up (down) quarks and correspond to dominantly
positive (negative) charges. ({\it Source:} From Ref. \cite{Carlson_Van}.) 
\label{p33_trans}}
\end{minipage}
\end{center}
\end{figure}

\begin{figure}[ht!]
\begin{center}
\begin{minipage}[t]{14.0 cm}
\epsfig{file=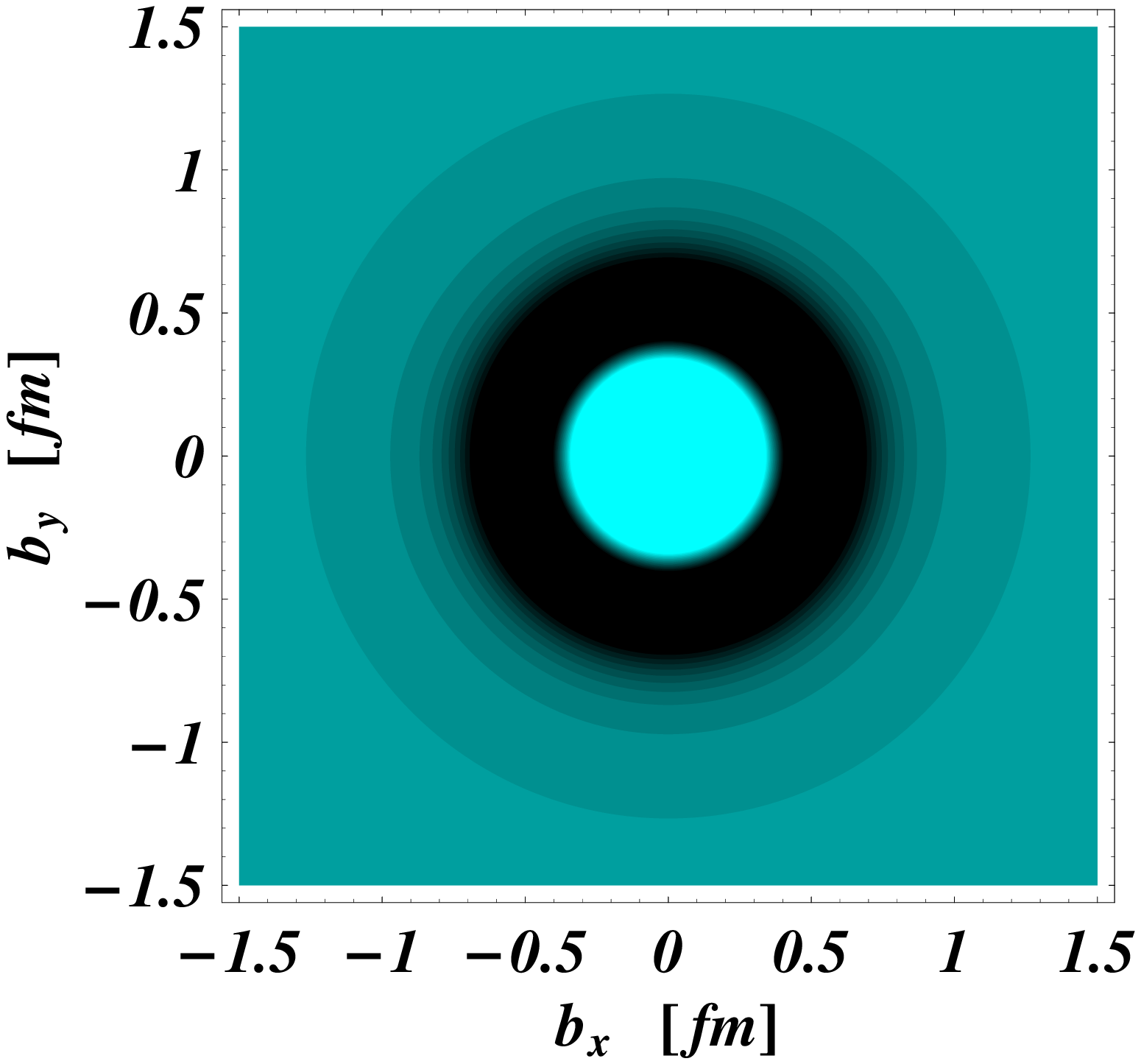,scale=0.4}
\epsfig{file=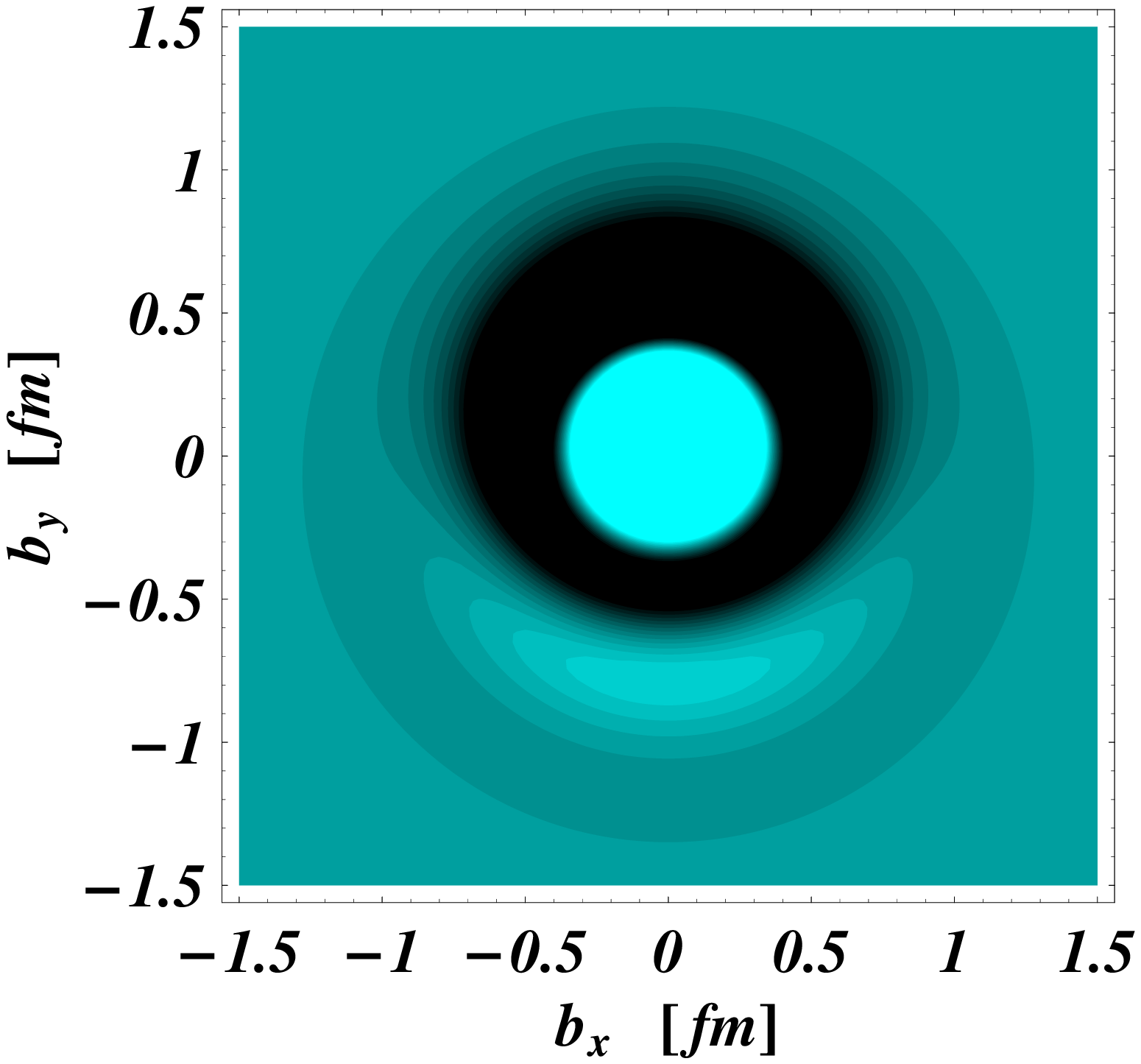,scale=0.4}
\end{minipage}
\begin{minipage}[t]{16.5 cm}
\caption{
Quark transverse charge density for the
$\gamma^* p \rightarrow~N(1440)P_{11}$
transition \cite{Tiator_Van}. The left panel corresponds to the 
unpolarized $p$ and $N(1440)P_{11}$.
In the right panel $p$ and $N(1440)P_{11}$ are
polarized along the $x$-axis.
The light (dark) areas are dominated by positive (negative) charges.
({\it Source:} From Ref. \cite{Tiator_Van}.)
\label{p11_trans}}
\end{minipage}
\end{center}
\end{figure}

\begin{figure}[ht!]
\begin{center}
\begin{minipage}[t]{14.0 cm}
\epsfig{file=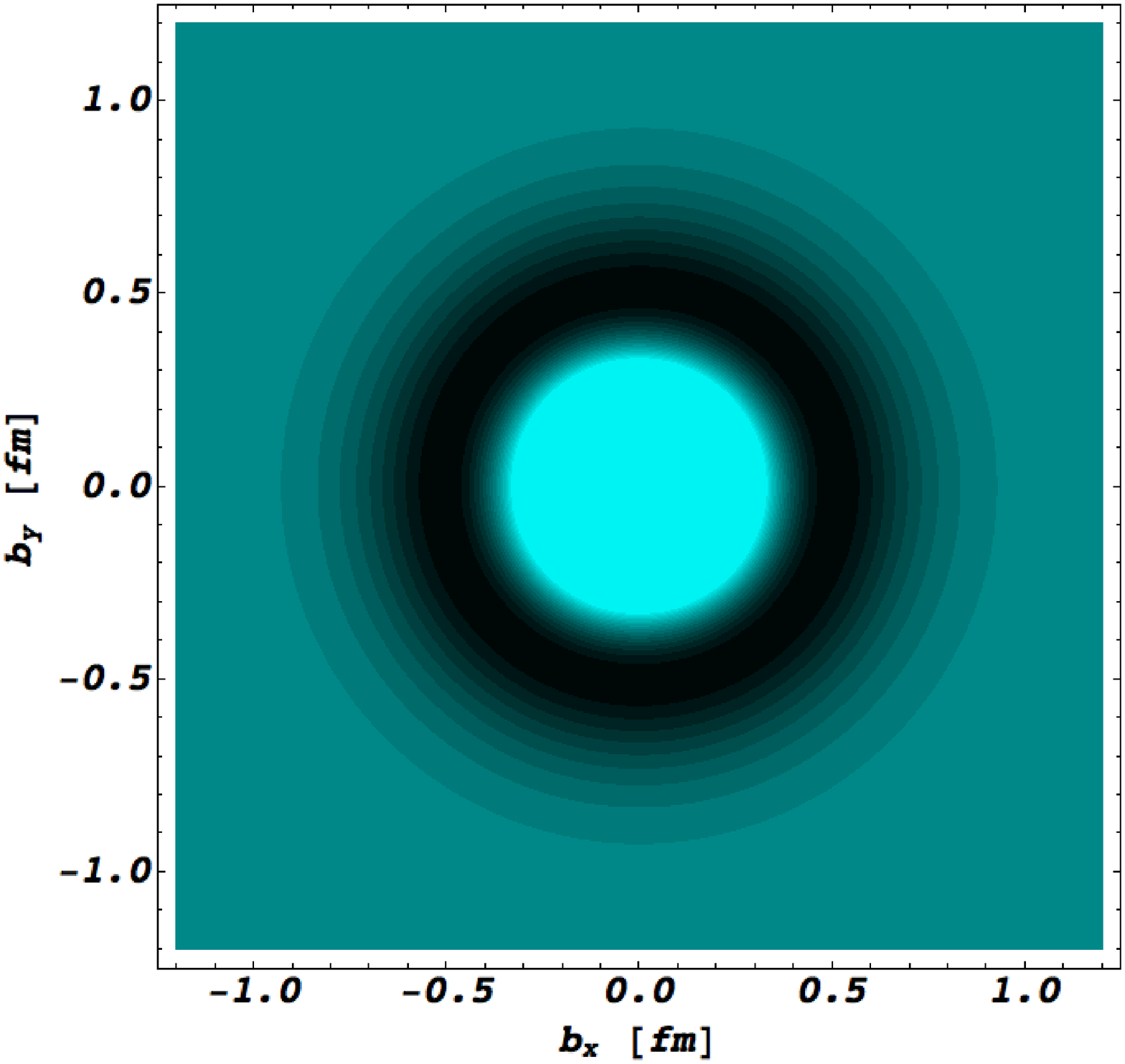,scale=0.25}
\epsfig{file=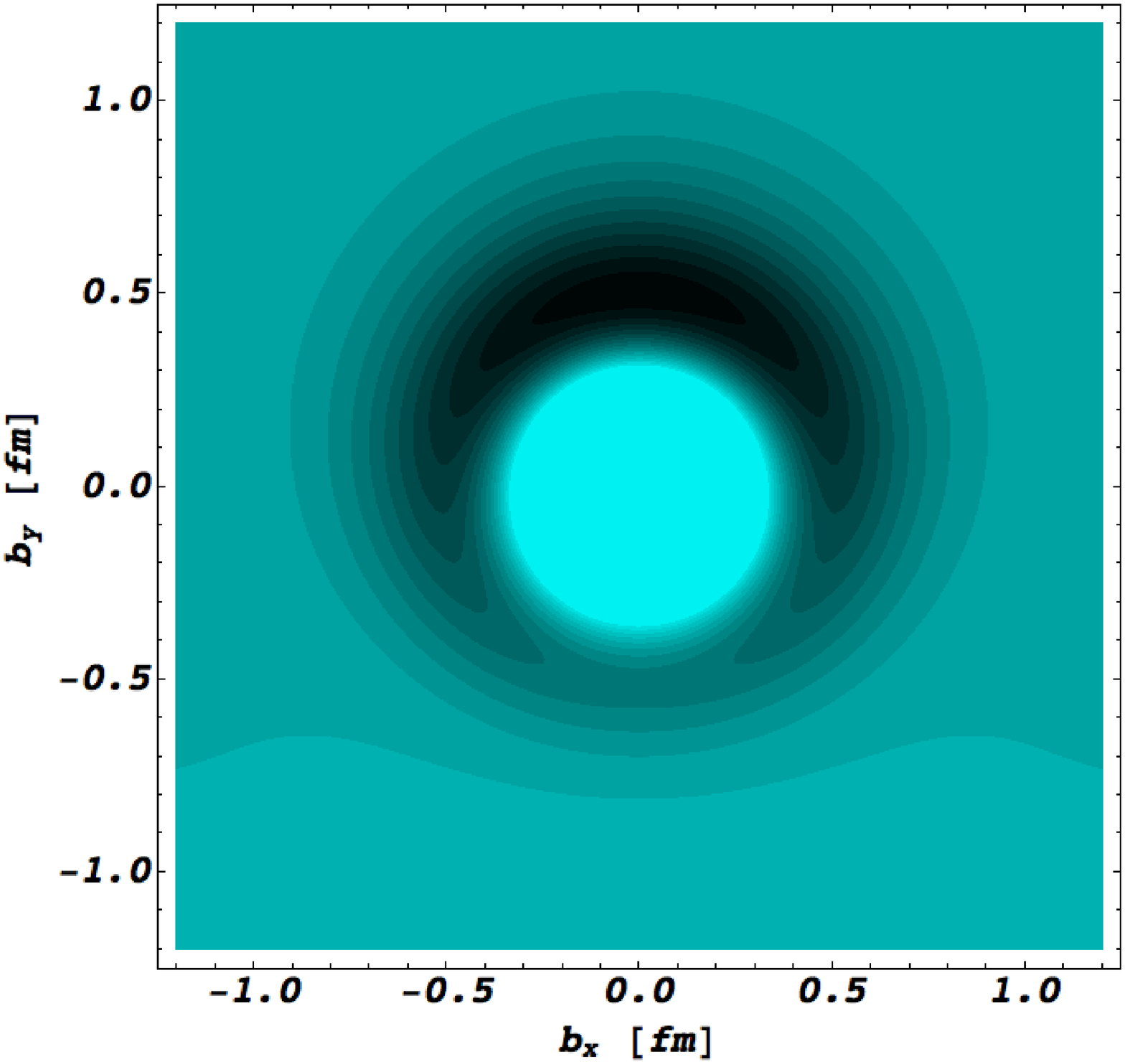,scale=0.25}
\end{minipage}
\begin{minipage}[t]{16.5 cm}
\caption{
Quark transverse charge density for the
$\gamma^* p \rightarrow N(1535)S_{11}$
transition \cite{Tiator_Van1}. 
The legend is as for Fig. \ref{p11_trans}.
({\it Source:} From Ref. \cite{Tiator_Van1}.)
\label{s11_trans}}
\end{minipage}
\end{center}
\end{figure}

\begin{figure}[ht!]
\begin{center}
\begin{minipage}[t]{14.0 cm}
\epsfig{file=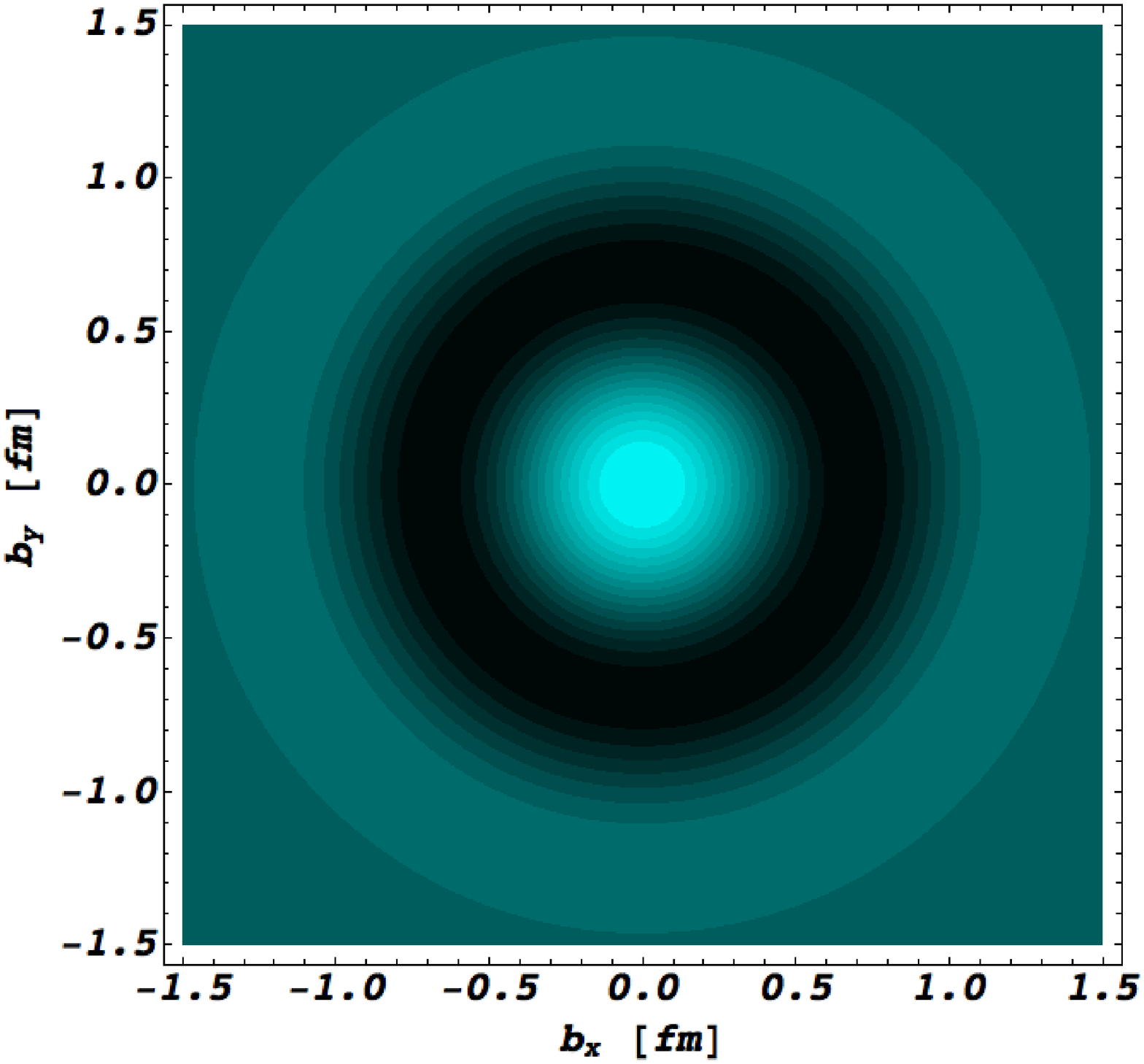,scale=0.25}
\epsfig{file=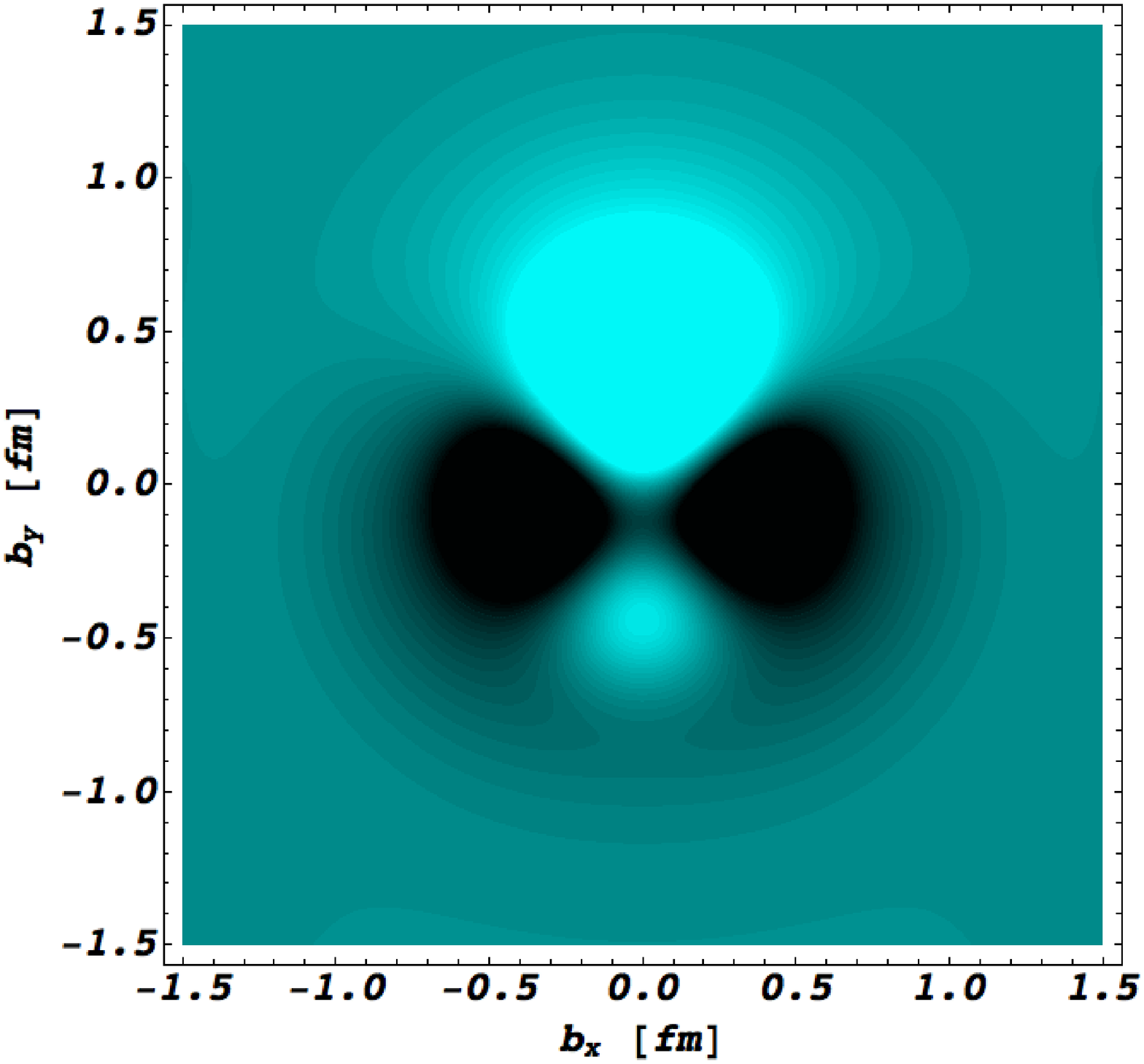,scale=0.25}
\end{minipage}
\begin{minipage}[t]{16.5 cm}
\caption{
Quark transverse charge density for the
$\gamma^* p \rightarrow N(1520)D_{13}$
transition \cite{Tiator_Van1}. 
The legend is as for Fig. \ref{p11_trans}.
({\it Source:} From Ref. \cite{Tiator_Van1}.)
\label{d13_trans}}
\end{minipage}
\end{center}
\end{figure}

In Figs. \ref{p33_trans}- \ref{d13_trans}, we present the results
obtained in Refs. \cite{Carlson_Van,Tiator_Van,Tiator_Van1}. They are
based on the parameterizations of the transition amplitudes
found in the global fit of the new electroproduction data by the 
Mainz group (MAID2007). 

For the $\Delta(1232)P_{33}$,  the analysis of 
Ref. \cite{Carlson_Van} shows (see Fig. \ref{p33_trans}) that
in the unpolarized case, the charge density has
a negative interior core and becomes 
positive for $b\geq 0.5~$fm. For the polarized baryons,
the density shows both dipole and quadrupole field patterns. 
The latter, shown separately in Fig. \ref{p33_trans},
provides a way of quantifying the deformation in the charge
distribution for this transition.

For the Roper resonance,  the analysis  
shows (Fig. \ref{p11_trans}) that
in the unpolarized case, 
there is an inner region of positive quark charge 
concentrated within a
0.5~fm radius accompanied by a relatively broad band of 
negative charge extending out to about 1~fm.
When both the ground state and the excited baryons are 
polarized in the transverse plane, the large value of the
magnetic transition strength at the real photon point 
yields a sizable shift
of the charge distribution, inducing an electric dipole moment. 

In the case of the $N(1535)S_{11}$ (Fig.
\ref{s11_trans}) and
the $N(1520)D_{13}$ (Fig. 
\ref{d13_trans}),
the unpolarized density is similar to the density 
for the Roper resonance with, 
however, more diffuse
boundaries between up and down quarks. 
For polarized baryons, 
the ring of down quarks 
for the $N(1535)S_{11}$ 
is less pronounced;
for the $N(1520)D_{13}$,
in addition to the dipole transition
density, there is also a quadrupole 
density.

We want to say a word of warning here. In order for these projections
to correctly reflect the charge distributions at small transverse 
distances, the 
transition form factors have to be known in a large $Q^2$ range. With
the currently available electroproduction data, large uncertainties 
are present in the Fourier integral, due
to model assumptions about the extrapolation of the form factors from 
finite $Q^2$ to $Q^2\rightarrow \infty$. Measurements at higher $Q^2$ are 
necessary to reduce those uncertainties. 

\section{Nucleon Resonances in the Mass Region Around 1.7 GeV
\label{sec:third}}
The available information on the electroexcitation 
amplitudes for the resonances in the so-called third resonance
region may be divided into three groups:
(i) The information that has been obtained in the global
MAID2007 fit \cite{MAID2007,MAID_China}. As mentioned
in section \ref{sec:results}, the data at $W>1.7~$GeV used
in this fit come mostly from older experiments \cite{SAID}.
These data are scarce, do not contain
polarization measurements, and have usually large statistical 
uncertainties. We therefore expect that the results from the analyses of 
these data may be significantly changed once new and precise
data at $W>1.7~$GeV become available. 
(ii) There is information obtained in the analyses 
\cite{Ripani,Mo091,Mokeev_2010}
of new $2\pi$ electroproduction data,
which extend up to $W=2.1~$GeV (see Table \ref{tab:2pi}).
(iii) There are also results at $Q^2=0.65~$GeV$^2$
found in the combined analysis \cite{Azn065} of
new $\pi$ and $2\pi$ electroproduction data.
The single pion electroproduction data were complemented 
in this analysis by the information
from older measurements.  

\begin{figure}[tb]
\begin{center}
\begin{minipage}[t]{14.0 cm}
\epsfig{file=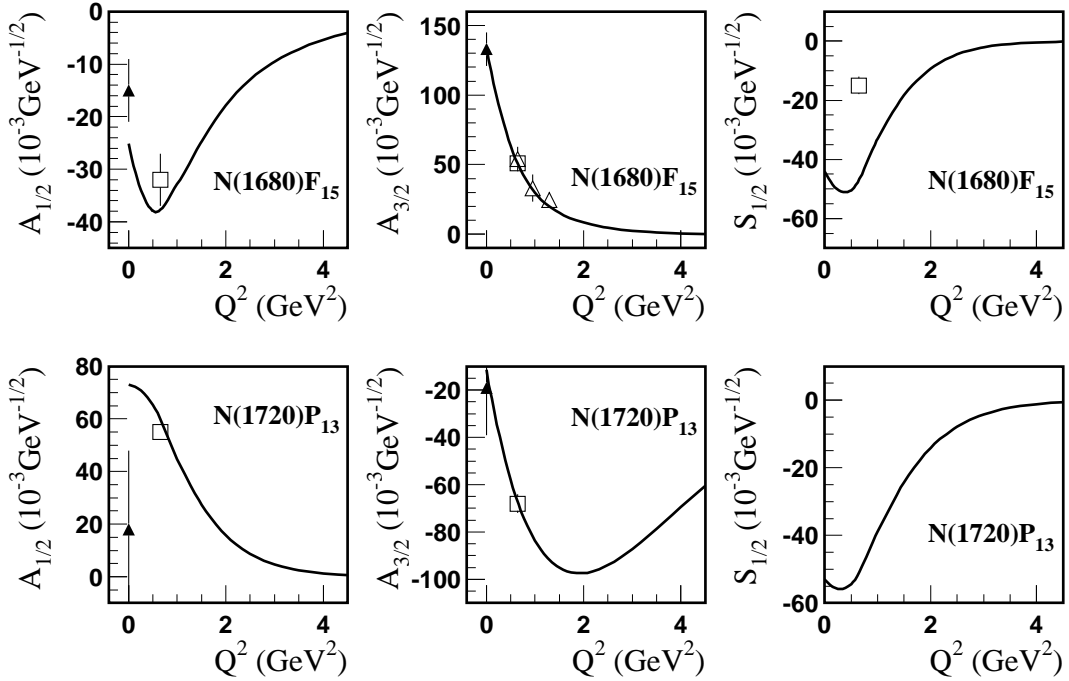,scale=0.8}
\end{minipage}
\begin{minipage}[t]{16.5 cm}
\caption{
The
$\gamma^* p \rightarrow N(1680)F_{15}$ and $N(1720)P_{13}$
helicity amplitudes.
The solid curves are from the 
global MAID2007 fit 
\cite{MAID2007,MAID_China}.
The data at $Q^2=0$ (solid triangles) are from RPP \cite{PDG},
the data at $Q^2=0.65~$GeV$^2$ (open boxes) are from the combined analysis
of the CLAS $\pi$ and $2\pi$ electroproduction data 
\cite{Azn065}, and the open triangles are the results extracted
from the CLAS $2\pi$ electroproduction data
\cite{Mo091,Mokeev_2010}. 
\label{fp}}
\end{minipage}
\end{center}
\end{figure}

In Fig. \ref{fp}, we show the 
electroexcitation amplitudes for the resonances
$N(1680)F_{15}$ and $N(1720)P_{13}$. 
An interesting observation is that $A_{1/2}$ and $A_{3/2}$ amplitudes
for the two states behave quite differently. The 
$\gamma^* p \rightarrow N(1680)F_{15}$
transition behaves similar to the $N(1520)D_{13}$. The $A_{3/2}$ amplitude drops 
rapidly with increasing $Q^2$ and is replaced by the $A_{1/2}$ amplitude as 
the dominant contribution to the total resonance strength. 
For the $N(1720)P_{13}$, the MAID2007 global analysis indicates
just the opposite behavior: $A_{1/2}$ is large at small $Q^2$, and 
$A_{3/2}$ becomes dominant at high $Q^2$. We remark that the $Q^2$ behavior of 
the $N(1720)P_{13}$ state is not consistent with the behavior predicted in pQCD.   

Other states in the third resonance region belong to the 
multiplet $[70,1^-]$ of the $SU(6)\otimes O(3)$ symmetry group.
We discuss the available information on these resonances in the 
following section along with the predictions of the 
single quark transition model (SQTM).

\subsection{\it SQTM predictions for the resonances of the multiplet
$[70,1^-]$ 
\label{sec:sqtm}}

In the approximation that only a single quark is involved in
a resonance transition, simple algebraic relations can be derived 
for the electroexcitation amplitudes for states assigned to the same
$[SU(6),L^P]$ multiplet of the $SU(6)\otimes O(3)$ symmetry
group \cite{Hey_Weyers,Babcock_Rosner,Cottingham,SQTM}.
In this section we present the predictions for
the $\gamma^* p \rightarrow~\Delta(1620)$
$S_{31}$, $N(1650)S_{11}$, $N(1675)D_{15}$, $\Delta(1700)D_{33}$,
and $N(1700)D_{13}$ transverse amplitudes that are
based on the information on the helicity amplitudes
for the resonances $N(1535)S_{11}$ and $N(1520)D_{13}$
reported in this review. All these resonances
belong to the multiplet $[70,1^-]$, and the predictions
are obtained from the relations between the transverse
amplitudes of their electroexcitation 
that follow from the SQTM. 

In the SQTM, the transverse component of the electromagnetic
current consists of four terms:
\be
J^+_{em}=AL^+_q +B\sigma^+L_{q,z} +C\sigma_z L^+_q + 
D\sigma^- L^+_qL^+_q,
\label{eq:sqtm0}\\
\ee
where $\sigma$ is the quark Pauli spin operator, and the terms
with coefficients $A,B,C$, and $D$ operate on the quark
wave function, changing its spin and orbital angular momentum $L_q$;
it is supposed that the $z$-axis is directed along
the direction of the momentum transfer. 
These terms lead, 
respectively, to the following selection rules for the 
transverse transition amplitudes:
\bea
&& \Delta S=0,~~\Delta S_z=0,~~\Delta L_z=\pm 1,
\label{eq:sqtm1}\\
&& \Delta S=1,~~\Delta S_z=\pm 1,~~\Delta L_z=0,
\label{eq:sqtm2}\\
&& \Delta S=1,~~\Delta S_z=0,~~\Delta L_z=\pm 1,
\label{eq:sqtm3}\\
&& \Delta S=1,~~\Delta S_z=\mp 1,~~\Delta L_z=\pm 2,
\label{eq:sqtm4}
\eea
where $S$ and $L$ are the total spin and orbital angular momentum
of the quarks. The non-relativistic CQM results contain
only two terms that correspond to the selection
rules (\ref{eq:sqtm1},\ref{eq:sqtm2}). 

For the multiplet $[70,1^-]$, only three selection rules 
(\ref{eq:sqtm1}-\ref{eq:sqtm3}) give non-zero contributions.
With this, the transitions from the proton to the states
with $S=\frac{3}{2}$ are forbidden. 
These are the states 
$N(1650)S_{11}$ (${}^48_{1/2}$), $N(1675)D_{15}$ (${}^48_{5/2}$), 
and $N(1700)D_{13}$ (${}^48_{3/2}$).
Here we use the notation
${}^{2S+1}SU(3)_{J}$, which gives the assignment of the
state according to the $SU(3)$ group, and $J$ is
the spin of the resonance. The relations between the coefficients 
$A, B$, and $C$  
and the transition helicity amplitudes for 
members of the multiplet $[70,1^-]$ with $S=\frac{1}{2}$ are given
in Table \ref{tab:sqtm}. Using
the empirical information on the 
$\gamma^* p \rightarrow N(1535)S_{11}$
and $N(1520)D_{13}$ transitions,
reported in this review, one can find these coefficients
as functions of $Q^2$.
With this goal, we have parameterized the transition
amplitudes extracted from the experimental
data by the JLab group \cite{Aznauryan2009} in the following way: 

\bea
&&    N(1535)S_{11}:~~~ A_{1/2}=
        \frac{92}{1-0.042\bar{Q}-0.135\bar{Q}^2+0.27\bar{Q}^3},
\label{eq:sqtm5}\\
&&    N(1520)D_{13}:~~~
A_{1/2}=-19.5
\frac{(1+7.12\bar{Q}^2+2.02\bar{Q}^4)}
        {(1+0.1\bar{Q}^3)(1+2.02\bar{Q}^4)},
\label{eq:sqtm6}\\
&&~~~~~~~~~~~~~~~~~~~~~~A_{3/2}=\frac{148}
        {1+2.69\bar{Q}^2+0.14\bar{Q}^4+0.39\bar{Q}^5}.
\label{eq:sqtm7}
\eea
In Eqs. (\ref{eq:sqtm5}-\ref{eq:sqtm7}), the amplitudes are 
given in the  $10^{-3}{\rm GeV}^{-1/2}$
units, and $\bar{Q}\equiv Q/{\rm GeV}$.

\begin{figure}[ht!]
\begin{center}
\begin{minipage}[t]{14.0 cm}
\epsfig{file=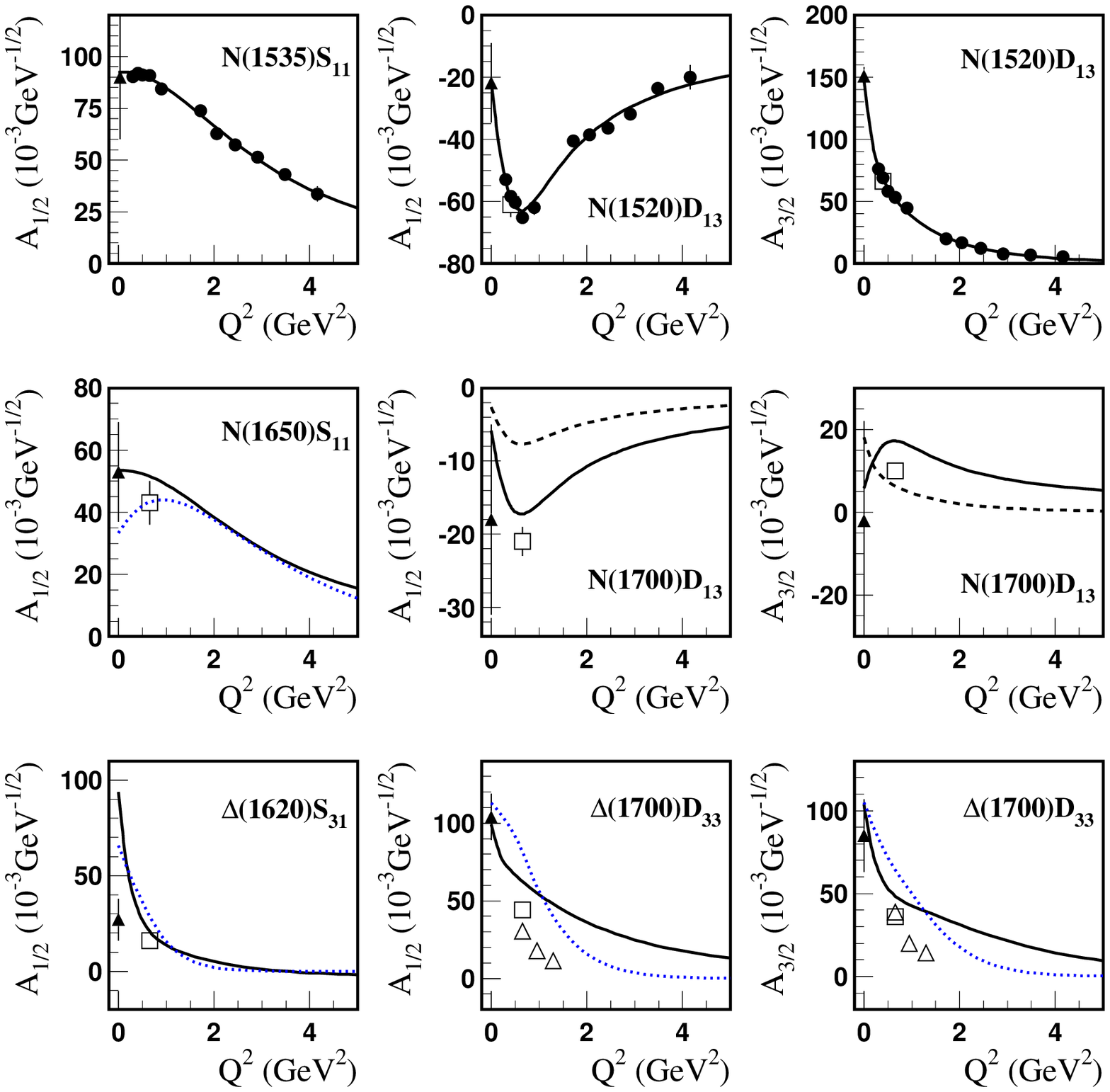,scale=0.8}
\end{minipage}
\begin{minipage}[t]{16.5 cm}
\caption{The
$\gamma^* p \rightarrow N(1535)S_{11}$ and $N(1520)D_{13}$
transitions: the results of the JLab group extracted
from $\pi$ electroproduction data \cite{Aznauryan2009} (solid circles)
and their parameterizations (\ref{eq:sqtm5}-\ref{eq:sqtm7}). 
The solid curves for other resonances are the SQTM predictions.
For the $N(1650)S_{11}$  
they correspond to the mixing
angle $\theta_S=-31^{\circ}$. For the 
$N(1700)D_{13}$, 
the predictions are obtained with the mixing angles 
$\theta_D=6^{\circ}$ (dashed 
lines)
and $\theta_D=14^{\circ}$ (solid lines).
The data at $Q^2=0$ (solid triangles) are from RPP \cite{PDG},
the data at $Q^2=0.65~$GeV$^2$ (open boxes) are from the combined analysis
of the CLAS $\pi$ and $2\pi$ electroproduction data 
\cite{Azn065}, and the open triangles are the results extracted
from the CLAS $2\pi$ electroproduction data
\cite{Mo091,Mokeev_2010}. The dotted curves are from the 
global MAID2007 fit 
\cite{MAID2007,MAID_China}.
\label{sqtm}}
\end{minipage}
\end{center}
\end{figure}

In Fig. \ref{sqtm}, we show the data on the
$\gamma^* p \rightarrow~N(1535)S_{11}$
and $N(1520)D_{13}$ amplitudes
along with the curves obtained using the parameterizations
of Eqs. (\ref{eq:sqtm5}-\ref{eq:sqtm7}).
The predictions obtained for other resonances are shown in Fig. \ref{sqtm}
in comparison with the available information.
For the states ${}^28_{1/2}$,
${}^48_{1/2}$ and ${}^28_{3/2}$, ${}^48_{3/2}$, we include the mixing
angles $\theta_S$ and $\theta_D$, respectively:
\bea
&S_{11}(1535)=\rm{cos}\theta_S |{}^28_{1/2}>-\rm{sin}\theta_S|{}^48_{1/2}>,
~~S_{11}(1650)=\rm{sin}\theta_S |{}^28_{1/2}>+\rm{cos}\theta_S|{}^48_{1/2}>,
\label{eq:sqtm8}\\
&D_{13}(1520)=\rm{cos}\theta_D |{}^28_{3/2}>-\rm{sin}\theta_D|{}^48_{3/2}>,
~~D_{13}(1700)=\rm{sin}\theta_D |{}^28_{3/2}>+\rm{cos}\theta_D|{}^48_{3/2}>.
\label{eq:sqtm9}
\eea
The mixing angle for the $S$-states is taken equal to
$\theta_S=-31^{\circ}$ as found from the hadronic
decays \cite{Isgur_Karl,Hey1975}. 
A much smaller mixing angle has been observed 
for the $D$-states. Here the predictions
are made for two angles.
The angle $\theta_D=6^{\circ}$
is found from the hadronic decays (dashed lines).
We consider also a larger angle $\theta_D=14^{\circ}$
that provides better agreement with the results of the combined
analysis of the
$\pi$ and $2\pi$ electroproduction data  
\cite{Azn065} (solid lines).
\begin{table}
\begin{center}
\begin{minipage}[t]{16.5 cm}
\caption{SQTM predictions for the 
$\gamma^* p\rightarrow [70,1^-]$ helicity amplitudes.
$A,B,C$ are coefficients at the terms that correspond 
to the selection rules (\ref{eq:sqtm1},\ref{eq:sqtm2},\ref{eq:sqtm3}),
respectively.
}
\label{tab:sqtm}
\end{minipage}
\begin{tabular}{lccc}
\hline
& & $$&\\
Resonance& ${}^{2S+1}SU(3)_{J}$& $A_{1/2}$&$A_{3/2}$\\
& & $$&\\
\hline
& & $$&\\
$N(1535)S_{11}$&${}^28_{1/2}$&$\frac{1}{6}(A+B-C)$&\\
& & $$&\\
$N(1520)D_{13}$&${}^28_{3/2}$&$\frac{1}{6\sqrt{2}}(A-2B-C)$&$\frac{1}{2\sqrt{6}}(A+C)$\\
& & $$&\\
$\Delta(1620)S_{31}$&${}^2 10_{1/2}$&$\frac{1}{18}(3A-B+C)$&\\
& & $$&\\
$\Delta(1700)D_{33}$&${}^210_{3/2}$&$\frac{1}{18\sqrt{2}}(3A+2B+C)$&$\frac{1}{6\sqrt{6}}(3A-C)$\\
& & $$&\\
\hline
\end{tabular}
\end{center}
\end{table}

\subsection{\it A new state with $J^P= {3 \over 2}^+$ ?
\label{sec:new_state}}

\begin{figure}[ht!]
\begin{center}
\begin{minipage}[l]{200pt}
\epsfig{file=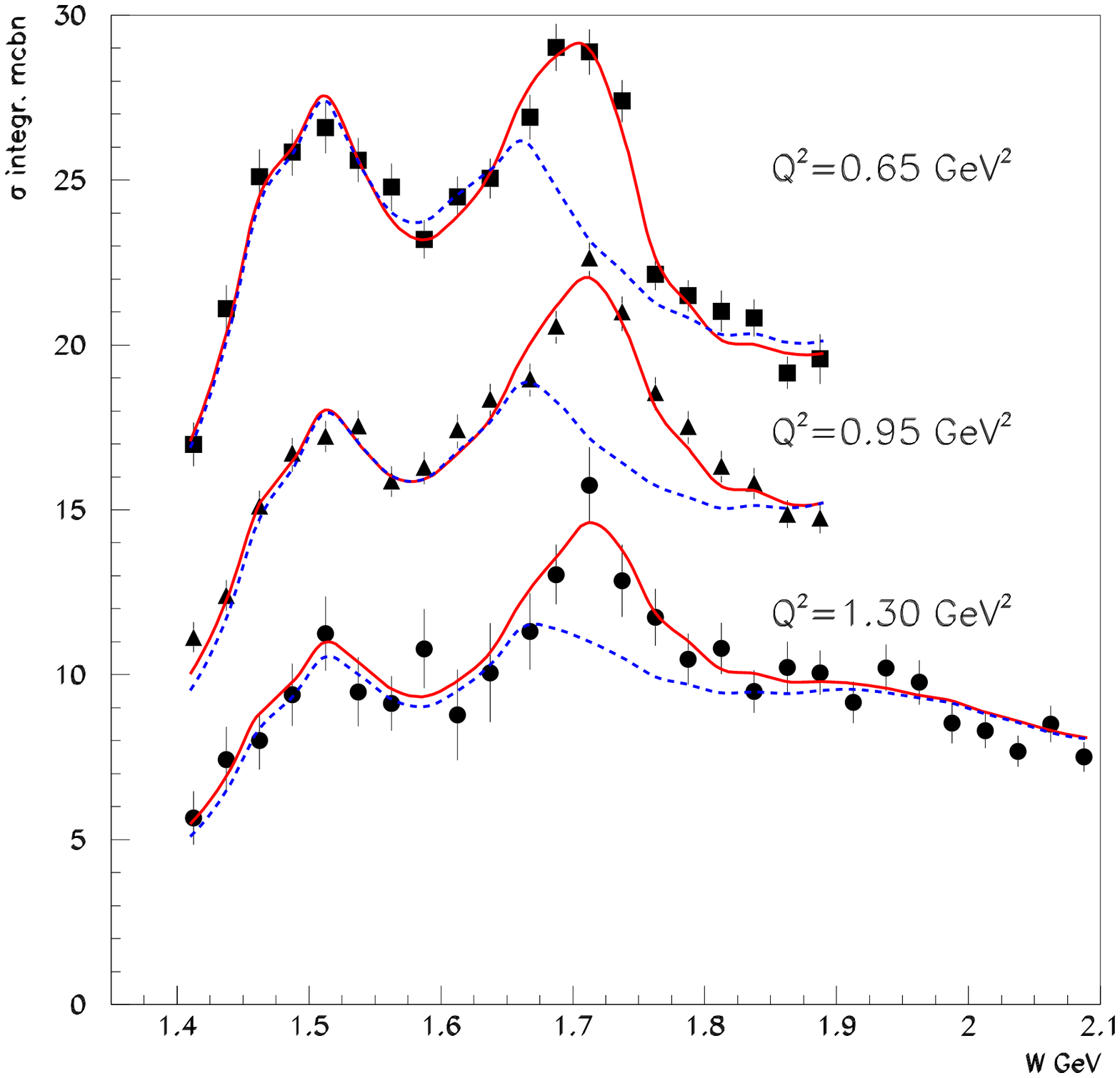,scale=0.6}
\end{minipage}
\hskip 150pt
\begin{minipage}[r]{150pt}
\caption{
Description of the CLAS  data
on fully integrated
$\gamma^*p\rightarrow \pi^-\pi^+p$ cross sections 
from Ref. \cite{Ripani} 
within the framework of
the JM reaction model \cite{Mokeev2009} with parameters fitted 
to the one-fold differential
cross sections.
The calculations,
taking into account the contributions from the conventional $N^*$'s only,
are shown by the dashed lines,
while the solid lines correspond to the fit after
implementation of the ${3\over 2}^+$
candidate state. The difference between the solid and dashed lines
represents a signal from the
possible new state.
\label{new_state}}
\end{minipage}
\end{center}
\end{figure}

In the analysis of the $ep\rightarrow ep\pi^-\pi^+$ data 
from CLAS~\cite{Ripani} possible evidence for a new state at 1720~MeV 
with $J^P={3\over 2}^+$ was found. This state is different
from the $N(1720)P_{13}$ reported by RPP \cite{PDG} in that 
it cannot be described using the resonance parameters from RPP. 
Although the best fit to the one-fold differential
cross sections required a prominent
$3/2^+$ partial wave, it could be attributed to the known
$N(1720)P_{13}$  resonance, if the hadronic couplings 
to the $\pi\Delta$ and $\rho N$ channels
are just the opposite to that given by RPP. Alternatively, a new state,
in addition to the RPP state, should be introduced with about
the same mass and width, but in contrast to the known
state, with large coupling to the 
$\pi\Delta$ channel and suppressed coupling to the $\rho N$ channel.
The signal from the state with such properties
in the fully integrated $\gamma^*p\rightarrow p\pi^-\pi^+$ cross sections
at different $Q^2$ is demonstrated in Fig. \ref{new_state}.

The large branching ratio of the known $N(1720)P_{13}$  state
to the $\rho N$ channel has been found in the analysis of hadronic data
on $\pi N\rightarrow \pi\pi N$ \cite{Longacre75,Longacre77,Manley92}.
Therefore, definite conclusions on the existence of the new state
require a combined analysis of the $\pi N\rightarrow \pi\pi N$
and $ep\rightarrow e\pi^-\pi^+p$ data, including
also the $\pi N\rightarrow \pi N$
and $ep\rightarrow e\pi N$ data.  
Such an analysis within the framework of the coupled-channel approach
is currently underway by the JLab-EBAC group.

It is interesting to add that
a second $P_{13}$ state is predicted in the CQM \cite{Capstick_Isgur},
however at a higher mass of $1870~$GeV. The new state may
have a different internal structure, such as a hybrid baryon
where the gluonic component would be excited. Such a hybrid $P_{13}$ state 
is predicted in the flux-tube model \cite{Capstick_Page}.

\section{Conclusions and Outlook \label{sec:outlook}}

The electroexcitation of baryon resonances is the next step after
measuring the elastic form factors of the proton and neutron.
Owing to the complexity of the excitation spectrum 
and high sensitivity of excited states to the properties of QCD
and to details of quark confinement,
it provides us with rich information 
for their understanding.

In this review we have presented the progress in the investigation
of electroexcitation of nucleon resonances
achieved due to experiments on the new generation
of electron beam facilities.
At this stage, accurate and complete information
has been obtained for the electroexcitation on the proton
of the four lowest excited states,
with the maximal range of $Q^2$ that is allowed 
presently. 
A consistent picture has emerged for the transition
amplitudes extracted 
from the data collected
on different facilities
and setups, and in different reactions. 
High precision of the data, rich experimental information,
and multiple techniques for the analyses of experimental data
allowed for accounting of the model and systematic
uncertainties of the amplitudes. 

The level of the precision and completeness of
the information, including the
nucleon electromagnetic form factors,  has challenged the 
theory in building the
bridge between QCD and the observed properties of the 
nucleon and its excited states.
In the continuation of twenty-years of effort,
there is significant progress in the lattice QCD  
results for the 
$\gamma^* N\rightarrow \Delta(1232)P_{33}$
transition amplitudes. 
However, in spite of qualitative agreement
with empirical amplitudes,
the errors are still large and
the degree of quantitative agreement is not satisfactory.
There are also first exploratory calculations of the
$\gamma^* p\rightarrow N(1440)P_{11}$ amplitudes \cite{Lin}. 
Now 
significant efforts are underway internationally 
to further improve
these results, to involve more excited states,
and to extend computation
of the form factors 
up to $Q^2\sim 10~$GeV$^2$ 
\cite{lattice_beijing,theory_support}.  

A different, continuum perspective 
in computing hadron properties from QCD
is given by the
Dyson-Schwinger equation framework, which has played a growing
role in the investigation of the excited nucleon states
(see reviews \cite{theory_support,roberts1} and references therein).  
It has been recognized that 
the dressed-quark mass dependence on its momentum
found within this framework \cite{roberts2}
and in lattice QCD \cite{bowman} had
an enormous impact on hadron physics.
Definitely, it is a primary goal to understand the role
of this phenomenon in the $Q^2$ dependence of the electromagnetic
form factors.
    
In the review we have discussed
the predictions for the $\Delta(1232)P_{33}$ 
derived in the large $N_c$ limit of QCD in conjuction
with GPD's \cite{Pascalutsa1,Pascalutsa2}
and with the idea of holography \cite{Grigoryan}. 
Good agreement with the amplitudes
extracted from experimental data
is obtained not only
at $Q^2=0$, but also in a wide range of $Q^2$.
Moreover, according to the predictions
\cite{Grigoryan}, $R_{SM}\rightarrow -100\%$
at $Q^2\rightarrow\infty$, 
in agreement with
the rapidly rising magnitude and negative 
value of $R_{SM}$ found
in experiment.  

The precise information on the 
transition amplitudes, extracted from experimental data
in a wide range of $Q^2$, allowed 
mapping out of the quark transverse charge distributions
that induce these transitions \cite{Carlson_Van,Tiator_Van,Tiator_Van1}.
In order for these projections
to correctly reflect the charge distributions at small 
distances, measurements at higher $Q^2$ are
necessary.

Very different results have been demonstrated
in the review when comparing
the $Q^2$ behavior of the transition amplitudes
for different resonances
with the pQCD predictions.
While the $\Delta(1232)P_{33}$
does not show
any tendency of approaching the asymptotic
QCD regime up to $Q^2= 7~$GeV$^2$, the 
$N(1440)P_{11}$, $N(1520)D_{13}$, and
$N(1535)S_{11}$ 
reveal the features specific for pQCD
starting with quite low values of $Q^2\approx 2-3~$GeV$^2$.
The reason of this difference is not clear. 
It can be the form of the asymptotic nucleon
and $\Delta(1232)P_{33}$ wave functions \cite{Carlson}
or the possibility
to describe some features of the pQCD behavior 
at moderate $Q^2$ within
non-perturbative approaches as is the case for the
nucleon \cite{Miller,Aznauryan93}. 
In this connection it should be mentioned
that there is a growing
concensus that soft non-perturbative 
contributions play the dominant
role at present energies (see, for example, Ref. \cite{RAD}). 
This needs further detailed investigation. In particular, 
new measurements at higher $Q^2$ are needed for reliable conclusions.

Comparison with CQM predictions shows that there are
additional non-3-quark
contributions at low $Q^2$  for
all resonances. 
The evaluations of these contributions are still 
limited and include mostly the $\pi N$ component 
in $\gamma^* N\rightarrow \Delta(1232)P_{33}$
\cite{Bermuth,Fiolhais,Thomas,Faessler1}
and $\sigma N$ in
$\gamma^* N\rightarrow N(1440)P_{11}$ \cite{Faessler2011}. 
We expect that a common picture of
the role of the hadronic component in transition amplitudes
will become clear
with the analyses based on the dynamical 
coupled-channel approaches that incorporate hadronic
and electromagnetic channels.
Much progress has recently been made in 
utilizing these approaches
\cite{Sato2007,
Sato2008,Kamano2009,Sato2009,Kamano2010,Sirca2009,Doring2010,Sirca2011}.
We have not included
this promising development
in the review, as much of it has focused on the analysis of
hadronic and photoproduction processes, while electroproduction
processes
are still in the development stage \cite{Sato2009,Sirca2009,
Sirca2011,suzuki}. It should be mentiond also
that the results of
the coupled-channel approaches on the
$\gamma^* N\rightarrow N^*$ transition amplitudes 
are related to
the $K$-matrix poles of the resonances.
Large efforts are necessary to compare these results
with the amplitudes at the $T$-matrix poles
presented in this review and with
model predictions.

Precision information is still limited to the lower
mass states and to low and moderate $Q^2$ values. However, more data
at higher masses will be available soon 
from CLAS that will allow for a significant extension 
of the analysis effort to masses of 2 GeV.

Currently, the study of resonance transition amplitudes has focused
on photon virtualities $Q^2 < 5-8~$GeV$^2$. But even in the long
distance regime open issues remain. 
The pion-cloud contribution of the
$\Delta(1232)P_{33}$
has not been explored to sufficiently small $Q^2$ to obtain a
complete picture of this transition. 
Especially the helicity $S_{1/2}$ amplitude,
which is strongly
$Q^2$ dependent, is not well known 
at $Q^2 < 0.06~$GeV$^2$. A recent experiment at
JLab Hall A ~\cite{Gilad} will shed new light on the low $Q^2$
behavior
of this amplitude. This is also the domain where lattice QCD can make
precise predictions
with realistic pion masses and where the effect of the still neglected
disconnected diagrams
can be tested.

At the highest $Q^2$ that are currently probed, 
the hadronic component is still
significant, but may be rapidly losing strength with increasing
$Q^2$. The domain
where the true quark core may reveal itself 
requires even higher $Q^2$. We will begin to probe the
transition to this
domain only with the Jefferson Lab 12 GeV upgrade. An
experiment \cite{Gothe}
to study several of the prominent nucleon 
resonances at 12 GeV using the
new CLAS12 spectrometer is currently in 
preparation and may take data in 2015.
This will open up a new era in the exploration of excited 
nucleons when the ground state and excited nucleon's
quark core may be more fully exposed to the electromagnetic probe.

 \section{Acknowledgments}
This work was supported by the US Department of Energy under
contract DE-AC05-06OR23177 and the Department of Education 
and Science of
Republic of Armenia, Grant-11-1C015.

\end{document}